\documentclass[a4paper,12pt]{article}
\pdfoutput=1
\usepackage[DIV=14,BCOR=0mm,headinclude=true,footinclude=false]{typearea}
\usepackage[utf8]{inputenc}
\usepackage[T1]{fontenc}
\usepackage{mathalfa,amssymb}

\usepackage[many]{tcolorbox}
\usepackage{empheq}
\usetikzlibrary{shadows}
\tcbset{
  highlight math style={
    colframe=black,
    colback=white,
    arc=2pt,
    boxrule=0.7pt,
  }
}

\usepackage{style}

\usepackage{subcaption}
\usepackage[english]{babel}
\usepackage[numbers,sort&compress]{natbib}

\usepackage{tipa}
\usepackage{graphicx}
\usepackage[bottom]{footmisc}
\usepackage[titletoc]{appendix}
\usepackage{hyperref}
\usepackage{listings}
\usepackage{float}
\usepackage{amsmath}
\usepackage{mathrsfs}
\usepackage{amssymb}
\usepackage{mathtools}
\usepackage{amsfonts}
\usepackage[T1]{fontenc}
\usepackage{braket}
\usepackage{slashed}
\usepackage{dashbox}
\usepackage[colorinlistoftodos]{todonotes}
\usepackage{tikz-feynman}   
\usepackage{tikz}
\usetikzlibrary{snakes}     
\usepackage{datetime}
\usepackage[normalem]{ulem}

\usepackage{enumitem}       

 \newcommand{\virg}{\hspace{1 mm}, \hspace{8 mm}}
\newcommand{\p}{\partial}
 \newcommand{\badat}{\begin{alignedat}}
 \newcommand{\eadat}{\end{alignedat}}
 
\usepackage{scalerel}

\usepackage{pict2e}         

\makeatletter
\DeclareRobustCommand{\loplus}{\mathbin{\mathpalette\dog@lsemi{+}}}
\DeclareRobustCommand{\lotimes}{\mathbin{\mathpalette\dog@lsemi{\times}}}
\DeclareRobustCommand{\roplus}{\mathbin{\mathpalette\dog@rsemi{+}}}
\DeclareRobustCommand{\rotimes}{\mathbin{\mathpalette\dog@rsemi{\times}}}

\DeclareSymbolFont{TOneChars}{T1}{\familydefault}{m}{it}
\SetSymbolFont{TOneChars}{bold}{T1}{\familydefault}{bx}{it} 
\DeclareMathSymbol{\mathdh}{\mathord}{TOneChars}{"F0}

\newcommand{\dog@rsemi}[2]{\dog@semi{#1}{#2}{-90,90}}
\newcommand{\dog@lsemi}[2]{\dog@semi{#1}{#2}{270,90}}
\newcommand{\dog@semi}[3]{%
	\begingroup
	\sbox\z@{$\m@th#1#2$}%
	\setlength{\unitlength}{\dimexpr\ht\z@+\dp\z@\relax}%
	\makebox[\wd\z@]{\raisebox{-\dp\z@}{%
			\begin{picture}(1,1)
				\linethickness{\variable@rule{#1}}
				\roundcap
				\put(0.5,0.5){\makebox(0,0){\raisebox{\dp\z@}{$\m@th#1#2$}}}
				\put(0.5,0.5){\arc[#3]{0.5}}
			\end{picture}%
	}}%
	\endgroup
}
\newcommand{\variable@rule}[1]{%
	\fontdimen8  
	\ifx#1\displaystyle\textfont3\else
	\ifx#1\textstyle\textfont3\else
	\ifx#1\scriptstyle\scriptfont3\else
	\scriptscriptfont3\relax
	\fi\fi\fi
}
\makeatother

\newcommand\reallywidehat[1]{\arraycolsep=0pt\relax%
	\begin{array}{c}
		\stretchto{
			\scaleto{
				\scalerel*[\widthof{\ensuremath{#1}}]{\kern-.5pt\bigwedge\kern-.5pt}
				{\rule[-\textheight/2]{1ex}{\textheight}} 
			}{\textheight} %
		}{0.5ex}\\           
		#1\\                 
		\rule{-1ex}{0ex}
	\end{array}
}

\def\be{\begin{equation}}
	\def\ee{\end{equation}}
\def\ba{\begin{aligned}}
	\def\ea{\end{aligned}}

\newcommand{\thorn}{\text{\textthorn}}
\newcommand{\vev}[1]{\ensuremath{\left\langle #1 \right\rangle}}

\newcommand{\tr}{\text{tr}}

\newcommand*{\defeq}{\mathrel{\vcenter{\baselineskip0.5ex \lineskiplimit0pt
			\hbox{\footnotesize.}\hbox{\footnotesize.}}}%
	=}
\newcommand{\scri}{\mathscr{I}}

\numberwithin{equation}{section}
\numberwithin{table}{section}

\title{Celestial $sw_{1+\infty}$ algebra in Einstein-Yang-Mills theory}

\author[a,b]{Shreyansh Agrawal\footnote{\texttt{sagrawal@sissa.it}}}
\author[a,b]{Panagiotis Charalambous\footnote{\texttt{pcharala@sissa.it}}}
\author[a,b]{Laura Donnay\footnote{\texttt{ldonnay@sissa.it}}}

\affiliation[a]{International School for Advanced Studies (SISSA), \\
Via Bonomea 265, 34136 Trieste, Italy}
\affiliation[b]{National Institute for Nuclear Physics (INFN), \\
Sezione di Trieste, Via Valerio 2, 34127, Italy}
\date{}

\abstract{
	From a study of the subleading structure of the asymptotic equations of motion in Einstein-Yang-Mills theory, we construct charges that are conserved up to quadratic order in non-radiative vacuum. We then show that these higher spin charges obey the celestial $sw_{1+\infty}$ symmetry algebra found earlier from the OPE of positive-helicity conformally soft gluons and gravitons.\\
}

\begin{document}

\maketitle


\section{Introduction}
While it has long been known that the self-dual sector of four-dimensional gravity is governed by the $w_{1+\infty}$ algebra \cite{Boyer:1985aj,Boyer:1983,Penrose:1976jq,Penrose:1976js,Park:1989fz}, it is only recently that the latter was related to universal features of gravitational scattering amplitudes \cite{Strominger:2021mtt}. In the context of celestial holography\footnote{See \cite{Pasterski:2021rjz,Raclariu:2021zjz,McLoughlin:2022ljp,Donnay:2023mrd} for reviews.}, the infinite tower of conformally soft graviton currents~\cite{Donnay:2018neh,Adamo:2019ipt,Puhm:2019zbl,Guevara:2019ypd} can be indeed organized, in the positive helicity sector and at tree level, in terms of the following algebra~\cite{Guevara:2021abz,Strominger:2021mtt}
\begin{equation}\label{eq:w_alg}
\{w_m^p, w_n^{q}\} = (m(q-1)-n(p-1))w_{m+n}^{p+q-2}\,.
\end{equation}
In the above, the mode indices $p,q$ run over the positive half-integers $p=1,\frac{3}{2}, 2, \frac 5 2,\dots$ and their value is inherited from the conformal dimension $\Delta$ of celestial graviton current operators via $\Delta_{\text{gr}}=4-2p$. The indices $m,n$ are required to belong to the range $1-p\leq m \leq p-1$, making \eqref{eq:w_alg} the wedge algebra of $w_{1+\infty}$ (see \cite{Pope:1991ig} for a review on $W$ algebras). The action of these symmetries on arbitrary spin massless celestial primaries and massive scalars was studied in \cite{Himwich:2021dau,Himwich:2023njb}.

For non-abelian gauge theories, an analog infinite tower of symmetries is organized into the so-called $s$-algebra~\cite{Guevara:2021abz,Strominger:2021mtt},
\begin{equation}
\badat{2}\label{eq:s_alg}
       & \{s_{n'}^{p,b}, s_n^{q,a}\} = i f^{ab}_{\,\,\,c}\,s^{q+p-1,c}_{n+n'}\,,
\eadat
\end{equation}
where $a,b$ are color indices and $p=1,\frac{3}{2}, 2, \frac 5 2,\dots$ is related to the conformal dimension of the celestial gluon operator as $\Delta_{\text{gl}}=3-2p$. For conformally soft photons, the above bracket vanishes, but both photon and gluon currents couple to gravitons and obey the following commutation relations with the $w$-generators
\begin{equation}
\badat{2}\label{eq:sw_alg}
       & \{w_m^p, s_n^{q,a}\} = (m(q-1)-n(p-1))s_{m+n}^{p+q-2,a}\,.
\eadat
\end{equation}

The above `$sw_{1+\infty}$ algebra', defined by the commutation relations \eqref{eq:w_alg}, \eqref{eq:s_alg} and \eqref{eq:sw_alg}, was originally derived from collinear limits of celestial operator product expansions (OPEs) of positive helicity gravitons and gluons~\cite{Pate:2019lpp,Guevara:2021abz,Strominger:2021mtt}. The goal of this paper is to show, by an analysis of the subleading structure of equations of motion, that the same algebra is encoded into the asymptotic phase space of Einstein-Maxwell and Einstein-Yang-Mills theory. 

Such an analysis is rooted in the works \cite{Freidel:2021ytz,Geiller:2024bgf} who showed, in the pure gravity case, how the $w_{1+\infty}$ algebra could be seen as arising from a tower of gravitational charges of spin weight $s=2p-3$ which encode the tower of sub$^s$-leading soft gravitons. 
The key observation made in \cite{Freidel:2021ytz} is that the $w_{1+\infty}$ algebra can be realized by an appropriate truncation of the asymptotic phase space which allows to build higher spin weight charges, denoted $\mathcal Q_s$, which evolve according to a set of recursion relations. However, the action of these higher spin charges on the gravitational data is generically divergent. The next key step requires to construct renormalized charges $\tilde q_s$ which have, on the one hand, a finite action as $u\to -\infty$ ($u$ denoting the retarded time at future null infinity) and, on the other hand, are conserved in the non-radiative vacuum at quadratic order (namely conserved up to corrections involving cubic or higher contributions in the fields). The smeared linearized bracket of the quasi-conserved higher spin charges is then seen to close the $w_{1+\infty}$ algebra given by the commutation relations \eqref{eq:w_alg}. In the case of pure Yang-Mills theory, analog higher spin celestial charges were constructed in \cite{Freidel:2023gue} and shown to realize the $s$-algebra brackets \eqref{eq:s_alg}; see also \cite{Nagy:2024jua} for a study of Yang-Mills recursion relations. The authors of \cite{Freidel:2023gue} also pushed their canonical analyses beyond the linear truncation and discussed the closure of the Yang-Mills algebra at quadratic order. For gravity, the nonlinear realization of the $w_{1+\infty}$ algebra, together with a Noether charge description of higher spin charges very recently studied in \cite{Cresto:2024fhd,Cresto:2024mne}.

In the first part of this paper, we show that the conformally soft photon and graviton mixed current algebra \eqref{eq:sw_alg} can be derived from an asymptotic charge algebra analysis through the construction of quasi-conserved quantities in Einstein-Maxwell theory. In the second part, we show that the generalization of our results to Einstein-Yang-Mills allows us to recover the $s$-algebra for gluons \eqref{eq:s_alg}, together with the extension of the mixed gauge-gravity bracket \eqref{eq:sw_alg} to the non-abelian case. 

The main results and outline of the paper can be summarized as follows. 
We start in Section \ref{sec:Setup} by setting up our notations for the Einstein-Maxwell theory and its associated asymptotic phase space. Using the Newman-Penrose (NP) formalism, we also study the evolution equations of the higher electromagnetic and Bondi aspects\footnote{They contain the so-called `Chthonian' degrees of freedom~\cite{Campoleoni:2023fug}.}, denoted $\Phi_0^n$ and $\Psi_0^n$, respectively. In Section \ref{sec:recursion_Max}, we show the existence of recursion relations \eqref{eq:rec1}, \eqref{eq:rec2} for the electromagnetic and gravitational higher spin charges  $\mathcal Q_s^{em}$, $\mathcal Q_s^{gr}$,  defined implicitly from $\Phi_0^n$ and $\Psi_0^n$ by means of  \eqref{eq:def_higher_spin}, \eqref{eq:Qgr}. In section \ref{sec:quasi_Max}, we construct electromagnetic and gravitational regularized charges \eqref{eq:DefRenormalized} that are quadratic in the radiative data, and show that they are quasi-conserved, namely conserved in the non-radiative vacuum (at quadratic order). The Poisson algebra is derived in Section \ref{sec:algebra_Max} by computing the linearized brackets of these quasi-conserved higher spin weight charges and using canonical relations on the radiative phase space. The final charge algebra obtained, written in \eqref{eq:final}, is shown to match with the $sw_{1+\infty}$ brackets in the abelian case. We generalize our results to the case of Einstein-Yang-Mills (EYM) in Section \ref{sec:w1pInftyEYM}, following the same protocol. Our main results are given by \eqref{eq:EvolutionQsEYM}, which provide unified expressions for the recursion relations for the full EYM system, together with the set of graviton-gluon charge brackets \eqref{EYM_bracket}, which is shown to led to the full $sw_{1+\infty}$ algebra. Section \ref{ref:summary} contains a discussion and further perspectives for this work. We collected details about the NP formalism in Appendix \ref{app:NPFormalism},  useful identities in Appendix \ref{app:MathIds}, and further computational details in Appendices \ref{app:tqsCons} and \ref{app:CanBraAntiDer}.

\section{Einstein-Maxwell NP scalars}
\label{sec:Setup}
In this Section, we set up the notations and review some formalism in the asymptotically flat spacetimes useful for our analysis. In Bondi coordinates $(u, r, x^A)$, we can write the line element as~\cite{Bondi:1962px,Sachs:1961zz},
\begin{equation}
    ds^2 = \frac{V}{r}e^{2B}du^2 - 2 e^{2B}dudr + g_{AB}(dx^A - U^Adu)(dx^B - U^Bdu)\,,
\end{equation}
where $V$, $U^A$ and $B$ are functions of all coordinates. We will work in the so-called partial Bondi gauge \cite{Geiller:2022vto}, namely only require the gauge conditions $g_{rr}=0=g_{rA}$. Further, once the fall-off conditions on the spatial metric are chosen\footnote{We are assuming here that no logarithms enter in this asymptotic expansion. In partial Bondi gauge, this requires the condition $D_{\vev{AB}} = \frac{1}{4}CC_{\vev{AB}}$, where $C\defeq q^{AB}C_{AB}$ and ``$\vev{...}$'' indicates the symmetric-tracefree part with respect to the boundary metric, e.g. $C_{\vev{AB}} = C_{AB} - \frac{1}{2}q_{AB}C$; this condition turns out to also ensure that the peeling theorem holds (see \cite{Geiller:2024ryw} and references therein for a discussion beyond peeling).},
\be
    g_{AB}(u,r,x^{C}) = r^2 q_{AB}(x^{C}) + r\,C_{AB}(u,x^{C}) + D_{AB} + \mathcal{O}\left(r^{-1}\right) \,,
\ee
where $C_{AB}$ is the gravitational shear, the conditions of asymptotic flatness $R_{rA},R_{ur}$, $g^{AB}R_{AB} = \mathcal{O}\left(r^{-3}\right)$ and the field equations determine the fall-off behaviors of the remaining metric components to be
\be
    g_{uu} = \mathcal{O}\left(1\right) \,,\quad  g_{ur} = -1 + \mathcal{O}\left(r^{-2}\right) \quad\text{and}\quad g_{uA} = \mathcal{O}\left(1\right) \,.
\ee
Similar fall-off conditions can also be taken for the  gauge field components (see e.g. \cite{Janis:1965tx,Exton:1969im}),
\be\label{eq:Maxwellbc}
\badat{2}
		&A_{u}(u,r,x^{A}) = \mathcal{O}\left(r^{-1}\right) \,,\quad A_{r}(u,r,x^{A}) = \mathcal{O}\left(r^{-2}\right) \,, \\
	&	A_{A}(u,r,x^{B}) = A_{A}^{\left(0\right)}(u,x^{B}) + \mathcal{O}\left(r^{-1}\right) \,.
\eadat
\ee
We will be imposing these boundary conditions for the action of Einstein-Maxwell theory\footnote{Compared to the Maxwell action using canonical field variables, $S_{\text{Maxwell}} = \int d^4x\sqrt{-g}\left[-\frac{1}{4}F_{\text{can}}^2\right]$, here we have performed the field redefinition $F_{\text{can}} = \frac{1}{\kappa}F$.},
    \be
        S = \frac{1}{2\kappa^2}\int d^4x \left[R - \frac{1}{2}F^2\right]
    \ee
with $\kappa^2 = 8\pi G$. The approach we will be taking is not to solve the complete Einstein-Maxwell equations for the fields but only to study the asymptotic forms of these equations near null infinity ($\mathscr I$). We will study these equations in the Newman-Penrose (NP) formalism to extract useful information efficiently; see Appendix \ref{app:NPFormalism} for a review. We will make the following choice of tetrad vectors,
\be\label{eq:tetradScri}
	\ell = \partial_{r} \,,\quad n = e^{-2B}\left(\partial_{u} + \frac{V}{2r}\partial_{r} + U^{A}\partial_{A}\right) \,,\quad m = \frac{1}{r}e^{A}\partial_{A} \,,\quad \bar{m} = \frac{1}{r}\bar{e}^{A}\partial_{A} \,,
\ee
where all the inner products among the tetrad vectors are zero except $\ell \cdot n = -1$ and $m\cdot\bar{m} = +1$. The dyad vector $e^{A} = e^{A}(u,r,x^{B})$ is a complex dyad for the $2d$ inverse spatial metric $r^2g^{AB}$, i.e. $r^2g^{AB} = e^{A}\bar{e}^{B} + \bar{e}^{A}e^{B}$, such that and $g_{AB}e^{A}e^{B} = 0$ and $r^{-2}g_{AB}e^{A}\bar{e}^{B} = +1$. This complex dyad itself admits an asymptotic expansion of the form
\be
	e^{A}(u,r,x^{B}) = \varepsilon^{A}(x^{B}) + \frac{1}{r}\sum_{n=0}^{\infty}\frac{e^{\left(n\right)A}(u,x^{B})}{r^{n}} \,,
\ee
with the zeroth order part satisfying $q^{AB} = 2\varepsilon^{(A}\bar{\varepsilon}^{B)}$. Once the choice of tetrad is made, the $6$ independent components of the field strength tensor $F_{\mu\nu}$ of a vector gauge field are traded for the $3$ complex Maxwell-NP scalars
\be
	\Phi_0 \defeq  F_{\ell m} \,,\quad \Phi_1 \defeq \frac{1}{2}\left(F_{\ell n} - F_{m\bar{m}}\right) \,,\quad \Phi_2 \defeq F_{\bar{m}n} \,,
\ee
while the $10$ independent components of the Weyl tensor $C_{\rho\sigma\mu\nu}$ are rearranged into the $5$ complex Weyl-NP scalars,
\be
\badat{2}
	&\Psi_0 \defeq -C_{\ell m \ell m} \,,\quad \Psi_1 \defeq -C_{\ell m \ell n} \,,\quad \Psi_2 \defeq -C_{\ell m \bar{m}n} \,,\\
    & \Psi_3 \defeq -C_{\ell n\bar{m} n} \,,\quad \Psi_4 \defeq -C_{\bar{m}n\bar{m}n} \,.
    \eadat
\ee
The boundary conditions amount to considering retarded multipole solutions for the highest-spin weight NP scalars of the form,
\be\ba\label{eq:expansions}
    \Phi_0\left(u,r,x^{A}\right) &= \sum_{s=0}^{\infty}\frac{1}{r^{3+s}}\Phi_0^{s}(u,x^{A}) \,, \\
    \Psi_0\left(u,r,x^{A}\right) &= \sum_{s=0}^{\infty}\frac{1}{r^{5+s}}\Psi_0^{s}(u,x^{A}) \,,
\ea\ee
which then imply `peeling' for the remaining NP scalars~\cite{Newman:1961qr,Newman:1962cia,Janis:1965tx,Exton:1969im},
\begin{equation}
\label{eq:npexpand}
    \begin{split}
        \Phi_1 &= r^{-2}\Phi_1^0 - r^{-3}\bar{\eth}\Phi_0^0 + \mathcal{O}\left(r^{-4}\right) \,, \\
        \Phi_2 &= r^{-1}\Phi_2^0 - r^{-2}\bar{\eth}\Phi_1^0 + \mathcal{O}\left(r^{-3}\right)\,, \\
        \Psi_1 &= r^{-4}\Psi_1^0 - r^{-5}\left(\bar{\eth}\Psi_0^0 - 3\bar{\Phi}_1^0\Phi_0^0\right) + \mathcal{O}\left(r^{-6}\right) \,, \\
        \Psi_2 &= r^{-3}\Psi_2^0 - r^{-4}\left(\bar{\eth}\Psi_1^0 - 2\bar\Phi_1^0\Phi_1^0\right) + \mathcal{O}\left(r^{-5}\right) \,, \\
        \Psi_3 &= r^{-2}\Psi_3^0 - r^{-3}\left(\bar{\eth}\Psi_2^0 - \bar\Phi_1^0\Phi_2^0\right) + \mathcal{O}\left(r^{-4}\right) \,, \\
        \Psi_4 &= r^{-1}\Psi_4^0 - r^{-2}\bar{\eth}\Psi_3^0 + \mathcal{O}\left(r^{-3}\right) \,. \\
    \end{split}
\end{equation}
In the above expressions, all near-$\scri$ modes are functions of $u$ and $x^{A}$. The subleading near-$\scri$ modes of the NP scalars of lower spin weights are fixed by the hypersurface equations of motion \eqref{eq:MaxwelFE2} and \eqref{eq:BianchiIds2}, while $\eth$ and $\bar{\eth}$ are the GHP ``edth'' operators with respect to the $2$-dimensional boundary transverse space $q_{AB}dx^{A}dx^{B}$\footnote{Note that, in our conventions, $\eth^{\text{here}} = \eth^{\text{\tiny\cite{Geiller:2024bgf}}} = -\eth^{\text{\tiny\cite{Exton:1969im}}}$.}.

 The leading near-$\scri$ modes of the lowest-spin weight NP scalars are related to the radiative fields according to
\be\ba
    \Phi_2^0 &= -\bar{F} = -\partial_{u}\bar{A}\,, \\
    \Psi_4^0 &= \partial_{u}\bar{N} = \partial_{u}^2\bar{C}\,,
\ea\ee
where we have defined the negative helicity photon field $\bar A$ (and field strength $\bar F=\p_u\bar A$) and the negative helicity gravitational shear $\bar C$ (of news $\bar N=\p_u \bar C$) as
\be
    \begin{gathered}
        \bar{A} \defeq \bar{\varepsilon}^{A}A_{A}^{\left(0\right)} \,,\quad \bar{F} \defeq \bar{\varepsilon}^A F_{uA}^{(0)} = \partial_{u}\bar{A} \,, \\
        \bar{C} \defeq \frac{1}{2}\bar{\varepsilon}^{A}\bar{\varepsilon}^{B}C_{AB} \,,\quad \bar{N} \defeq \frac{1}{2}\bar{\varepsilon}^{A}\bar{\varepsilon}^{B}N_{AB} = \partial_{u}\bar{C} \,.
    \end{gathered}
\ee
Similarly, the positive helicity counterparts, $A$, $F$, $C$ and $N$, are obtained by replacing $\bar{\varepsilon}^{A}$ with $\varepsilon^{A}$ in the above expressions\footnote{We remark here that, compared to~\cite{Freidel:2021ytz,Geiller:2024bgf},
\begin{equation*}
    C^{\text{here}} = \frac{1}{2}C^{\text{\tiny\cite{Freidel:2021ytz}}} = -C^{\text{\tiny\cite{Geiller:2024bgf}}} \quad\text{and}\quad N^{\text{here}} = -\frac{1}{2}\bar{N}^{\text{\tiny\cite{Freidel:2021ytz}}} = -N^{\text{\tiny\cite{Geiller:2024bgf}}} \,.
\end{equation*}}.

\section{Recursion equations}
\label{sec:recursion_Max}

We start with the pure Maxwell theory. The Maxwell-NP equations~\eqref{eq:MaxwelFE1}-\eqref{eq:MaxwelFE2} can be combined to get the following separated second order equation for $\Phi_0$
\be
    \frac{1}{r}\partial_{u}\partial_{r}\left(r^3\Phi_0\right) = \eth\bar{\eth}\Phi_0 + \frac{R\left[q\right]}{4r}\partial_{r}\left(r^2\partial_{r}\left(r\Phi_0\right)\right) \,,
\ee
where $R\defeq R\left[q\right]$ is the Ricci scalar of the boundary metric, which we take to be constant, and we are working in the absence of Maxwell sources. In terms of the near-$\scri$ modes $\Phi_0^{s}$ introduced in the previous section, these read~\cite{Newman:1968uj}, 
\begin{equation}
    \dot{\Phi}_0^{s+1} = -\frac{1}{s+1}\left(\bar{\eth}\eth+\frac{R}{4}s\left(s+3\right)\right)\Phi_0^{s}\,,
\end{equation}
where the dot stands for the $u$-derivative and we have made use of the property
\be\label{eq:ethComm}
    \left[\bar{\eth},\eth\right]\varphi_{s} = s\frac{R}{2}\varphi_{s}
\ee
for an object object $\varphi_{s}$ of spin weight $s$. To proceed, we rewrite the expansion \eqref{eq:expansions} of $\Phi_0$ as
\be\label{eq:def_higher_spin}
    \Phi_0 = \frac{\mathcal{Q}^{em}_1}{r^3} + \sum_{s=2}^\infty\frac{1}{r^{s+2}}\frac{(-1)^{s+1}}{\left(s-1\right)!}\left(\bar{\eth}^{s-1} \mathcal{Q}^{em}_s + \tilde{\Phi}_0^{s-1}\right) ,
\ee
where in the first term $\mathcal{Q}^{em}_1=\Phi_0^0$.
The second term defines (in an implicit way) what will be referred to as the electromagnetic higher spin charges $\mathcal{Q}_{s}^{em}$, whose gravitational analogs were first introduced in \cite{Freidel:2021ytz,Geiller:2024bgf}. These charges are of higher spin weight, namely, $\mathcal{Q}_{s}^{em}$ has spin weight $s$. The last term in \eqref{eq:def_higher_spin} contains the remaining pieces of $\Phi_0^{s}$; for pure Maxwell theory, all $\tilde \Phi_0$'s are linear in the fields. Recall that the operator $\eth$ (resp. $\bar{\eth}$) raises (resp. lowers) the spin weight by one unit, hence all the terms in this expansion have the correct spin weight.

\noindent The decomposition of $\Phi_0^{s}$ in terms of spin weighted spherical harmonics starts from $\ell=1$. Then at linear order, \eqref{eq:def_higher_spin} can be viewed as splitting this decomposition into a part that starts from $\ell=s$ (the higher spin charges $\mathcal{Q}_{s}^{em}$) and a remaining part with $1\le\ell\le s-1$ (the $\tilde{\Phi}_0^{s-1}$ part). The above allows us to write a recursion relation for the higher spin charges separately\footnote{Note also that decomposing the equations for $\tilde{\Phi}_0^{s}$ we get $\dot{\tilde{\Phi}}_{0(s,m)}^{s}=0$, where $\tilde{\Phi}_{0(\ell,m)}^{s}$ are the spherical harmonic modes of $\tilde{\Phi}_0^{s}$; these are the linearized NP constants for electrodynamics \cite{Newman:1968uj}.},
\begin{equation}
        \bar{\eth}^{s+1}\dot{\mathcal{Q}}_{s+2}^{em} = \left(\bar{\eth}\eth+\frac{R}{4}s\left(s+3\right)\right)\bar{\eth}^{s}\mathcal{Q}_{s+1}^{em} = \bar{\eth}^{s+1}\eth\mathcal{Q}_{s+1}^{em} \, ,
\end{equation}
where we have repeatedly used~\eqref{eq:ethComm}. This then implies the linear recursion relation for the pure Maxwell sector,
\begin{equation}
    \label{eq:linearqem}
    \dot{\mathcal{Q}}_{s}^{em} = \eth \mathcal{Q}_{s-1}^{em} \,,\quad s\ge+2 \,.
\end{equation}
This relation can in fact be extended to also include the cases $0\le s\le+1$ after defining the corresponding electromagnetic charges as
\be
    \mathcal{Q}_{-1}^{em} \defeq \Phi_2^0 \,,\quad \mathcal{Q}_0^{em} \defeq \Phi_1^0 \,,
\ee
\begin{equation}
\label{eq:fprrecursion}
    \dot{\mathcal{Q}}_{s}^{gr} = \eth \mathcal{Q}_{s-1}^{gr} + \left(s+1\right)C\mathcal{Q}_{s-2}^{gr}\,.
\end{equation}
The above equation is modified once electromagnetic interactions are included, as we shall now discuss.

We now go beyond the (linear) pure Maxwell case and consider the (non-linear) Einstein-Maxwell system. The relevant equations of motion for the Maxwell field coupled to gravity are~\eqref{eq:MaxwelFE1} in the absence of electromagnetic sources, while, for the gravitational field, one looks at the Bianchi identities~\eqref{eq:BianchiIds1} with the gravitational sources coming from the Maxwell stress-energy-momentum tensor according to the Einstein field equations~\eqref{eq:EFE_Max}. Expanding all of these equations in terms of the near-$\scri$ modes of the NP scalars, the first few orders read~\cite{Exton:1969im},
\begin{equation}
\label{eq:npEvolutionEquations}
    \badat{2}
       \dot{\Psi}_0^0 - \eth \Psi_1^0 - 3 C \Psi_2^0 + 3\Phi_0^0 \bar{\Phi}_2^0 &= 0\,, \\
       \dot{\Psi}_0^1 + 4 \bar{\eth} \left(C \Psi_1^0\right) + \bar{\eth}\eth\Psi_0^0 + 4 \bar{\Phi_1^0} \eth \Phi_0^0 + 8 C \Phi_1^0 \bar{\Phi}_1^0 + 4 \Phi_0^1 \bar{\Phi}_2^0 &= 0 \,,\\
        \dot{\Psi}_1^0 - \eth \Psi_2^0 - 2 C \Psi_3^0 + 2 \Phi_1^0 \bar{\Phi}_2^0&= 0 \,,\\
     \dot{\Psi}_2^0 - \eth \Psi_3^0 - C \Psi_4^0 + \Phi_2^0 \bar{\Phi}_2^0 &= 0 \,,\\
     \dot{\Phi}_0^0 - \eth\Phi_1^0 - C \Phi_2^0 &= 0 \,,\\
       \dot{\Phi}_0^1 + \bar{\eth}\eth\Phi_0^0 + 2\bar{\eth}\left(C \Phi_1^0\right) &= 0 \,,\\
       \dot{\Phi}_1^0 - \eth \Phi_2^0 &= 0\,.
\eadat
\end{equation}
The data needed to determine a solution of the Einstein-Maxwell system are $\Psi_1^0$, $\Psi_2^0$, $\Psi_0^{s}$ and $\Phi_1^0$, $\Phi_2^0$, $\Phi_0^{s}$, plus the free data of the gravitational shear $C$ and the asymptotic photon field $F$ \cite{Janis:1965tx}. 
Now, by inspecting the form of subleading terms in \eqref{eq:npexpand} and following the treatment of \cite{Geiller:2024bgf,Freidel:2021ytz}, we write the expansion of $\Psi_0$ in terms of higher spin charges $\mathcal{Q}_s$ as,
\begin{equation}\label{eq:Qgr}
    \begin{split}
        \Psi_0 &= \frac{\mathcal{Q}^{gr}_2}{r^5} + \sum_{s=3}^\infty \frac{1}{r^{s+3}}\frac{(-1)^s}{\left(s-2\right)!}\left(\bar{\eth}^{s-2} \mathcal{Q}^{gr}_s - \left(s+1\right)\bar{\mathcal{Q}}_0^{em}\bar{\eth}^{s-3}\mathcal{Q}_{s-1}^{em} + \tilde{\Psi}_0^{s-2}\right) \,,
    \end{split}
\end{equation}
which reduces to the expression in \cite{Geiller:2024bgf} in the absence of electromagnetic charges.
In the full Einstein-Maxwell theory, the $\tilde{\Psi}_0^s$ and $\tilde{\Phi}_0^s$ in \eqref{eq:def_higher_spin} are some generic functions, including possibly non-linear and non-local quantities. We will however see that these do not contribute to the first few charge aspects but might appear at further order. For the remaining NP near-$\scri$ modes, we also write
\be\badat{2}\label{eq:QsMin}
   &\Phi^0_{1-s}=\mathcal Q_{s}^{em} \virg -1\leq s \leq 1\,,\\
   &\Psi^0_{2-s}=\mathcal Q_{s}^{gr} \virg -2\leq s \leq 2\,.
\eadat\ee
With this, one can study the evolution equations \eqref{eq:npEvolutionEquations} in terms of the higher spin charges. We see that, for $s=0,1,2$, the following equations hold
\begin{empheq}[box=\tcbhighmath]{align}
\label{eq:rec1}
\dot{\mathcal{Q}}_s^{em} = \eth \mathcal{Q}_{s-1}^{em} + s C \mathcal{Q}_{s-2}^{em}\,,
\end{empheq}
where the extra term compared to \eqref{eq:linearqem} encodes a non-linear contribution due to the gravitational interactions.
We also find that, for $s=-1,0,1,2,3$, 
\begin{empheq}[box=\tcbhighmath]{align} \label{eq:rec2}
\dot{\mathcal{Q}}^{gr}_s = \eth\mathcal{Q}^{gr}_{s-1} + \left(s+1\right) C \mathcal{Q}^{gr}_{s-2} + \left(s+1\right)F \mathcal{Q}^{em}_{s-1}\,,
\end{empheq}
where the last term generalizes \eqref{eq:fprrecursion} to include the presence of Maxwell fields.

The insight of \cite{Freidel:2021ytz} was the realization that the $w_{1+\infty}$ algebra appears when extending the validity of the higher spin charges recursion relations for \emph{all} values of $s$. In the Maxwell-Einstein case, the role of the $\tilde{\Psi}_0^s$ and $\tilde{\Phi}_0^s$ quantities\footnote{As noted in \cite{Geiller:2024bgf}, a clever choice of tetrad might result in simplification of the expression for these auxiliary functions.} is thus precisely to absorb all relevant terms in such a way that the electromagnetic and gravitational charges obey, by definition, the recursions relations~\eqref{eq:rec1} for all $s\geq0$ and~\eqref{eq:rec2} for all $s\ge-1$, respectively. While a systematic understanding of this mechanism is still lacking,  the relation between the gravitational recursion relations and equations for self-dual gravity in twistor space was shown in \cite{Kmec:2024nmu}.
In the next sections, we will show that the definition of recursion higher spin charges in Einstein-Maxwell satisfying \eqref{eq:rec1}, \eqref{eq:rec2} (and, in Section \ref{sec:w1pInftyEYM}, their generalization for Einstein-Yang-Mills) allows to derive the $sw_{1+\infty}$ algebra.

\section{Quadratic quasi-conserved charges}
\label{sec:quasi_Max}
The radiative degrees of freedom are encoded in the negative spin weight charges,
\be \label{eq:rad}
\badat{2}
&\mathcal Q_{-1}^{em}=-\bar F\virg 
\mathcal Q_{-1}^{gr}=\eth \bar N\virg \mathcal Q_{-2}^{gr}=\p_u \bar N \,.
\eadat
\ee 
In the absence of radiation, namely, when the above quantities vanish, only the spin weight-zero charges $\mathcal{Q}_0^{em}$, $\mathcal{Q}_0^{gr}$ are conserved, according to the evolution equations~\eqref{eq:rec1},\eqref{eq:rec2},\footnote{When dropping the $^{em}$ and $^{gr}$ superscripts we mean that the relation applies for both the electromagnetic and gravitational charges.}
\be
	\partial_{u}\mathcal{Q}_0\overset{\text{non-rad}}{=} 0 \,.
\ee
Nevertheless, one can perturbatively construct combinations of the recursion charges and the boundary gauge/metric fields that satisfy quasi-conservation equations, i.e., that are conserved in the absence of radiation, for all $s\ge0$. Building on the works of~\cite{Freidel:2021ytz,Geiller:2024bgf}, let us then consider the following `renormalized charge aspects'
    \begin{equation}
    \begin{split}
\label{eq:DefRenormalized}
   \tilde{q}_s^{em} &\!= \sum_{n=0}^s \frac{(-u)^n}{n!}\eth^n \mathcal Q^{em}_{s-n} -\sum_{\ell=2}^s \sum_{n=0}^{\ell-2} \frac{(-1)^n \ell}{(s-\ell)!} \eth^{s-\ell}(\partial_u^{-(n+1)}((-u)^{s-\ell}C)\eth^n \mathcal{Q}^{em}_{\ell-2-n})+\mathcal{O}(\mathbb{F}^3), \\
   \tilde{q}_s^{gr} &\!= \sum_{n=0}^s \frac{(-u)^n}{n!}\eth^n \mathcal{Q}^{gr}_{s-n} - \sum_{\ell=2}^s \sum_{n=0}^{\ell-2} \frac{(-1)^{n}(\ell+1)}{(s-\ell)!} \eth^{s-\ell}(\partial_u^{-(n+1)}((-u)^{s-\ell}C)\eth^n \mathcal{Q}^{gr}_{\ell-2-n}) \\
   &-\sum_{\ell=1}^s \sum_{n=0}^{\ell-1} \frac{(-1)^{n}(\ell+1)}{(s-\ell)!} \eth^{s-\ell}(\partial_u^{-(n+1)}((-u)^{s-\ell}F)\eth^n \mathcal{Q}^{em}_{\ell-1-n})+\mathcal{O}(\mathbb{F}^3) \,,
    \end{split}
    \end{equation}
where $s\ge0$ and the notation $\mathcal{O}(\mathbb{F}^3)$ emphasizes that we are working up to quadratic order, suppressing potential terms that are of cubic or higher order in the fields.
As we demonstrate in Appendix C, \eqref{eq:rec1} and \eqref{eq:rec2} imply that these charges evolve in time as 
\begin{equation}
\label{eq:RrnormalizedEM}
    \begin{split}
    &\partial_u\tilde{q}_s^{em} = \frac{(-u)^s}{s!}\eth^{s+1}\mathcal{Q}_{-1}^{em} - \sum_{\ell=1}^s \frac{(-1)^{\ell}\ell}{(s-\ell)!}\eth^{s-\ell}((\partial_u^{1-\ell}((-u)^{s-\ell}C)\eth^{\ell-1}\mathcal{Q}_{-1}^{em}) + \mathcal{O}(\mathbb{F}^3) \,, \\
    &\partial_u \tilde{q}_s^{gr} = \frac{(-u)^s}{s!}\eth^s(\eth \mathcal{Q}^{gr}_{-1} + C \mathcal{Q}_{-2}^{gr}) - \sum_{\ell=1}^s \frac{(-1)^{\ell}\left(\ell+1\right)}{(s-\ell)!}\eth^{s-\ell}(\partial_u^{1-\ell}((-u)^{s-\ell}C)\eth^{\ell-1}\mathcal{Q}_{-1}^{gr}) \\
    &\quad + \frac{(-u)^s}{s!}\eth^s (F \mathcal{Q}^{em}_{-1}) + \sum_{\ell=1}^s \frac{(-1)^{\ell}\left(\ell+1\right)}{(s-\ell)!}\eth^{s-\ell}(\partial_u^{-\ell}((-u)^{s-\ell}F)\eth^{\ell}\mathcal{Q}_{-1}^{em}) + \mathcal{O}(\mathbb{F}^3)\,,
    \end{split}
\end{equation}
and are thus indeed conserved at quadratic order in the absence of radiation.

We can now iteratively integrate the recursion relations \eqref{eq:rec1}, \eqref{eq:rec2} at each order, starting from the linear order. This is achieved employing the anti-derivative the iterated anti-derivative operator $\partial_{u}^{-n}$, $n\ge0$, defined as \cite{Geiller:2024bgf,Freidel:2021dfs}
\be\label{eq:anti-u}
	\partial_{u}^{-n}\mathcal{F}\left(u\right) \defeq \int_{+\infty}^{u}du_{1}\int_{+\infty}^{u_{1}}du_{2}\dots \int_{+\infty}^{u_{n-1}}du_{n}\,\mathcal{F}\left(u_{n}\right)\,,
\ee
for any function $\mathcal{F}\left(u\right)$ (see also Appendix \ref{app:MathIds}). This is a well-defined operation for integrating the evolution equations for $\mathcal Q_s$ as long as
\be\ba
	&\bar{N} = \mathcal{O}\left(\left|u\right|^{-\left(1+s+\varepsilon\right)}\right) =\bar F \quad\text{as $u\rightarrow\pm\infty$} \quad\text{and}\quad \lim_{u\rightarrow +\infty}\mathcal{Q}_{s} = 0 \,,
\ea\ee
with $0<\varepsilon<1$. The former fall-off conditions on the news tensor and the radiative part of the field strength tensor are the minimum conditions needed to access the $\text{sub}^{s}$-leading soft theorem~\cite{Strominger:2013jfa,Campiglia:2020qvc}, while the condition that $\varepsilon\notin\mathbb{N}$ ensures the absence of logarithmic corrections~\cite{Laddha:2017ygw,Laddha:2018myi}.

\subsection*{Soft/hard decomposition} At each order and for arbitrary spin $s$, the recursion charges can be perturbatively expanded as\footnote{The upper limits of the sums follow from the observation that the terms in the recursion relations that increase the number of insertions of the radiative data come from charge aspects whose spin-weight is offset by $2$ units, e.g. $\partial_{u}\mathcal{Q}_{s}^{em} \supset s\,C\mathcal{Q}_{s-2}^{em}$. We thank Marc Geiller for bringing to our attention this pattern.}
\begin{equation}
\label{decomp}
    \mathcal{Q}_s^{em}(u,x^A) = \sum_{k=1}^{\left\lfloor \frac{s+1}{2}\right\rfloor + 1}\mathcal{Q}_s^{k,em} \virg \mathcal{Q}_s^{gr} = \sum_{k=1}^{\left\lfloor \frac{s+2}{2}\right\rfloor + 1}\mathcal{Q}_s^{k,gr}(u,x^A)\,,
\end{equation}
where each term $\mathcal{Q}_s^{k,em}$ and $\mathcal{Q}_s^{k,gr}$ contains one insertion of the radiation fields $\bar F$, $\bar N$ and $k-1$ insertions of the radiative data $A$, $C$. For our purposes, it will be sufficient to work out the soft ($k=1$) and leading hard ($k=2$) contributions. Starting from \eqref{eq:rad}, the equations \eqref{eq:rec1}, \eqref{eq:rec2} can then be solved recursively. Up to the quadratic order, we get
\begin{subequations}
\label{eq:RecursionSolution}
\begin{align}
       &\mathcal{Q}_{s\geq -1}^{1,em} = -(\partial_u^{-1}\eth)^{s+1}\bar{F}\,,\\
       &\mathcal{Q}_{s\geq 0}^{2,em} = -\sum_{\ell=0}^s \ell\,(\partial_u^{-1}\eth)^{s-\ell}\partial_u^{-1}(C(\partial_u^{-1}\eth)^{\ell-1} \bar{F})\,,\\
       &\mathcal{Q}_{s\geq -2}^{1,gr} = (\partial_u^{-1}\eth)^{s+2}\partial_u\bar{N}\,,\\
       &\mathcal{Q}_{s\geq -1}^{2,gr} = \sum_{\ell=0}^s \left(\ell+1\right)(\partial_u^{-1}\eth)^{s-\ell}\partial_u^{-1}\left(C(\partial_u^{-1}\eth)^{\ell}\partial_u \bar{N} - F (\partial_u^{-1}\eth)^{\ell}\bar{F}\right)\,.
    \end{align}
\end{subequations}

Similarly, the quasi-conserved charge aspects \eqref{eq:DefRenormalized} can be split into soft, leading-order hard and higher order terms,
\begin{equation}\label{eq:decomp}
    \tilde{q}_s^{em}(u,x^A) = \sum_{k=1}^{\left\lfloor\frac{s+1}{2}\right\rfloor+1}\tilde{q}_s^{k,em}(u,x^A) \virg \tilde{q}_s^{gr}(u,x^A) = \sum_{k=1}^{\left\lfloor\frac{s+2}{2}\right\rfloor+1}\tilde{q}_s^{k,gr}(u,x^A)\,.
\end{equation}
Let us start with the soft ($k=1$) contributions. Using \eqref{eq:RecursionSolution} and the integral Leibniz rule~\eqref{eq:IntLR2}, we can write the soft charges as
\begin{equation}
    \tilde{q}_s^{1,em} =- \partial_u^{-1}\left( \frac{(-u)^s}{s!} \eth^{s+1}\bar{F}\right) \virg   \tilde{q}_s^{1,gr} = \partial_u^{-1}\left( \frac{(-u)^s}{s!} \eth^{s+2}\bar{N}\right)\,.
\end{equation}

\noindent For the leading hard pieces ($k=2$), following the strategy of~\cite{Geiller:2024bgf}, it turns out sufficient to find their evolution equations, since this will be the only part of the quadratic charges that contribute in the derivation of the $sw_{1+\infty}$ algebra, as we will see in the next section. Using the quasi-conservation equations~\eqref{eq:RrnormalizedEM}, we arrive at
\begin{subequations}
\label{evoeq}
    \begin{align}
       \partial_u \tilde{q}_s^{2,em} &= \sum_{\ell=0}^s \frac{(-1)^{\ell}\ell}{(s-\ell)!} \eth^{s-\ell}(\partial_u^{1-\ell}((-u)^{s-\ell}C)\eth^{\ell-1}\bar{F})\,,\\
       \partial_u \tilde{q}_s^{2,gr} &= \frac{(-u)^s}{s!}\eth^s(C\partial_u \bar{N}) - \sum_{\ell=1}^s \frac{(-1)^{\ell}(\ell+1)}{(s-\ell)!}\eth^{s-\ell}(\partial_u^{1-\ell}((-u)^{s-\ell}C)\eth^{\ell} \bar{N}) \nonumber\\
       &\quad- \sum_{\ell=0}^s \frac{(-1)^{\ell}(\ell+1)}{(s-\ell)!}\eth^{s-\ell}(\partial_u^{-\ell}((-u)^{s-\ell}F)\eth^{\ell} \bar{F})\,.
    \end{align}
\end{subequations}

\section{Celestial $sw_{1+\infty}$ algebra}
\label{sec:algebra_Max}
In this section, we derive the linearized charge algebra of electromagnetic and gravitational higher spin charges. This will be achieved by computing the Poisson bracket among the celestial charges, defined from the quasi-conserved quantities \eqref{eq:DefRenormalized} as
\be
	q_{s}(x^{A}) \defeq \lim_{u\rightarrow-\infty}\tilde{q}_{s}(u,x^{A})\,,
\ee
at linear order in the fields~\cite{Freidel:2021ytz,Geiller:2024bgf}. Performing again a perturbative expansion over the asymptotic fields as in \eqref{eq:decomp}, we see that the linearized bracket only involves the soft and the leading order hard parts of these charges,
\be\label{eq:2bra}
	\left\{q_{s}(x^{A}),q_{s^{\prime}}(x^{\prime A})\right\} = \left\{q_{s}^{1}(x^{A}),q_{s^{\prime}}^{2}(x^{\prime A})\right\} + \left\{q_{s}^{2}(x^{A}),q_{s^{\prime}}^{1}(x^{\prime A})\right\} + \mathcal{O}\left(\mathbb{F}^2\right) \,.
\ee
The goal is thus to obtain the expressions for the two above bracket contributions, and the final algebra will simply be a smeared version of \eqref{eq:2bra}.

The electromagnetic soft celestial charge is easily seen to be given by
\be\label{eq:q1em}
	q_{s}^{1,em} = \eth^{s+1}\bar{F}_{s} \,,\quad \bar{F}_{s} \defeq \int_{-\infty}^{+\infty}du\, \frac{(-u)^s}{s!}\bar{F} \,, 
\ee
namely it descends from the negative helicity (sub)$^s$-leading soft photon operator.
Similarly, the gravitational soft celestial charge,
\begin{equation}\label{eq:q1gr}
	q_{s}^{1,gr} = -\eth^{s+2}\bar{N}_{s} \,,\quad \bar{N}_{s} \defeq \int_{-\infty}^{+\infty}du\, \frac{(-u)^s}{s!}\bar{N} \,,
\end{equation}
descends from the negative helicity (sub)$^s$-leading soft graviton operator~\cite{Freidel:2021ytz}.

Strictly speaking, the bracket $\left\{q_{s}^{2}(x^{A}),\mathcal{F}(u^{\prime},x^{\prime A})\right\}$, for some $\mathcal{F}(u,x^{A})$, in general diverges, because of the divergent behavior of $q_{s}^{2}(x^{A}) = \lim_{u\rightarrow-\infty}\tilde{q}_{s}^{2}(u,x^{A})$. It will be regularized  according to the prescription
\be\label{eq:prescription}
	\left\{q_{s}^{2}(x^{A}),\mathcal{F}(u^{\prime},x^{\prime A})\right\} = \lim_{u\rightarrow-\infty}\partial_{u}^{-1}\left\{\partial_{u}\tilde{q}_{s}^{2}(u,x^{A}),\mathcal{F}(u^{\prime},x^{\prime A})\right\} \,.
\ee

We are now ready to derive the current algebra, starting from the canonical Poisson brackets on the radiative phase space at $\mathscr I^+$~(see e.g. \cite{Ashtekar:1987tt})\footnote{Note again here the unconventional normalization we are using for the field strength, namely $F=\kappa F_{\text{can}}$.},
\begin{subequations}\label{eq:canonical}
    \begin{align}
    \left\{A(u,z), \bar{F}\left(u^{\prime}, z^{\prime}\right)\right\} &= \kappa^2\,\delta(u-u^{\prime}) \delta(z,z^{\prime}) \,,\\
    \left\{C(u,z), \bar{N}(u^{\prime},z^{\prime})\right\} &= \kappa^2\,\delta(u-u^{\prime})\delta(z,z^{\prime}) \,.
    \end{align}
\end{subequations}
In the above relations, we charted as usual the $2$-dimensional transverse metric $q_{AB}$ with complex coordinates $\left(z,\bar{z}\right)$ such that $q_{AB}\left(x^{C}\right)dx^{A}dx^{B} = 2q_{z\bar{z}}\left(z,\bar{z}\right)\,dzd\bar{z} \,$\footnote{The $2$-dimensional $\delta$-function density is defined as $\delta(z,z^{\prime}) \defeq q_{z\bar{z}}^{-1}\delta\left(z-z^{\prime}\right)\delta\left(\bar{z}-\bar{z}^{\prime}\right)$.}.

From the canonical relations \eqref{eq:canonical}, the Poisson brackets for the soft charges can be readily obtained using \eqref{eq:q1em}, \eqref{eq:q1gr},
\begin{subequations}
    \begin{align}
    \left\{q_s^{1,em}(z), F(u^{\prime},z^{\prime})\right\} &= \kappa^2\,\frac{(-u^{\prime})^{s-1}}{(s-1)!}\eth^{s+1}_{z}\delta(z,z^{\prime}) \,,\\ 
    \left\{q_s^{1,gr}(z), N(u^{\prime},z^{\prime})\right\} &= -\kappa^2\,\frac{(-u^{\prime})^{s-1}}{(s-1)!}\eth^{s+2}_{z}\delta(z,z^{\prime}) \,.
    \end{align}
\end{subequations}
The case of the hard part is slightly more involved. Let us detail the computation for one of the brackets involving the action of the renormalized gravitational charges on the electromagnetic potential. Using \eqref{eq:prescription}, the $u\rightarrow-\infty$ limit is well-defined and by means of \eqref{evoeq} gives the following hard action,
\begin{equation}
    \badat{4}
    &\left\{{q}^{2,gr}_{s}(z),\bar{A}(u^{\prime},z^{\prime})\right\} = \lim_{u\rightarrow-\infty}\partial_{u}^{-1} \left\{\partial_{u}\tilde{q}^{2,gr}_{s}(u,z),\bar{A}(u^{\prime},z^{\prime})\right\} \\
    &= -\kappa^2\sum_{\ell=0}^s  \frac{(-1)^{\ell}\left(\ell+1\right)}{(s-\ell)!}\eth^{s-\ell}_z\lim_{u\rightarrow-\infty}\partial_u^{-1}\left(\left\{\partial_u^{-\ell}((-u)^{s-\ell}F(u,z)),\bar{A}(u^{\prime},z^{\prime})\right\}\eth_z^\ell \bar{F}(u,z)\right) \\
\eadat
\end{equation}
To massage this, let us rewrite it by defining
\be
    \mathcal{A}_{s\ell}\left(z;u^{\prime},z^{\prime}\right) \defeq \lim_{u\rightarrow-\infty}\partial_{u}^{-1}\left(\mathcal{B}_{s\ell}\left(u,z;u^{\prime},z^{\prime}\right)\eth_{z}^{\ell}\bar{F}(u,z)\right) \,,
\ee
with
\be
    \mathcal{B}_{s\ell}\left(u,z;u^{\prime},z^{\prime}\right) \defeq \left\{\partial_{u}^{-\ell}\left(\frac{\left(-u\right)^{s-\ell}}{\left(s-\ell\right)!}F(u,z)\right),\bar{A}(u^{\prime},z^{\prime})\right\} \,,
\ee
such that
\be
    \left\{{q}^{2,gr}_{s}(z),\bar{A}(u^{\prime},z^{\prime})\right\} = -\sum_{\ell=0}^{s}\frac{\left(-1\right)^{\ell}\left(\ell+1\right)}{\left(s-\ell\right)!}\eth_{z}^{s-\ell}\mathcal{A}_{s\ell}\left(z;u^{\prime},z^{\prime}\right) \,.
\ee
The repeated integral $\partial_{u}^{-\ell}$ in $\mathcal{B}_{s\ell}$ can be distributed using~\eqref{eq:AntiDervProdL} to get
\be
    \mathcal{B}_{s\ell}\left(u,z;u^{\prime},z^{\prime}\right) = \sum_{n=0}^{s-\ell}\binom{s-n-1}{\ell-1}\frac{\left(-u\right)^{n}}{n!}\left\{\partial_{u}^{-\left(s-n\right)}F(u,z),\bar{A}(u^{\prime},z^{\prime})\right\} \,.
\ee
To proceed, we borrow from Appendix~\ref{app:CanBraAntiDer} the following bracket
\be\ba
    \left\{\partial_{u}^{-n}F(u,z),\bar{A}(u^{\prime},z^{\prime})\right\} &= -\frac{\kappa^2}{2}\left(\partial_{u}^{-n}+\left(-1\right)^{n}\partial_{u^{\prime}}^{-n}\right)\delta\left(u-u^{\prime}\right)\delta(z,z^{\prime}) \\
    &= -\frac{\kappa^2}{2}\frac{\left(u-u^{\prime}\right)^{n-1}}{\left(n-1\right)!}\Theta\left(u-u^{\prime}\right)\delta(z,z^{\prime}) \,,
\ea\ee
where $\Theta\left(t\right) = \theta\left(t\right)-\theta\left(-t\right)$ is the antisymmetrized Heaviside function, satisfying $\frac{d}{dt}\Theta\left(t\right) = 2\delta\left(t\right)$. Using this in $\mathcal{B}_{s\ell}$, the sum over $n$ turns out to be a binomial expansion sum with the end result
\be
    \mathcal{B}_{s\ell}\left(u,z;u^{\prime},z^{\prime}\right) = -\frac{\kappa^2}{2}\frac{\left(-u^{\prime}\right)^{s-\ell}}{\left(s-\ell\right)!}\frac{\left(u-u^{\prime}\right)^{\ell-1}}{\left(\ell-1\right)!}\Theta\left(u-u^{\prime}\right)\delta(z,z^{\prime}) \,.
\ee
Next, using the definition of the repeated integral from Appendix~\ref{app:MathIds}, one can show that
\be\ba
    \partial_{u}^{-1}\left(\frac{\left(u-u^{\prime}\right)^{\ell-1}}{\left(\ell-1\right)!}f\left(u\right)\Theta\left(u-u^{\prime}\right)\right) &= \left(-1\right)^{\ell-1}\bigg\{2\partial_{u^{\prime}}^{-\ell}f\left(u^{\prime}\right)\theta\left(u^{\prime}-u\right) \\
    &\kern-1em+ \Theta\left(u-u^{\prime}\right)\int_{+\infty}^{u}du^{\prime\prime}\frac{\left(u^{\prime}-u^{\prime\prime}\right)^{\ell-1}}{\left(\ell-1\right)!}f\left(u^{\prime\prime}\right)\bigg\} \,.
\ea\ee
Plugging the expression of $\mathcal{B}_{s\ell}$ into $\mathcal{A}_{s\ell}$ and taking the $u\rightarrow-\infty$ limit of the above equation, we then find
\be\ba
    {}&\mathcal{A}_{s\ell}\left(z;u^{\prime},z^{\prime}\right) \\
    &= \kappa^2\delta(z,z^{\prime})\left(-1\right)^{\ell}\frac{\left(-u^{\prime}\right)^{s-\ell}}{\left(s-\ell\right)!}\bigg\{\partial_{u^{\prime}}^{1-\ell}\eth_{z^{\prime}}^{\ell}\bar{A}(u^{\prime},z^{\prime}) + \frac{1}{2}\eth_{z^{\prime}}^{\ell}\int_{-\infty}^{+\infty}du^{\prime\prime}\frac{\left(u^{\prime}-u^{\prime\prime}\right)^{\ell-1}}{\left(\ell-1\right)!}\bar{F}\left(u^{\prime\prime},z^{\prime}\right) \bigg\} \\
    &= \kappa^2\delta(z,z^{\prime})\frac{\left(-1\right)^{s}}{\left(s-\ell\right)!}\left(\Delta_{u^{\prime}}-1\right)_{s-\ell}\partial_{u^{\prime}}^{1-s}\eth_{z^{\prime}}^{\ell}\bar{A}(u^{\prime},z^{\prime}) \,.
\ea\ee
In the last line we have used the identity $u^{n} = \left(\Delta_{u}-1\right)_{n}\partial_{u}^{-n}$, where $\Delta_{u}\defeq u\partial_{u}+1$ and $(x)_{n}$ is the falling factorial operation, that follows from~\eqref{eq:manip}. We also realized there that the integral over the entire real line that appears turns out to vanish. This is because we can write the integrand as a total derivative since it is the same as the integrand of an anti-derivative,
\be
    \frac{\left(u^{\prime}-u\right)^{\ell-1}}{\left(\ell-1\right)!}\bar{F}\left(u,z^{\prime}\right) = \partial_{u}\left(\partial_{u}^{-\ell}\bar{F}\left(u,z^{\prime}\right)\right) \,.
\ee
Then, the fact that $0\le \ell\le s$ and the decaying conditions for the radiative fields are sufficient to conclude that
\be
    \int_{-\infty}^{+\infty}du^{\prime\prime}\frac{\left(u^{\prime}-u^{\prime\prime}\right)^{\ell-1}}{\left(\ell-1\right)!}\bar{F}\left(u^{\prime\prime},z^{\prime}\right) = -\lim_{u\rightarrow-\infty}\partial_{u}^{-\ell}\bar{F}\left(u,z^{\prime}\right) = 0 \,.
\ee
Finally, the bracket $\left\{q_{s}^{2,gr}\left(z\right),\bar{A}(u^{\prime},z^{\prime})\right\}$ can be extracted from the above to be
\be\label{eq:wActionA}
    \left\{{q}^{2,gr}_{s}(z),\bar{A}(u^{\prime},z^{\prime})\right\} = -\kappa^2\sum_{\ell=0}^s \frac{(-1)^{s-\ell}(\ell+1)}{(s-\ell)!}(\Delta_{u^{\prime}} - 1)_{s-\ell} \partial_{u^{\prime}}^{1-s}\eth_{z^{\prime}}^{\ell}\bar{A}(u^{\prime},z^{\prime})\eth_z^{s-\ell}\delta(z,z^{\prime}) \,.
\ee

A similar computation can be performed to obtain the action of the hard electromagnetic charges on the gravitational shear,
\begin{equation}
    \left\{{q}^{2,em}_s(z),\bar{C}(u^{\prime},z^{\prime})\right\} = \kappa^2 \sum_{n=0}^s \frac{(-1)^{s-n}n}{(s-n)!}(\Delta_{u^{\prime}} - 2)_{s-n} \partial_{u^{\prime}}^{1-s}\eth_{z^{\prime}}^{n-1}\bar{A}(u^{\prime},z^{\prime})\eth_z^{s-n}\delta(z,z^{\prime}) \,.
\end{equation}
The remaining brackets are found to be given by 
\be \badat{2}
  & \left\{q_{s}^{2,gr}(z),A(u^{\prime},z^{\prime})\right\} = -\kappa^2 \sum_{n=0}^s \frac{(-1)^{s-n}(n+1)}{(s-n)!}(\Delta_{u^{\prime}} + 1)_{s-n} \partial_{u^{\prime}}^{1-s}\eth_{z^{\prime}}^n A(u^{\prime},z^{\prime})\eth_z^{s-n}\delta(z,z^{\prime}), \\
  & \left\{q^{2,em}_s(z),C(u^{\prime},z^{\prime})\right\} = 0 ,\\
   &\left\{q_{s}^{2,em}(z),A(u^{\prime},z^{\prime})\right\} = -\kappa^2\sum_{n=0}^{s}\frac{(-1)^{s-n}n}{(s-n)!}(\Delta_{u^{\prime}}+1)_{s-n}\partial_{u^{\prime}}^{1-s}\eth_{z^{\prime}}^{n-1}C(u^{\prime},z^{\prime})\eth_{z}^{s-n}\delta(z,z^{\prime}), \\
  &  \left\{q^{2,em}_s(z),\bar{A}(u^{\prime},z^{\prime})\right\} = 0, \\
   & \left\{q_{s}^{2,gr}(z),C(u^{\prime},z^{\prime})\right\} = -\kappa^2\sum_{n=0}^{s}\frac{(-1)^{s-n}(n+1)}{(s-n)!}(\Delta_{u^{\prime}}+2)_{s-n}\partial_{u^{\prime}}^{1-s}\eth_{z^{\prime}}^{n}C(u^{\prime},z^{\prime})\eth_{z}^{s-n}\delta(z,z^{\prime}), \\
   & \left\{q_{s}^{2,gr}(z),\bar{C}(u^{\prime},z^{\prime})\right\} = -\kappa^2\sum_{n=0}^{s}\frac{(-1)^{s-n}(n+1)}{(s-n)!}(\Delta_{u^{\prime}}-2)_{s-n}\partial_{u^{\prime}}^{1-s}\eth_{z^{\prime}}^{n}\bar{C}(u^{\prime},z^{\prime})\eth_{z}^{s-n}\delta(z,z^{\prime}).
\eadat \ee
The action of the charges on the negative helicity photon field or gravitational follows using the same methods as above. For their action on the positive helicity fields, the corresponding brackets are derived using the steps outlined in~\cite{Geiller:2024bgf}\footnote{See, in particular, Section 6.4.2 there, baring in mind that the canonical brackets of~\cite{Geiller:2024bgf} come with an opposite sign compared to what we use here.}

In Section \ref{sec:w1pInftyEYM}, along the way with generalizing these formulae for the Yang-Mills case, we will also repackage these expressions in a more compact form by identifying the generic spin structure (see  \eqref{eq:compact1}, \eqref{eq:compact2}).

From the above action of the charges on the radiative data, one can deduce their action on the sub$^s$-leading soft operators $\bar F_s$, $\bar N_s$ defined in \eqref{eq:q1em} by means of pseudo-differential calculus identities (see Appendix \ref{app:MathIds}), as shown in the gravity case in \cite{Freidel:2021ytz}. Starting from \eqref{eq:wActionA}, we find, using identities \eqref{eq:manip},
    \begin{align}
        &\left\{{q}^{2,gr}_s (z), \bar{F}_{s^{\prime}} (z^{\prime})\right\} \nonumber \\
        &= -\kappa^2\int_{-\infty}^{+\infty}   du \, \frac{(-u)^{s^{\prime}}}{s^{\prime}!}\partial_{u}\sum_{n=0}^{s}\frac{(-1)^{s-n}(n+1)}{(s-n)!}(\Delta_{u}-1)_{s-n}\partial_{u}^{1-s}\eth_{z^{\prime}}^n \bar{A}(u,z^{\prime})\eth_z^{s-n}\delta(z,z^{\prime})\nonumber  \\
        &=-\kappa^2\sum_{n=0}^s \int_{-\infty}^{+\infty}\hspace{-0.5em}  du\, \frac{(-1)^{n+1}(n+1)}{s^{\prime}!(s-n)!} \frac{(\Delta_u - s^{\prime})_{s-n}}{(\Delta_u - s^{\prime}-1)_{s-1}} (-u)^{s^{\prime}+s-1}\eth_{z^{\prime}}^{n}\bar{F}(u,z^{\prime})\eth_z^{s-n}\delta(z,z^{\prime})\nonumber \\
        &=-\kappa^2\sum_{n=0}^s \begin{pmatrix}
        s^{\prime}+s-n-1 \\
        s^{\prime}-1
    \end{pmatrix} (n+1)\,\eth_{z^{\prime}}^{n}\bar{F}_{s+s^{\prime}-1}(z^{\prime})\eth^{s-n}_z\delta(z,z^{\prime})\,.
    \end{align}
To arrive at the last line, one uses some combinatorial identities and assumes that the field strength $\bar{F}$ falls off as $\mathcal{O}(u^{-1-s-\epsilon})$ at large $u$ for every $s$, such that one can set $\Delta_{u} \approx 0$ inside the integral (see e.g. Appendix B in \cite{Freidel:2021ytz}).

Similarly, we compute the action of the electromagnetic charge on soft graviton operators,
\begin{equation}
    \left\{{q}^{2,em}_s (z), \bar{N}_{s^{\prime}} (z^{\prime})\right\} =\kappa^2 \sum_{n=0}^s \begin{pmatrix}
        s^{\prime}+s-n \\
        s^{\prime}
    \end{pmatrix} n \,\eth_{z^{\prime}}^{n-1}\bar{F}_{s+s^{\prime}-1}(z^{\prime})\eth^{s-n}_z\delta(z,z^{\prime})\,.
\end{equation}
It is then straightforward, from \eqref{eq:q1em},  \eqref{eq:q1gr}, to obtain the brackets between the linear and quadratic charges,
    \begin{align}\label{eq:bare_brackets}
    &     \{{q}^{2,gr}_s (z), {q}^{1,em}_{s^{\prime}} (z^{\prime})\} = -\kappa^2\sum_{n=0}^s \begin{pmatrix}
        s^{\prime}+s-n-1 \\
        s^{\prime}-1
    \end{pmatrix} (n+1)\eth_{z^{\prime}}^{s^{\prime}+1}(\eth_{z^{\prime}}^{n}\bar{F}_{s+s^{\prime}-1}(z^{\prime})\eth^{s-n}_z\delta(z,z^{\prime}))\,,\nonumber\\
       & \{{q}^{1,gr}_s (z), {q}^{2,em}_{s^{\prime}} (z^{\prime})\} = \kappa^2\sum_{n=0}^{s^{\prime}} \begin{pmatrix}
        s^{\prime}+s-n \\
        s
    \end{pmatrix} n \eth^{s+2}_z(\eth^{n-1}_z\bar{F}_{s+s^{\prime}-1}(z)\eth^{s^{\prime}-n}_{z^{\prime}}\delta(z^{\prime},z)),\nonumber\\
    & \{{q}^{1,em}_s (z), {q}^{2,em}_{s^{\prime}} (z^{\prime})\} =0\,,\\
    &     \{{q}^{2,gr}_s (z), {q}^{1,gr}_{s^{\prime}} (z^{\prime})\} = \kappa^2\sum_{n=0}^s \begin{pmatrix}
        s^{\prime}+s-n \\
        s^{\prime}
    \end{pmatrix} (n+1)\eth_{z^{\prime}}^{s^{\prime}+2}(\eth_{z^{\prime}}^{n}\bar{N}_{s+s^{\prime}-1}(z^{\prime})\eth^{s-n}_z\delta(z,z^{\prime}))\,.\nonumber
    \end{align}

Finally, we derive the algebra at the level of smeared charges \cite{Geiller:2024bgf}, defined as\footnote{The trasnverse $2$-dimensional integral here is abbreviated as $\oint \defeq \int d^2z\,q_{z\bar{z}}\left(z,\bar{z}\right)$.}
\begin{subequations}
    \begin{align}
       & Q_s^{em}(\sigma_s) = \frac{1}{\kappa^2}\oint\, \sigma_s(z,\bar{z})\, {q}^{em}_s (z,\bar{z})\,,\\
    &  Q_s^{gr}(\tau_s) =\frac{1}{\kappa^2}\oint\, \tau_s(z,\bar{z}) \,{q}^{gr}_s (z,\bar{z})\,.
    \end{align}
\end{subequations}
They involve the pairing between the higher spin charges $q_s$ and higher spin symmetry parameters $\sigma_s, \tau_s$ of helicity $-s$.
The linearized algebra involves the quadratic brackets between the smeared charges. For the mixed gravity-gauge field bracket,
\begin{equation}
\label{eq:DefLinearBrackets}
    \left\{Q_s^{gr}(\tau_s),Q_{s^{\prime}}^{em}(\sigma_{s^{\prime}})\right\}^{(1)} = \left\{\mathcal{Q}_s^{1,gr}(\tau_s),\mathcal{Q}_{s^{\prime}}^{2,em}(\sigma_{s^{\prime}})\right\} + \left\{Q_s^{2,gr}(\tau_s),Q_{s^{\prime}}^{1,em}(\sigma_{s^{\prime}})\right\}\,,
\end{equation}
the two above brackets can then be calculated using the bare charge brackets \eqref{eq:bare_brackets} and integrating by parts, as demonstrated, for instance, in Appendix~I of~\cite{Geiller:2024bgf} for the case of pure gravity. We find,
\begin{align}
        &\left\{Q_{s}^{1,gr}(\tau_s),Q_{s^{\prime}}^{2,em}(\sigma_{s^{\prime}})\right\} = -s^{\prime}Q_{s+s^{\prime}-1}^{1,em}(\sigma_{s^{\prime}}\eth \tau_s)  \nonumber\\
            & -\frac{1}{\kappa^2}\oint \footnotesize\sum_{k=0}^{s^{\prime}-1}\sum_{p=1}^{s+k+1}(-1)^{k}  \frac{(s+s^{\prime}+1)_2}{(k+s+2)_2}
            \begin{pmatrix}
                s + s^{\prime} - 1 \\
                s
            \end{pmatrix}
            \begin{pmatrix}
                s^{\prime}-1 \\
                k
            \end{pmatrix}
            \begin{pmatrix}
                s+k+1 \\
                p
            \end{pmatrix}  \normalsize\eth \tau_s \eth^p \sigma_{s^{\prime}} \eth^{s+s^{\prime}-p}\bar{F}_{s+s^{\prime}-1},\nonumber\\[10pt]
    &        \left\{Q_{s}^{2,gr}(\tau_s),Q_{s^{\prime}}^{1,em}(\sigma_{s^{\prime}})\right\}= (s+1)Q_{s+s^{\prime}-1}^{1,em}(\tau_s\eth \sigma_{s^{\prime}}) \\
            &+ \frac{1}{\kappa^2}\oint \sum_{k=0}^{s}\sum_{p=1}^{s^{\prime}+k}(-1)^{k}\frac{(s+s^{\prime}+1)_2}{(k+s^{\prime}+1)_2}
           \footnotesize	\begin{pmatrix}
                s + s^{\prime} - 1 \\
                s^{\prime}-1
            \end{pmatrix}
            \begin{pmatrix}
                s \\
                k
            \end{pmatrix}
            \begin{pmatrix}
                s^{\prime}+k \\
                p
            \end{pmatrix} \normalsize \eth \sigma_{s^{\prime}} \eth^p \tau_s \eth^{s+s^{\prime}-p}\bar{F}_{s+s^{\prime}-1}\,.\nonumber
        \end{align}
Substituting the expressions back in \eqref{eq:DefLinearBrackets}, the second lines of each term remarkably combine to give an integral over a total derivative. Together with the pure gravity bracket which can be derived in a similar way \cite{Freidel:2021ytz,Geiller:2024bgf} and the vanishing pure Maxwell bracket, we thus arrive at the final algebra
\begin{empheq}[box=\tcbhighmath]{align}
\badat{2}\label{eq:final}&\left\{Q_{s}^{gr}(\tau_{s}),Q_{s^{\prime}}^{gr}(\tau^{\prime}_{s^{\prime}})\right\}^{(1)} = Q_{s+s^{\prime}-1}^{1,gr}((s+1)\tau_{s}\eth \tau^{\prime}_{s^{\prime}} - (s^{\prime}+1)\tau^{\prime}_{s^{\prime}}\eth \tau_{s})\,,\\[7pt]
  &  \left\{Q_{s}^{gr}(\tau_{s}),Q_{s^{\prime}}^{em}(\sigma_{s^{\prime}})\right\}^{(1)} = Q_{s+s^{\prime}-1}^{1,em}((s+1)\tau_{s}\eth \sigma_{s^{\prime}} - s^{\prime} \sigma_{s^{\prime}}\eth \tau_{s})\,,\\[7pt]
    &  \left\{Q_{s}^{em}(\sigma_{s}),Q_{s^{\prime}}^{em}(\sigma'_{s^{\prime}})\right\}^{(1)} = 0\,.
    \eadat
\end{empheq}

\noindent To make contact with the notation of \cite{Strominger:2021mtt}, let us introduce the modes
\be
    \badat{2}
       & Q^{s,gr}_{k,l} \defeq Q_{s}^{gr}\left(\tau^s_{k,l}\right) \,,\quad \tau^s_{k,l} = z^{1+s-k}\bar{z}^{1-s-l} \,, \\
      &  Q^{s,em}_{k,l} \defeq Q_{s}^{em}\left(\sigma^s_{k,l}\right) \,,\quad \sigma^s_{k,l} = z^{\frac{1}{2}+s-k}\bar{z}^{\frac{1}{2}-s-l} \,,
    \eadat
\ee
in terms of which \eqref{eq:final} reads
\begin{equation}
\badat{2}
    &\left\{Q^{s,gr}_{k,l},Q^{s^{\prime},em}_{k^{\prime},l^{\prime}}\right\} = \left( k\,s^{\prime} + \left(k^{\prime}-\tfrac{1}{2}\right)\left(s+1\right)  \right)Q^{s+s^{\prime}-1,em}_{k+k^{\prime}-1,l+l^{\prime}}\,,\\
    &\left\{Q^{s,em}_{k,l},Q^{s^{\prime},em}_{k^{\prime},l^{\prime}}\right\} = 0\,,\\
    &\left\{Q^{s,gr}_{k,l},Q^{s^{\prime},gr}_{k^{\prime},l^{\prime}}\right\} = \left(k\left(s^{\prime}+1\right) - k^{\prime}\left(s+1\right)\right)Q^{s+s^{\prime}-1,gr}_{k+k^{\prime}-1,l+l^{\prime}}\,.
    \eadat
\end{equation}
With the mode re-labeling
\begin{subequations}
    \begin{equation}
        q^{s,gr}_k(\bar{z}) = - 2w^{\frac{s+3}{2}}_{\frac{s+1}{2}-k}(\bar z) \virg 
        q^{s,em}_k(\bar{z}) = -2is^{\frac{s+2}{2}}_{\frac{s+1}{2}-k}(\bar{z})\,,
    \end{equation}
\end{subequations}
we finally get,
\begin{equation}
\badat{2}
  &\left\{w_m^p, w_n^{q}\right\} = (m(q-1)-n(p-1))w_{m+n}^{p+q-2}\,,\\
    &\left\{w_m^p, s_n^{q}\right\} = (m(q-1)-n(p-1))s_{m+n}^{p+q-2}\,,\\
    &\left\{s_m^p, s_n^{q}\right\}=0\,,\\
    \eadat
\end{equation}
which is the Einstein-Maxwell $sw_{1+\infty}$ algebra we wished to derive. Notice that the indices-independent numerical factor ($-2i$) relating $q_{k}^{s,em}$ and $s_{m}^{p}$ can in principle be any number in the current Einstein-Maxwell setup. However, as we will see in the next section, the Einstein--Yang-Mills $sw_{1+\infty}$ algebra precisely fixes this normalization constant to be as above.

\section{Generalization to Einstein-Yang-Mills}
\label{sec:w1pInftyEYM}

In this section, we will equip the gauge field with an internal Lie algebra structure and consider the Einstein--Yang-Mills system. This consists of an (inverse) metric field $g^{\mu\nu}$ and a non-abelian $1$-form gauge field $\text{A}=A_{\mu}^{a}dx^{\mu}\otimes T_{a}$, where $T_{a}\in\mathfrak{g}$ generate the Lie algebra $\mathfrak{g}$ that corresponds to the gauge symmetry Lie group $G$,
\be
	\left[T_{a},T_{b}\right] = if_{ab}^{\,\,\,\,\,c}T_{c} \,,\quad a,b,c=1,\dots,\dim G \,.
\ee
The action describing this system is then\footnote{We remark here that the gauge fields $A_{\mu}^{a}$ and YM coupling constant $g_{\text{YM}}$ we are using here are not the ones appearing when using canonical variables $A_{\mu}^{\text{can}}$ and $g_{\text{YM}}^{\text{can}}$, for which $S_{\text{YM}}\left[g,A^{\text{can}}\right] = -\frac{1}{2}\int d^4x \sqrt{-g}\,\tr\left(\text{F}^{\text{can}2}\right)$, with $\text{F}^{\text{can}} = \text{d}\text{A}^{\text{can}} - ig_{\text{YM}}^{\text{can}}\left[\text{A}^{\text{can}},\text{A}^{\text{can}}\right]$. Rather, we have performed the redefinition $A_{\mu}^{\text{can}} = \frac{1}{\kappa}A_{\mu}$ and $g_{\text{YM}}^{\text{can}} = \kappa g_{\text{YM}}$, such that $\text{F}^{\text{can}} = \frac{1}{\kappa}\text{F}$, for future convenience.}
\be\label{eq:ActionEYM}
	S\left[g,A\right] = \frac{1}{2\kappa^2}\int d^4x\sqrt{-g}\left[R-\frac{1}{2}\tr\left(\text{F}^2\right)\right] \,,
\ee
where the field strength tensor $\text{F}=\frac{1}{2}F_{\mu\nu}^{a}dx^{\mu}\wedge dx^{\nu}\otimes T_{a}$ is related to the gauge $1$-form field according to $\text{F} = \text{d}\text{A} - ig_{\text{YM}}\left[\text{A},\text{A}\right]$ or, in components,
\be
	F_{\mu\nu}^{a} = 2\partial_{[\mu}A_{\nu]}^{a} + g_{\text{YM}}f_{bc}^{\,\,\,\,\,a}A_{\mu}^{b}A_{\nu}^{c} \,.
\ee
The trace entering the Yang-Mills (YM) action refers to the trace in color space and defines a notion of ``color metric'' $c_{ab}\defeq \tr\left(T_{a}T_{b}\right)$ such that
\be
	\tr\left(\text{F}^2\right) = c_{ab}g^{\mu\rho}g^{\nu\sigma}F_{\mu\nu}^{a}F_{\rho\sigma}^{b} \,.
\ee
The only algebraic requirement here is that the Lie algebra $\mathfrak{g}$ associated with the gauge symmetry group $G$ is a direct sum of commuting simple and $\mathfrak{u}\left(1\right)$ algebras; this is a necessary condition for the gauge invariance of the YM action and is equivalent to the statement that $f_{cb}^{\,\,\,\,\,d}c_{da} = -f_{ca}^{\,\,\,\,\,d}c_{ab}$. On top of this, we will assume here that the gauge group is a compact and simple Lie group, and henceforth choose to normalize the generators $T_{a}$ in the fundamental representation such that $c_{ab} = \delta_{ab}$, and such that the structure constants $f_{abc} \defeq f_{ab}^{\,\,\,\,\,d}c_{cd}$ are totally antisymmetric in their color indices. Nevertheless the analysis and results reported here are straightforward to generalize for a generic invertible $c_{ab}$ compatible with gauge invariance.

The equations of motion for the YM field in the absence of sources are the following
\be
	D^{\nu}F_{\nu\mu} = 0 \,,\quad D_{[\rho}F_{\mu\nu]} = 0 \,,
\ee
where $D_{\mu} = \nabla_{\mu} - ig_{\text{YM}}A_{\mu}^{a}\rho\left(T_{a}\right)$ are the spacetime components of the gauge covariant derivative, with $\rho\left(T_{a}\right)$ the representation of the generators $T_{a}$ under which the field on which $D_{\mu}$ acts on transforms. These are translated into the NP formalism in Appendix~\ref{app:NPFormalism}, after introducing the $3\times\dim G$ complex YM-NP scalars $\Phi_{1-s}^{a}$, with $s=-1,0,+1$, according to
\be
	\Phi_0^{a} \defeq F_{\ell m}^{a} \,,\quad \Phi_1^{a} \defeq \frac{1}{2}\left(F_{\ell n}^{a}-F_{m\bar{m}}^{a}\right) \,,\quad \Phi_2^{a} \defeq F_{\bar{m}n}^{a} \,.
\ee

\subsection{Evolution equations of YM-NP and Weyl-NP scalars}

For the choice of null tetrad vectors used in Section~\ref{sec:Setup}, and from the fact that the YM gauge field satisfies the same fall-off conditions \eqref{eq:Maxwellbc} near-$\scri$ as the Maxwell field, the YM-NP scalars still exhibit a peeling behavior near $\scri$,
\be\ba
	\Phi_{1-s}(u,r,x^{A}) &= \frac{1}{r^{2+s}}\left[\Phi_{1-s}^{0}(u,x^{A}) + \mathcal{O}\left(r^{-1}\right)\right] \,,\quad -1 \le s \le +1 \,,
\ea\ee
with the lowest-spin weight YM-NP scalar capturing radiation according to
\be
    \Phi_2^0(u,x^{A}) = -\bar{F}(u,x^{A}) \,.
\ee
In the above expression, $\bar{F}(u,x^{A}) = \bar{F}^{a}(u,x^{A})T_{a} \defeq \partial_{u}\bar{A}(u,x^{A})$ is the negative helicity YM field strength, with $\bar{A}(u,x^{A}) = \bar{A}^{a}(u,x^{A})T_{a} \defeq \bar{\varepsilon}^{A}(x^{B})A_{A}^{\left(0\right)a}\left(u,x^{B}\right)T_{a}$ the negative helicity gluon field. We also remark the usage of colored YM-NP scalars
\be
	\Phi_{1-s} \defeq \Phi_{1-s}^{a}T_{a} \,,
\ee
a practice that we will frequently apply in this section.

We can now derive the recursion relations analogous to the ones of Section~\ref{sec:recursion_Max}. On top of the $5$ Weyl-NP scalars $\mathcal{Q}_{s}^{gr}$ of spin weight $-2\le s\le +2$ defined in~\eqref{eq:QsMin}, one also defines the $3\times\dim G$ YM-NP scalars $\mathcal{Q}_{s}^{gl,a}$ of spin weight $-1 \le s \le +1$ analogously to the Maxwell case, i.e.  according to
\be
	\mathcal{Q}_{s}^{gl} \defeq \Phi_{1-s}^{0} \,,\quad -1 \le s \le +1 \,.
\ee

Plugging these asymptotic expansions of the NP scalars into the equations of motion, the leading order evolution equations can be worked out to be
\begin{subequations}
	\begin{align}
		\begin{split}
			\partial_{u}\mathcal{Q}_{s}^{gl} &= \eth \mathcal{Q}_{s-1}^{\left(1\right)} - ig_{\text{YM}}\left[A,\mathcal{Q}_{s-1}^{gl}\right] + s C\mathcal{Q}_{s-2}^{gl} \\
			&= \left\{\eth \mathcal{Q}_{s-1}^{gl,a} + g_{\text{YM}}f_{bc}^{\,\,\,\,\,a}A^{b}\mathcal{Q}_{s-1}^{gl,c} + sC\mathcal{Q}_{s-2}^{gl,a}\right\}T_{a} \,,
		\end{split} \label{eq:EvolutionQsEYM1} \\
		\begin{split}
			\partial_{u}\mathcal{Q}_{s}^{gr} &= \eth \mathcal{Q}_{s-1}^{gr} + \left(s+1\right) C\mathcal{Q}_{s-2}^{gr} + \left(s+1\right)\tr\left(F\mathcal{Q}_{s-1}^{gl}\right) \\
			&= \eth \mathcal{Q}_{s-1}^{gr} + \left(s+1\right) C\mathcal{Q}_{s-2}^{gr)} + \left(s+1\right)\delta_{ab}F^{a}\mathcal{Q}_{s-1}^{gl,b} \,.
		\end{split} \label{eq:EvolutionQsEYM2}
	\end{align}
\end{subequations}
The above expressions are exact for $0 \le s \le +1$ in the YM case and for $-1 \le s \le +2$ for the gravitational case, while they can be extended to all higher spin weights by means of redefining the subleading terms in the asymptotic expansions of $\Phi_0$ and $\Psi_0$ as per~\eqref{eq:def_higher_spin} and~\eqref{eq:Qgr}.

For future convenience in collectively deriving various formulae, we will now introduce a collective notation that will be able to address both spacetime spin cases at once. This is achieved by introducing an index $j$ that captures the spacetime spin\footnote{The spacetime spin index $j$ should not be confused with the spin weight $s$ of a NP scalar.} of the relevant type of fields. Then, the spin weight $s$ Weyl-NP scalars $\Psi_{2-s}$ and YM-NP scalars $\Phi_{1-s}$ are collected into the single spin weight $s$ NP scalar
\be
    \psi_{j-s} \defeq \begin{cases}
        \Phi_{1-s} & \text{for $j=1$} \,; \\
        \Psi_{2-s} & \text{for $j=2$} \,;
    \end{cases} \,,\quad -j \le s \le +j \,.
\ee
In terms of these, the peeling property of both types of fields, for instance, reads
\be
	\psi_{j-s}(u,r,x^{A}) = \frac{1}{r^{j+1+s}}\left[\psi_{j-s}^{0}(u,x^{A}) + \mathcal{O}\left(r^{-1}\right)\right] \,,\quad -j \le s \le +j \,.
\ee
Furthermore, the higher spin charge aspects $\mathcal{Q}_{s}^{gl}$, with $-1\le s\le +1$, and $\mathcal{Q}_{s}^{gr}$, with $-2\le s\le +2$, can be collected into
\be
    \mathcal{Q}_{s}^{\left(j\right)} \defeq \begin{cases}
        \mathcal{Q}_{s}^{gl} & \text{for $j=1$} \,; \\
        \mathcal{Q}_{s}^{gr} & \text{for $j=2$} \,;
    \end{cases} \,,\quad -j \le s\le +j
\ee
In particular, the above $2j+1$ NP scalars $\mathcal{Q}_{s}^{\left(j\right)}$ of spin weight $-j \le s \le +j$ area defined from the NP scalars $\psi_{j-s}$ according to
\be
	\mathcal{Q}_{s}^{\left(j\right)} \defeq \psi_{j-s}^{0} \,,\quad -j \le s \le +j \,,
\ee
while $\mathcal{Q}_{-j}^{\left(j\right)}$, and, in fact, all $\mathcal{Q}_{-j\le s<0}^{\left(j\right)}$, encode the radiative information of the spacetime-spin-$j$ configuration.

Last, the evolution equations for the higher spin charge aspects can be compactly written as
\begin{empheq}[box=\tcbhighmath]{align}\label{eq:EvolutionQsEYM}
	\partial_{u}\mathcal{Q}_{s}^{\left(j\right)} = \eth \mathcal{Q}_{s-1}^{\left(j\right)} - ig_{\text{YM}}\left[A,\mathcal{Q}_{s-1}^{\left(j\right)}\right]  + \left(j+s-1\right)\sum_{i=0}^{j-1}\sigma^{\left(2-i\right)}\bullet \mathcal{Q}_{s-2+i}^{\left(j-i\right)} \,,
\end{empheq}
with $-j+1 \le s \le +j$, where
\be
	\sigma^{\left(2\right)} \defeq C \quad\text{and}\quad \sigma^{\left(1\right)}  \defeq F^{a}T_{a} \,,
\ee
and we have furthermore introduced a ``bullet'' operation which contracts indices in the color space, i.e.
\be
	\begin{gathered}
		\sigma^{\left(2\right)}\bullet \mathcal{Q}_{s}^{\left(2\right)} = \sigma^{\left(2\right)} \mathcal{Q}_{s}^{\left(2\right)} \,,\quad \sigma^{\left(2\right)}\bullet \mathcal{Q}_{s}^{\left(1\right)} = \sigma^{\left(2\right)} \mathcal{Q}_{s}^{\left(1\right)a}T_{a} \,, \\
		\sigma^{\left(1\right)}\bullet \mathcal{Q}_{s}^{\left(2\right)} = \sigma^{\left(1\right)a} \mathcal{Q}_{s}^{\left(2\right)}T_{a} \,,\quad \sigma^{\left(1\right)}\bullet \mathcal{Q}_{s}^{\left(1\right)} = \delta_{ab}\sigma^{\left(1\right)a} \mathcal{Q}_{s}^{\left(1\right)b} \,.
	\end{gathered}
\ee
The above recursion relations are exact for $-j\le s\le +j$, but can be extended to all $s\ge-j$ by redefining the subleading modes of the spin weight $+j$ NP scalar $\psi_0$.

\subsection{Quasi-conserved charges up to quadratic order}
We now turn to the construction of quasi-conserved charges, namely charges that are conserved when $\mathcal{Q}_{-j\le s<0}^{\left(j\right)}$ are zero. Let us consider the following quadratic charges,
\begin{subequations}
	\begin{align}
		\begin{split}
			\tilde{q}_{s}^{gl} &= \sum_{n=0}^{s}\alpha_{n}\eth^{n}\mathcal{Q}_{s-n}^{\left(1\right)} -ig_{\text{YM}}\sum_{\ell=1}^{s}\sum_{n=0}^{\ell-1}\eth^{s-\ell}[(-\partial_{u})^{-(n+1)}(\alpha_{s-\ell}A),\eth^{n}\mathcal{Q}_{\ell-n-1}^{\left(1\right)}] \\
			&\quad+ \sum_{\ell=2}^{s}\sum_{n=0}^{\ell-2}\ell\,\eth^{s-\ell}((-\partial_{u})^{-(n+1)}(\alpha_{s-\ell}C)\eth^{n}\mathcal{Q}_{\ell-n-2}^{\left(1\right)}) + \mathcal{O}\left(\mathbb{F}^3\right) \\
			&= \bigg\{ \sum_{n=0}^{s}\alpha_{n}\eth^{n}\mathcal{Q}_{s-n}^{\left(1\right)a} + g_{\text{YM}}f_{bc}^{\,\,\,\,\,a}\sum_{\ell=1}^{s}\sum_{n=0}^{\ell-1}\eth^{s-\ell}((-\partial_{u})^{-(n+1)}(\alpha_{s-\ell}A^{b})\eth^{n}\mathcal{Q}_{\ell-n-1}^{\left(1\right)c}) \\
			&\quad+ \sum_{\ell=2}^{s}\sum_{n=0}^{\ell-2}\ell\,\eth^{s-\ell}((-\partial_{u})^{-(n+1)}(\alpha_{s-\ell}C)\eth^{n}\mathcal{Q}_{\ell-n-2}^{\left(1\right)a}) \bigg\} T_{a} + \mathcal{O}\left(\mathbb{F}^3\right) \,,
		\end{split} \\
		\begin{split}
			\tilde{q}_{s}^{gr} &= \sum_{n=0}^{s}\alpha_{n}\eth^{n}\mathcal{Q}_{s-n}^{\left(2\right)} + \sum_{\ell=2}^{s}\sum_{n=0}^{\ell-2}(\ell+1)\,\eth^{s-\ell}((-\partial_{u})^{-(n+1)}(\alpha_{s-\ell}C)\eth^{n}\mathcal{Q}_{\ell-n-2}^{\left(2\right)}) \\
			&\quad+\sum_{\ell=1}^{s}\sum_{n=0}^{\ell-1}(\ell+1)\,\eth^{s-\ell}\tr\left((-\partial_{u})^{-(n+1)}(\alpha_{s-\ell}F)\eth^{n}\mathcal{Q}_{\ell-n-1}^{\left(1\right)}\right) + \mathcal{O}\left(\mathbb{F}^3\right) \\
			&=\sum_{n=0}^{s}\alpha_{n}\eth^{n}\mathcal{Q}_{s-n}^{\left(2\right)} + \sum_{\ell=2}^{s}\sum_{n=0}^{\ell-2}(\ell+1)\,\eth^{s-\ell}((-\partial_{u})^{-(n+1)}(\alpha_{s-\ell}C)\eth^{n}\mathcal{Q}_{\ell-n-2}^{\left(2\right)}) \\
			&\quad+\delta_{ab}\sum_{\ell=1}^{s}\sum_{n=0}^{\ell-1}(\ell+1)\,\eth^{s-\ell}((-\partial_{u})^{-(n+1)}(\alpha_{s-\ell}F^{a})\eth^{n}\mathcal{Q}_{\ell-n-1}^{\left(1\right)b}) + \mathcal{O}\left(\mathbb{F}^3\right) \,,
		\end{split}
	\end{align}
\end{subequations}
where $s\ge0$ and we have introduced the quantity
\be
    \alpha_{n}\left(u\right) \defeq \frac{(-u)^{n}}{n!}
\ee
to make the expressions shorter.  The above charge aspects can be expressed more collectively as
\be\ba
	{}&\tilde{q}_{s}^{\left(j\right)} = \sum_{n=0}^{s}\alpha_{n}\eth^{n}\mathcal{Q}_{s-n}^{\left(j\right)} -ig_{\text{YM}}\sum_{\ell=1}^{s}\sum_{n=0}^{\ell-1}\eth^{s-\ell}[(-\partial_{u})^{-(n+1)}(\alpha_{s-\ell}A),\eth^{n}\mathcal{Q}_{\ell-n-1}^{\left(j\right)}] \\
	&+\sum_{i=0}^{j-1}\sum_{\ell=2-i}^{s}\sum_{n=0}^{\ell-2+i}(\ell+j-1)\,\eth^{s-\ell}((-\partial_{u})^{-(n+1)}(\alpha_{s-\ell}\sigma^{\left(2-i\right)})\bullet\eth^{n}\mathcal{Q}_{\ell-n-2+i}^{\left(j-i\right)}) + \mathcal{O}\left(\mathbb{F}^3\right) \,,
\ea\ee
where $\tilde{q}_{s}^{\left(1\right)}=\tilde{q}_{s}^{gl}$, $\tilde{q}_{s}^{\left(2\right)}=\tilde{q}_{s}^{gr}$.
In Appendix~\ref{app:tqsCons}, we show that these charges are conserved in non-radiative configurations up to quadratic order in the fields. More explicitly, they satisfy the following evolution equation
\be\ba\label{eq:EvolutionstqsEYM}
	{}&\partial_{u}\tilde{q}_{s}^{\left(j\right)} = \alpha_{s}\eth^{s}\partial_{u}\mathcal{Q}_0^{\left(j\right)} - ig_{\text{YM}}\sum_{\ell=1}^{s}\eth^{s-\ell}[(-\partial_{u})^{-\ell}(\alpha_{s-\ell}A),\eth^{\ell}\mathcal{Q}_{-1}^{\left(j\right)}] \\
	&+\sum_{i=0}^{j-1}\sum_{\ell=1}^{s}(\ell+j-1)\,\eth^{s-\ell}((-\partial_{u})^{1-\ell-i}(\alpha_{s-\ell}\sigma^{\left(2-i\right)})\bullet\eth^{\ell-1+i}\mathcal{Q}_{-1}^{\left(j-i\right)}) + \mathcal{O}\left(\mathbb{F}^3\right) \,.
\ea\ee

\subsubsection*{Soft/hard decomposition}
Let us now write the charges explicitly in terms of the radiative degrees of freedom $\bar{F}$ and $\bar{N}$. Through this section, we will use the collective notation
\be
	\bar{N}^{\left(j\right)}(u,x^{A}) \defeq \begin{cases} \bar{F}(u,x^{A}) & \text{for $j=1$}\,; \\ \bar{N}(u,x^{A}) & \text{for $j=2$}\,; \end{cases}
\ee
for the spacetime-spin-$j$ radiation field. Similarly, $N^{\left(j\right)}$ will be used for the positive helicity radiation fields.
The relation between $\mathcal{Q}_{-j}^{\left(j\right)}$ and the radiation fields can then be written as
\be\label{eq:QmjNj}
	\mathcal{Q}_{-j}^{\left(j\right)}(u,x^{A}) = -\left(-\partial_{u}\right)^{j-1}\bar{N}^{\left(j\right)}(u,x^{A}) \,.
\ee
For future convenience, we also introduce the following notation
\be
	C^{\left(j\right)}(u,x^{A}) \defeq \begin{cases} A(u,x^{A}) & \text{for $j=1$}\,; \\ C(u,x^{A}) & \text{for $j=2$}\,; \end{cases}
\ee
for the collection of the positive helicity gravitational shear and positive helicity gluon field in a single object, while $\bar{C}^{\left(j\right)}$ will denote the corresponding negative helicity quantities. Then, $N^{\left(j\right)} = \partial_{u}C^{\left(j\right)}$ by definition, and, furthermore,
\be
	\sigma^{\left(j\right)}(u,x^{A}) = \partial_{u}^{2-j}C^{\left(j\right)}(u,x^{A}) \,,\quad j=1,2 \,.
\ee

We can now start iteratively integrating the evolution equations~\eqref{eq:EvolutionQsEYM} for all $s\ge-j+1$. We obtain
\be\label{eq:AntiEvolQsEYM}
	\mathcal{Q}_{s}^{\left(j\right)} = \partial_{u}^{-1}\left\{\eth\mathcal{Q}_{s-1}^{\left(j\right)} - ig_{\text{YM}}\left[A,\mathcal{Q}_{s-1}^{\left(j\right)}\right] + \left(j+s-1\right)\sum_{i=0}^{j-1}\partial_{u}^{i}C^{\left(2-i\right)}\bullet\mathcal{Q}_{s-2+i}^{\left(j-i\right)}\right\} \,,
\ee
starting from $s=-j+1$, with the iterated anti-derivative operator $\partial_{u}^{-n}$ ($n\ge0$) defined in \eqref{eq:anti-u}.

We will iteratively apply~\eqref{eq:AntiEvolQsEYM}, while also performing a soft/hard decomposition of the charge aspects, that is, we will perturbatively expand the charge aspects in powers of the fields\footnote{For gravity or non-self-interacting gauge fields of other spacetime spin, the fact that the $k$-increasing terms in the recursion relations come from charge aspects of spin-weight offset by $2$ units change the upper limits of the sums from $s+j+1$ to $\left\lfloor\frac{s+j}{2}\right\rfloor+1$. This is also true for the soft/hard decomposition of $\tilde{q}_{s}^{\left(j\right)}$ and $q_{s}^{\left(j\right)}$ below.},
\be
	\mathcal{Q}_{s}^{\left(j\right)}(u,x^{A}) = \sum_{k=1}^{s+j+1}\mathcal{Q}_{s}^{k\left(j\right)}(u,x^{A}) \,,
\ee
with each term $\mathcal{Q}_{s}^{k\left(j\right)}$ containing $1$ insertion of the radiation fields $\bar{N}^{\left(j\right)}$ and $k-1$ insertions of $C^{\left(j\right)}$. As in the Einstein-Maxwell case, it suffices to work out the soft ($k=1$) and the leading order hard ($k=2$) charge aspects. Starting from $s=-j+1$ and using that the relation between $\mathcal{Q}_{-j}^{\left(j\right)}$ and $\bar{N}^{\left(j\right)}$ from~\eqref{eq:QmjNj}, the algorithm~\eqref{eq:AntiEvolQsEYM} allows to work our way up and eventually arrive at the following expressions
\begin{subequations}
\begin{align}
	&\mathcal{Q}_{s\ge-j}^{1\left(j\right)} = -(\partial_{u}^{-1}\eth)^{s+j}(-\partial_{u})^{j-1}\bar{N}^{\left(j\right)} \,, \\[5pt]
	&\mathcal{Q}_{s\ge-j+1}^{2\left(j\right)} = \partial_{u}^{-1}\bigg\{ig_{\text{YM}}\sum_{\ell=1-j}^{s}(\partial_{u}^{-1}\eth)^{s-\ell} [A,(\partial_{u}^{-1}\eth)^{\ell+j-1}(-\partial_{u})^{j-1}\bar{N}^{\left(j\right)}] \\[-2pt]
	&-\sum_{i=0}^{j-1}\sum_{\ell=2-j}^{s}(\ell+j-1)(\partial_{u}^{-1}\eth)^{s-\ell}(\partial_{u}^{i}C^{\left(2-i\right)}\bullet(\partial_{u}^{-1}\eth)^{\ell+j-2}(-\partial_{u})^{j-i-1}\bar{N}^{\left(j-i\right)}) \bigg\} \nonumber \,.
\end{align}
\end{subequations}
More explicitly, for each $j=1,2$, these become respectively the following Einstein--Yang-Mills charges,
\begin{subequations}
    \be\begin{gathered}
        \mathcal{Q}_{s\ge-1}^{1,gl} = -(\partial_{u}^{-1}\eth)^{s+1}\bar{F} \,, \\
	   \begin{split}
            \mathcal{Q}_{s\ge0}^{2,gl} &= \partial_{u}^{-1}\bigg\{ ig_{\text{YM}}\sum_{\ell=0}^{s}(\partial_{u}^{-1}\eth)^{s-\ell}[A,(\partial_{u}^{-1}\eth)^{\ell}\bar{F}] - \sum_{\ell=1}^{s}\ell\,(\partial_{u}^{-1}\eth)^{s-\ell}(C(\partial_{u}^{-1}\eth)^{\ell-1}\bar{F}) \bigg\} 
        \end{split}
    \end{gathered}\ee 
    \be\begin{gathered}
        \mathcal{Q}_{s\ge-2}^{1,gr} = (\partial_{u}^{-1}\eth)^{s+2}\partial_{u}\bar{N} \,, \\
        \begin{split}
            \mathcal{Q}_{s\ge-1}^{2,gr} &= \partial_{u}^{-1}\sum_{\ell=0}^{s}(\ell+1)(\partial_{u}^{-1}\eth)^{s-\ell}\bigg\{C(\partial_{u}^{-1}\eth)^{\ell}\partial_{u}\bar{N} - \tr\left(F(\partial_{u}^{-1}\eth)^{\ell}\bar{F}\right) \bigg\}\,. 
        \end{split}
    \end{gathered}\ee
\end{subequations}

Then, performing a similar decomposition for the quasi-conserved renormalized charges,
\be
	\tilde{q}_{s}^{\left(j\right)}(u,x^{A}) = \sum_{k=1}^{s+j+1}\tilde{q}_{s}^{k\left(j\right)}(u,x^{A}) \,,
\ee
we find, for the soft part,
\be
	\tilde{q}_{s}^{1\left(j\right)} = (-1)^{j}\partial_{u}^{-1}\left\{\alpha_{s}\eth^{s+j}\bar{N}^{\left(j\right)} \right\} \,,
\ee
where, in deriving the above expression, we have made use of the integral Leibniz rule~\eqref{eq:IntLR2} to write $\tilde{q}_{s}^{1\left(j\right)}$ as a total anti-derivative. 
Next, for the leading order part of the hard charges, $\tilde{q}_s^{2\left(j\right)}$, it is once again sufficient to find $\partial_{u}q_{s}^{2\left(j\right)}$ since this will be the only part of the quadratic charges that will enter the derivation of the $sw_{1+\infty}$ algebra, as we will see at the end of this section. Using the quasi-conservation equation~\eqref{eq:EvolutionstqsEYM} for $\tilde{q}_{s}^{\left(j\right)}$, we then find
\begin{equation}\badat{3}
	\partial_{u}\tilde{q}_{s}^{2\left(j\right)} &= ig_{\text{YM}}(-1)^{j-1}\sum_{\ell=0}^{s}\eth^{s-\ell}[(-\partial_{u})^{-\ell}(\alpha_{s-\ell}A),\eth^{\ell+j-1}\bar{N}^{\left(j\right)}] \\
	&\quad-\sum_{i=0}^{j-1}(-1)^{j-i-1}\bigg\{ (j-1)\,\alpha_{s}\eth^{s}(\partial_{u}^{i}C^{\left(2-i\right)}\bullet\partial_{u}^{1-i}\eth^{j-2}\bar{N}^{\left(j-i\right)}) \\
	&\quad+\sum_{\ell=1}^{s}(\ell+j-1)\,\eth^{s-\ell}((-\partial_{u})^{1-\ell-i}(\alpha_{s-\ell}\partial_{u}^{i}C^{\left(2-i\right)})\bullet\eth^{\ell+j-2}\bar{N}^{\left(j-i\right)}) \bigg\} \,.
\eadat\end{equation}
More explicitly, for each $j=1,2$, these read
\begin{subequations}
	\begin{align}
			&\partial_{u}\tilde{q}_{s}^{2,gl} = \sum_{\ell=0}^{s}\eth^{s-\ell}\bigg\{ ig_{\text{YM}}[(-\partial_{u})^{-\ell}(\alpha_{s-\ell}A),\eth^{\ell}\bar{F}] - \ell\,(-\partial_{u})^{1-\ell}(\alpha_{s-\ell}C)\eth^{\ell-1}\bar{F} \bigg\} \,,\\
		&	\partial_{u}\tilde{q}_{s}^{2,gr} = \alpha_{s}\eth^{s}(C\partial_{u}\bar{N}) + \sum_{\ell=1}^{s}(\ell+1)\,\eth^{s-\ell}((-\partial_{u})^{1-\ell}(\alpha_{s-\ell}C)\eth^{\ell}\bar{N}) \nonumber\\
			&\quad \quad\quad \quad-\tr\bigg\{\sum_{\ell=0}^{s}(\ell+1)\,\eth^{s-\ell}((-\partial_{u})^{-\ell}(\alpha_{s-\ell}F)\eth^{\ell}\bar{F}) \bigg\} \,.
	\end{align}
\end{subequations}

\subsection{Einstein-YM celestial algebra}

In this final section, we derive the $sw_{1+\infty}$ charge-algebra brackets, using the Einstein--Yang-Mills charges constructed above. Similarly to the previous section, the goal is to compute the linearized Poisson bracket
\be
	\left\{q_{s}^{\left(j\right)}(x^{A}),q_{s^{\prime}}^{\left(j^{\prime}\right)}(x^{\prime A})\right\}^{(1)} = \left\{q_{s}^{1\left(j\right)}(x^{A}),q_{s^{\prime}}^{2\left(j^{\prime}\right)}(x^{\prime A})\right\} + \left\{q_{s}^{2\left(j\right)}(x^{A}),q_{s^{\prime}}^{1\left(j^{\prime}\right)}(x^{\prime A})\right\}\,,
\ee
among the celestial charges defined as~\cite{Freidel:2021ytz,Geiller:2024bgf}
\be
	q_{s}^{\left(j\right)}(x^{A}) \defeq \lim_{u\rightarrow-\infty}\tilde{q}_{s}^{\left(j\right)}(u,x^{A})\,.
\ee
The soft part $q_{s}^{1\left(j\right)}$ is easily found to be
\be
	q_{s}^{1\left(j\right)} = (-1)^{j-1}\eth^{s+j}\bar{N}^{\left(j\right)}_{s} \,,
\ee
where
\be
	\bar{N}^{\left(j\right)}_{s}(x^{A}) \defeq \int_{-\infty}^{+\infty}du\,\alpha_{s}(u)\bar{N}^{\left(j\right)}(u,x^{A})
\ee
is the negative helicity $\left(\text{sub}\right)^{s}$-leading soft operator of spacetime spin $j$. 

In order to compute $\left\{q_{s}^{2\left(j\right)}(x^{A}),q_{s^{\prime}}^{1\left(j^{\prime}\right)}(x^{\prime} A)\right\}$, we first need the canonical brackets between the fundamental fields. As demonstrated in Appendix~\ref{app:CanBraAntiDer}, these can be collectively written as
\be
	\left\{C^{\left(j\right)}(u,z),\bar{N}^{\left(j^{\prime}\right)}(u^{\prime},z^{\prime})\right\} = \kappa^2\,\delta(u-u^{\prime})\delta(z,z^{\prime})\delta^{j,j^{\prime}}\mathbf{1}_{\text{c}}^{-1} \,,
\ee
where $\mathbf{1}_{\text{c}}^{-1}$ refers to the color space inverse metric structure as dictated by the color of the objects involved in the bracket.

Then, for instance, the action of the soft celestial charges on the radiative data can be computed to be
\be
    \left\{q_{s}^{1\left(j\right)}(z),N^{\left(j^{\prime}\right)}(u^{\prime},z^{\prime})\right\} = -\kappa^2(-1)^{j}\alpha^{\prime}_{s-1}\eth_{z}^{s+j}\delta(z,z^{\prime}) \,,
\ee
where it is understood that $\alpha_{n}^{\prime} \defeq \alpha_{n}(u^{\prime})$.

Next, we wish to find the action of the hard celestial charges $q_{s}^{2\left(j\right)}(z)$ on the gravitational shear and on the gluon field. For the moment, we will focus on the piece that is independent of the color structure of the fields, i.e. on the part that is zeroth order in the YM coupling constant. We then have, using in the previous section the prescription \eqref{eq:prescription} and the manipulations mentioned there,
\be\ba\label{eq:compact1}
	{}&\left\{q_{s}^{2\left(j\right)}(z),C^{\left(j^{\prime}\right)}(u^{\prime},z^{\prime})\right\} = \lim_{u\rightarrow-\infty}\partial_{u}^{-1}\left\{\partial_{u}\tilde{q}_{s}^{2\left(j\right)}(u,z),C^{\left(j^{\prime}\right)}(u^{\prime},z^{\prime})\right\} \\
	&= -\kappa^2\,\delta_{1 \le j^{\prime} \le j}\mathbf{1}_{c}^{-1} \sum_{n=2-j}^{s}\frac{(-1)^{s-n}(n+j-1)}{(s-n)!}\times \\
	&\quad(\Delta_{u^{\prime}}+j^{\prime})_{s-n}\partial_{u^{\prime}}^{1-s}\eth_{z^{\prime}}^{n+j-2}C^{\left(2-j+j^{\prime}\right)}(u^{\prime},z^{\prime})\eth_{z}^{s-n}\delta(z,z^{\prime}) + \mathcal{O}\left(g_{\text{YM}}\right)\,,
\ea\ee
and
\be\ba\label{eq:compact2}
	{}&\left\{q_{s}^{2\left(j\right)}(z),\bar{C}^{\left(j^{\prime}\right)}(u^{\prime},z^{\prime})\right\} = \lim_{u\rightarrow-\infty}\partial_{u}^{-1}\left\{\partial_{u}\tilde{q}_{s}^{2\left(j\right)}(u,z),\bar{C}^{\left(j+j^{\prime}-2\right)}(u^{\prime},z^{\prime})\right\} \\
	&= -\kappa^2(-1)^{j}\,\delta_{3-j \le j^{\prime} \le 2}\mathbf{1}_{c}^{-1}\sum_{n=2-j}^{s}\frac{(-1)^{s-n}(n+j-1)}{(s-n)!}\times \\
	&\quad(\Delta_{u^{\prime}}-j^{\prime})_{s-n}\partial_{u^{\prime}}^{1-s}\eth_{z^{\prime}}^{n+j-2}\bar{C}^{\left(j+j^{\prime}-2\right)}(u^{\prime},z^{\prime})\eth_{z}^{s-n}\delta(z,z^{\prime}) + \mathcal{O}\left(g_{\text{YM}}\right)\,.
\ea\ee
As for the $\mathcal{O}\left(g_{\text{YM}}\right)$ terms, these only enter for $j=j^{\prime}=1$ and can be worked out to be
\be\ba
	{}&\left\{q_{s}^{2\left(1\right)a}(z),A^{b}(u^{\prime},z^{\prime})\right\}\bigg|_{\mathcal{O}\left(g_{\text{YM}}\right)} \\
    &= -\kappa^2g_{\text{YM}}f_{cd}^{\,\,\,\,\,a}\delta^{bd}\sum_{n=0}^{s}\frac{(-1)^{s-n}}{(s-n)!}(\Delta_{u^{\prime}})_{s-n}\partial_{u^{\prime}}^{-s}\eth_{z^{\prime}}^{n}A^{c}(u^{\prime},z^{\prime})\eth_{z}^{s-n}\delta(z,z^{\prime}) \,, \\
	{}&\left\{q_{s}^{2\left(1\right)a}(z),\bar{A}^{b}(u^{\prime},z^{\prime})\right\}\bigg|_{\mathcal{O}\left(g_{\text{YM}}\right)} \\
    &= -\kappa^2g_{\text{YM}}f_{cd}^{\,\,\,\,\,a}\delta^{bd}\sum_{n=0}^{s}\frac{(-1)^{s-n}}{(s-n)!}(\Delta_{u^{\prime}}-2)_{s-n}\partial_{u^{\prime}}^{-s}\eth_{z^{\prime}}^{n}\bar{A}^{c}(u^{\prime},z^{\prime})\eth_{z}^{s-n}\delta(z,z^{\prime}) \,.
\ea\ee

Summarizing the results, for each $j,j^{\prime}=1,2$, we have found
\be\ba
	&\left\{q_{s}^{2\left(1\right)a}(z),A^{b}(u^{\prime},z^{\prime})\right\} = -\kappa^2\sum_{n=0}^{s}\frac{(-1)^{s-n}}{(s-n)!}\bigg\{ \\
	& \quad \quad\quad\quad\quad\quad\quad\quad\quad\quad +g_{\text{YM}}f_{cd}^{\,\,\,\,\,a}\delta^{bd}(\Delta_{u^{\prime}})_{s-n}\partial_{u^{\prime}}^{-s}\eth_{z^{\prime}}^{n}A^{c}(u^{\prime},z^{\prime})\eth_{z}^{s-n}\delta(z,z^{\prime}) \\
	&\quad \quad\quad\quad\quad\quad\quad\quad\quad\quad +\delta^{ab}n\,(\Delta_{u^{\prime}}+1)_{s-n}\partial_{u^{\prime}}^{1-s}\eth_{z^{\prime}}^{n-1}C(u^{\prime},z^{\prime})\eth_{z}^{s-n}\delta(z,z^{\prime}) \bigg\}\,, \\
	&\left\{q_{s}^{2\left(1\right)a}(z),\bar{A}^{b}(u^{\prime},z^{\prime})\right\} = -\kappa^2g_{\text{YM}}f_{cd}^{\,\,\,\,\,a}\delta^{bd}\sum_{n=0}^{s}\frac{(-1)^{s-n}}{(s-n)!}(\Delta_{u^{\prime}}-2)_{s-n}\partial_{u^{\prime}}^{-s}\eth_{z^{\prime}}^{n}\bar{A}^{c}(u^{\prime},z^{\prime})\eth_{z}^{s-n}\delta(z,z^{\prime}) \,, \\
	&\left\{q_{s}^{2\left(1\right)a}(z),C(u^{\prime},z^{\prime})\right\} = 0 \,, \\
	&\left\{q_{s}^{2\left(1\right)a}(z),\bar{C}(u^{\prime},z^{\prime})\right\} = +\kappa^2\sum_{n=1}^{s}\frac{(-1)^{s-n}n}{(s-n)!}(\Delta_{u^{\prime}}-2)_{s-n}\partial_{u^{\prime}}^{1-s}\eth_{z^{\prime}}^{n-1}\bar{A}^{a}(u^{\prime},z^{\prime})\eth_{z}^{s-n}\delta(z,z^{\prime}) \,,
\ea\ee
\be\ba\label{eq:qs2bra}
	&\left\{q_{s}^{2\left(2\right)}(z),A^{a}(u^{\prime},z^{\prime})\right\} = -\kappa^2\sum_{n=0}^{s}\frac{(-1)^{s-n}(n+1)}{(s-n)!}(\Delta_{u^{\prime}}+1)_{s-n}\partial_{u^{\prime}}^{1-s}\eth_{z^{\prime}}^{n}A^{a}(u^{\prime},z^{\prime})\eth_{z}^{s-n}\delta(z,z^{\prime}) \,, \\
	&\left\{q_{s}^{2\left(2\right)}(z),\bar{A}^{a}(u^{\prime},z^{\prime})\right\} = -\kappa^2 \sum_{n=0}^{s}\frac{(-1)^{s-n}(n+1)}{(s-n)!}(\Delta_{u^{\prime}}-1)_{s-n}\partial_{u^{\prime}}^{1-s}\eth_{z^{\prime}}^{n}\bar{A}^{a}(u^{\prime},z^{\prime})\eth_{z}^{s-n}\delta(z,z^{\prime}) \,, \\
	&\left\{q_{s}^{2\left(2\right)}(z),C(u^{\prime},z^{\prime})\right\} = -\kappa^2 \sum_{n=0}^{s}\frac{(-1)^{s-n}(n+1)}{(s-n)!}(\Delta_{u^{\prime}}+2)_{s-n}\partial_{u^{\prime}}^{1-s}\eth_{z^{\prime}}^{n}C(u^{\prime},z^{\prime})\eth_{z}^{s-n}\delta(z,z^{\prime}) \,, \\
	&\left\{q_{s}^{2\left(2\right)}(z),\bar{C}(u^{\prime},z^{\prime})\right\} = -\kappa^2 \sum_{n=0}^{s}\frac{(-1)^{s-n}(n+1)}{(s-n)!}(\Delta_{u^{\prime}}-2)_{s-n}\partial_{u^{\prime}}^{1-s}\eth_{z^{\prime}}^{n}\bar{C}(u^{\prime},z^{\prime})\eth_{z}^{s-n}\delta(z,z^{\prime}) \,.
\ea\ee

Moving on, we want to compute the bracket $\left\{q_{s}^{2\left(j\right)}(z),q_{s^{\prime}}^{1\left(j^{\prime}\right)}(z^{\prime})\right\}$, an intermediate step being the extraction of the action of the quadratic charges on the $\left(\text{sub}\right)^{s}$-leading soft operator. One can show that
\be\ba
	{}&\left\{q_{s}^{2\left(j\right)}(z),\bar{N}^{\left(j^{\prime}\right)}_{s^{\prime}}(z^{\prime})\right\} = -\kappa^2(-1)^{j}\delta_{3-j \le j^{\prime} \le 2}\mathbf{1}_{c}^{-1} \\
	&\times\sum_{n=2-j}^{s}\binom{s+s^{\prime}-n+j^{\prime}-2}{s^{\prime}+j^{\prime}-2}(n+j-1)\,\eth_{z^{\prime}}^{n+j-2}\bar{N}^{\left(j+j^{\prime}-2\right)}_{s+s^{\prime}-1}(z^{\prime})\eth_{z}^{s-n}\delta(z,z^{\prime}) + \mathcal{O}\left(g_{\text{YM}}\right) \,,
\ea\ee
and, hence,
\be\ba
	{}&\left\{q_{s}^{2\left(j\right)}(z),q_{s^{\prime}}^{1\left(j^{\prime}\right)}(z^{\prime})\right\} = +\kappa^2(-1)^{j+j^{\prime}}\delta_{3-j \le j^{\prime} \le 2}\mathbf{1}_{c}^{-1} \\
	&\times\sum_{n=2-j}^{s}\binom{s+s^{\prime}-n+j^{\prime}-2}{s^{\prime}+j^{\prime}-2}(n+j-1)\,\eth_{z^{\prime}}^{s^{\prime}+j^{\prime}}(\eth_{z^{\prime}}^{n+j-2}\bar{N}^{\left(j+j^{\prime}-2\right)}_{s+s^{\prime}-1}(z^{\prime})\eth_{z}^{s-n}\delta(z,z^{\prime})) + \mathcal{O}\left(g_{\text{YM}}\right)  \,.
\ea\ee
The new terms arising from the color structure of the YM fields only contribute for $j=j^{\prime}=1$ and can be computed to be
\be\ba
	&\left\{q_{s}^{2\left(1\right)a}(z),\bar{F}^{b}_{s^{\prime}}(z^{\prime})\right\}\bigg|_{\mathcal{O}\left(g_{\text{YM}}\right)} = -\kappa^2g_{\text{YM}}f_{cd}^{\,\,\,\,\,a}\delta^{bd}\sum_{n=0}^{s}\binom{s+s^{\prime}-n}{s^{\prime}}\eth_{z^{\prime}}^{n}\bar{F}^{c}_{s+s^{\prime}}(z^{\prime})\eth_{z}^{s-n}\delta(z,z^{\prime}) \,, \\
	&\left\{q_{s}^{2\left(1\right)a}(z),q_{s^{\prime}}^{1\left(1\right)b}(z^{\prime})\right\}\bigg|_{\mathcal{O}\left(g_{\text{YM}}\right)} = -\kappa^2g_{\text{YM}}f_{cd}^{\,\,\,\,\,a}\delta^{bd}\sum_{n=0}^{s}\binom{s+s^{\prime}-n}{s^{\prime}}\eth_{z^{\prime}}^{s^{\prime}+1}\left(\eth_{z^{\prime}}^{n}\bar{F}^{c}_{s+s^{\prime}}(z^{\prime})\eth_{z}^{s-n}\delta(z,z^{\prime})\right) \,.
\ea\ee
Putting everything together, we arrive at the following list of brackets
\be\ba
	&\left\{q_{s}^{2\left(1\right)a}(z),\bar{F}^{b}_{s^{\prime}}(z^{\prime})\right\} = -\kappa^2g_{\text{YM}}f_{cd}^{\,\,\,\,\,a}\delta^{bd}\sum_{n=0}^{s}\binom{s+s^{\prime}-n}{s^{\prime}}\eth_{z^{\prime}}^{n}\bar{F}^{c}_{s+s^{\prime}}(z^{\prime})\eth_{z}^{s-n}\delta(z,z^{\prime}) \,, \\
	&\left\{q_{s}^{2\left(1\right)a}(z),\bar{N}_{s^{\prime}}(z^{\prime})\right\} = +\kappa^2\sum_{n=1}^{s}\binom{s+s^{\prime}-n}{s^{\prime}}n\,\eth_{z^{\prime}}^{n-1}\bar{F}^{a}_{s+s^{\prime}-1}(z^{\prime})\eth_{z}^{s-n}\delta(z,z^{\prime}) \,, \\
	&\left\{q_{s}^{2\left(2\right)}(z),\bar{F}^{a}_{s^{\prime}}(z^{\prime})\right\} = -\kappa^2\sum_{n=0}^{s}\binom{s+s^{\prime}-n-1}{s^{\prime}-1}(n+1)\eth_{z^{\prime}}^{n}\bar{F}^{a}_{s+s^{\prime}-1}(z^{\prime})\eth_{z}^{s-n}\delta(z,z^{\prime}) \,, \\
	&\left\{q_{s}^{2\left(2\right)}(z),\bar{N}_{s^{\prime}}(z^{\prime})\right\} = -\kappa^2\sum_{n=0}^{s}\binom{s+s^{\prime}-n}{s^{\prime}}(n+1)\eth_{z^{\prime}}^{n}\bar{N}_{s+s^{\prime}-1}(z^{\prime})\eth_{z}^{s-n}\delta(z,z^{\prime}) \,,
\ea\ee
and
\be\ba
	&\left\{q_{s}^{2\left(1\right)a}(z),q_{s^{\prime}}^{1\left(1\right)b}(z^{\prime})\right\} = -\kappa^2g_{\text{YM}}f_{cd}^{\,\,\,\,\,a}\delta^{bd}\sum_{n=0}^{s}\binom{s+s^{\prime}-n}{s^{\prime}}\eth_{z^{\prime}}^{s^{\prime}+1}\left(\eth_{z^{\prime}}^{n}\bar{F}^{c}_{s+s^{\prime}}(z^{\prime})\eth_{z}^{s-n}\delta(z,z^{\prime})\right), \\
	&\left\{q_{s}^{2\left(1\right)a}(z),q_{s^{\prime}}^{1\left(2\right)}(z^{\prime})\right\} = -\kappa^2\sum_{n=1}^{s}\binom{s+s^{\prime}-n}{s^{\prime}}n\,\eth_{z^{\prime}}^{s^{\prime}+2}\left(\eth_{z^{\prime}}^{n-1}\bar{F}^{a}_{s+s^{\prime}-1}(z^{\prime})\eth_{z}^{s-n}\delta(z,z^{\prime})\right), \\
	&\left\{q_{s}^{2\left(2\right)}(z),q_{s^{\prime}}^{1\left(1\right)a}(z^{\prime})\right\} = -\kappa^2 \sum_{n=0}^{s}\binom{s+s^{\prime}-n-1}{s^{\prime}-1}(n+1)\eth_{z^{\prime}}^{s^{\prime}+1}\left(\eth_{z^{\prime}}^{n}\bar{F}^{a}_{s+s^{\prime}-1}(z^{\prime})\eth_{z}^{s-n}\delta(z,z^{\prime})\right), \\
	&\left\{q_{s}^{2\left(2\right)}(z),q_{s^{\prime}}^{1\left(2\right)}(z^{\prime})\right\} = +\kappa^2\sum_{n=0}^{s}\binom{s+s^{\prime}-n}{s^{\prime}}(n+1)\eth_{z^{\prime}}^{s^{\prime}+2}\left(\eth_{z^{\prime}}^{n}\bar{N}_{s+s^{\prime}-1}(z^{\prime})\eth_{z}^{s-n}\delta(z,z^{\prime})\right).
\ea\ee

We can now go ahead and compute the Poisson brackets between the $sw_{1+\infty}$ charges $q_{s}^{\left(j\right)}$ up to quadratic order. To be more precise, we wish to compute the Poisson brackets between the smeared charges
\be
	Q_{s}^{\left(j\right)}\left(Z_{s}\right) = \frac{1}{\kappa^2}\oint\,Z_{s}(z)q_{s}^{\left(j\right)}(z) \,,
\ee
with $Z_{s}(z)$ the spin weight $-s$ smearing function and we are reminding here that we are using the abbreviation of the spatial arguments, e.g. $Z_{s}(z) \defeq Z_{s}\left(z,\bar{z}\right)$. After some integrations by parts, one finds
\be\ba
	{}&\left\{Q_{s}^{2\left(j\right)}\left(Z_{s}\right),Q_{s^{\prime}}^{1\left(j^{\prime}\right)}\left(Z^{\prime}_{s^{\prime}}\right)\right\} = \delta_{3-j \le j^{\prime} \le 2}\mathbf{1}_{c}^{-1}\bigg\{ \\
	&\quad\quad\quad\left(s+j-1\right)Q_{s+s^{\prime}-1}^{1\left(j+j^{\prime}-2\right)}\left(Z_{s}\eth Z^{\prime}_{s^{\prime}}\right) + \oint\, \slashed{\eth}B^{jj^{\prime}}_{ss^{\prime}}\left(Z_{s},Z^{\prime}_{s^{\prime}}\right) \bigg\} + \mathcal{O}\left(g_{\text{YM}}\right) \,,
\ea\ee
where we have isolated the following terms entering the bracket above
\be\ba
	{}&\quad\slashed{\eth}B^{jj^{\prime}}_{ss^{\prime}}\left(Z_{s},Z^{\prime}_{s^{\prime}}\right) \defeq \frac{(-1)^{j+j^{\prime}}}{\kappa^2}\binom{s+s^{\prime}+j+j^{\prime}-4}{s^{\prime}+j^{\prime}-2}\sum_{k=0}^{s+j-2}\sum_{p=1}^{\ell+s^{\prime}+j^{\prime}-1}(-1)^{k+1} \times \\
	&\frac{(s+s^{\prime}+j+j^{\prime}-2)_2}{(k+s^{\prime}+j^{\prime})_2}\binom{s+j-2}{\ell}\binom{k+s^{\prime}+j^{\prime}-1}{p}\eth^{p}Z_{s}\eth Z^{\prime}_{s^{\prime}}\eth^{s+s^{\prime}+j+j^{\prime}-3-p}\bar{N}^{\left(j+j^{\prime}-2\right)}_{s+s^{\prime}-1} \,.
\ea\ee
This is because this quantity turns out to become a total derivative when combined with $\left\{Q_{s}^{1\left(j\right)}\left(Z_{s}\right),Q_{s^{\prime}}^{2\left(j^{\prime}\right)}\left(Z^{\prime}_{s^{\prime}}\right)\right\}$, due to the property
\be
	\slashed{\eth}B^{jj^{\prime}}_{ss^{\prime}}\left(Z_{s},Z^{\prime}_{s^{\prime}}\right) - \slashed{\eth}B^{j^{\prime}j}_{s^{\prime}s}\left(Z^{\prime}_{s^{\prime}}Z_{s}\right) = \eth\left(\dots\right)
\ee
where do no write down the explicit form of the right hand side, the only relevant property of it being that it is a total derivative that does not contribute when integrated over the boundary $2$-dimensional space. As a result, we arrive at
\be\ba
	{}&\left\{Q_{s}^{\left(j\right)}\left(Z_{s}\right),Q_{s^{\prime}}^{\left(j^{\prime}\right)}\left(Z^{\prime}_{s^{\prime}}\right)\right\}^{\left(1\right)} = \left\{Q_{s}^{2\left(j\right)}\left(Z_{s}\right),Q_{s^{\prime}}^{1\left(j^{\prime}\right)}\left(Z^{\prime}_{s^{\prime}}\right)\right\} + \left\{Q_{s}^{1\left(j\right)}\left(Z_{s}\right),Q_{s^{\prime}}^{2\left(j^{\prime}\right)}\left(Z^{\prime}_{s^{\prime}}\right)\right\} \\
	&= (1-\delta_{j1}\delta_{j^{\prime}1})\mathbf{1}_{c}^{-1}Q_{s+s^{\prime}-1}^{1\left(j+j^{\prime}-2\right)}\left((s+j-1)Z_{s}\eth Z^{\prime}_{s^{\prime}} - (s^{\prime}+j^{\prime}-1)Z^{\prime}_{s^{\prime}}\eth Z_{s}\right) + \mathcal{O}\left(g_{\text{YM}}\right) \,.
\ea\ee

Similarly, the $\mathcal{O}\left(g_{\text{YM}}\right)$ contributions enter only for $j=j^{\prime}=1$ and read
\be\ba
	{}&\left\{Q_{s}^{2\left(1\right)a}\left(Z_{s}\right),Q_{s^{\prime}}^{1\left(1\right)b}\left(Z^{\prime}_{s^{\prime}}\right)\right\} = g_{\text{YM}}f_{cd}^{\,\,\,\,\,a}\delta^{bd}\bigg\{Q_{s+s^{\prime}}^{1\left(1\right)c}\left(Z_{s}Z^{\prime}_{s^{\prime}}\right) + \oint\,\slashed{\eth}\tilde{B}^{c}_{ss^{\prime}}\left(Z_{s},Z^{\prime}_{s^{\prime}}\right) \bigg\} \,,
\ea\ee
with
\be\ba
	{}&\slashed{\eth}\tilde{B}^{c}_{ss^{\prime}}\left(Z_{s},Z^{\prime}_{s^{\prime}}\right) \defeq \frac{1}{\kappa^2}\binom{s+s^{\prime}}{s^{\prime}}\sum_{\ell=0}^{s}\sum_{n=1}^{\ell+s^{\prime}+1}(-1)^{\ell}\frac{s+s^{\prime}+1}{\ell+s^{\prime}+1}\binom{s}{\ell}\binom{\ell+s^{\prime}+1}{n}\eth^{n}Z_{s}Z^{\prime}_{s^{\prime}}\eth^{s+s^{\prime}+1-n}\bar{F}^{c}_{s+s^{\prime}}
\ea\ee
now satisfying
\be
	\slashed{\eth}\tilde{B}^{c}_{ss^{\prime}}\left(Z_{s},Z^{\prime}_{s^{\prime}}\right) + \slashed{\eth}\tilde{B}^{c}_{s^{\prime}s}\left(Z^{\prime}_{s^{\prime}},Z_{s}\right) = \eth\left(\dots\right) \,,
\ee
where we are again refraining from writing explicitly the right hand side and only emphasize that it is a total derivative. As a result,
\be\ba
	\left\{Q_{s}^{\left(1\right)a}\left(Z_{s}\right),Q_{s^{\prime}}^{\left(1\right)b}\left(Z^{\prime}_{s^{\prime}}\right)\right\}^{\left(1\right)} &= \left\{Q_{s}^{2\left(1\right)a}\left(Z_{s}\right),Q_{s^{\prime}}^{1\left(1\right)b}\left(Z^{\prime}_{s^{\prime}}\right)\right\} + \left\{Q_{s}^{1\left(1\right)a}\left(Z_{s}\right),Q_{s^{\prime}}^{2\left(1\right)b}\left(Z^{\prime}_{s^{\prime}}\right)\right\} \\
	&= 2g_{\text{YM}}f_{cd}^{\,\,\,\,\,a}\delta^{bd}Q_{s+s^{\prime}}^{1\left(1\right)c}\left(Z_{s}Z^{\prime}_{s^{\prime}}\right) \,.
\ea\ee

Summarizing, we have found the following $sw_{1+\infty}$ algebra
\begin{empheq}[box=\tcbhighmath]{align}\badat{2}\label{EYM_bracket}
	&\left\{Q_{s}^{\left(1\right)a}\left(Z_{s}\right),Q_{s^{\prime}}^{\left(1\right)b}\left(Z^{\prime}_{s^{\prime}}\right)\right\}^{\left(1\right)} = 2g_{\text{YM}}f_{cd}^{\,\,\,\,\,a}\delta^{bd}Q_{s+s^{\prime}}^{1\left(1\right)c}\left(Z_{s}Z^{\prime}_{s^{\prime}}\right) \,, \\
&	\left\{Q_{s}^{\left(2\right)}\left(Z_{s}\right),Q_{s^{\prime}}^{\left(1\right)a}\left(Z^{\prime}_{s^{\prime}}\right)\right\}^{\left(1\right)} = Q_{s+s^{\prime}-1}^{1\left(1\right)a}\left(\left(s+1\right)Z_{s}\eth Z^{\prime}_{s^{\prime}}-s^{\prime}Z^{\prime}_{s^{\prime}}\eth Z_{s}\right) \,, \\
	&\hspace{-0cm}\left\{Q_{s}^{\left(2\right)}\left(Z_{s}\right),Q_{s^{\prime}}^{\left(2\right)}\left(Z^{\prime}_{s^{\prime}}\right)\right\}^{\left(1\right)} = Q_{s+s^{\prime}-1}^{1\left(2\right)}\left(\left(s+1\right)Z_{s}\eth Z^{\prime}_{s^{\prime}}-\left(s^{\prime}+1\right)Z^{\prime}_{s^{\prime}}\eth Z_{s}\right).
\eadat\end{empheq}
Last, let us define the modes $Q_{s;k,\ell}^{\left(j\right)}$ in the holomorphic basis $Z_{s;k,\ell}^{\left(j\right)}$ for each spacetime spin $j$ according to
\be
	Q_{s;k,\ell}^{\left(j\right)} \defeq Q_{s}^{\left(j\right)}\left(Z_{s;k,\ell}^{\left(j\right)}\left(z,\bar{z}\right)\right) \,,\quad Z_{s;k,\ell}^{\left(j\right)}\left(z,\bar{z}\right) = z^{\frac{j}{2}+s-k}\bar{z}^{\frac{j}{2}-s-\ell} \,.
\ee
Then, the $sw_{1+\infty}$ algebra reads
\be\ba
	{}&\left\{Q_{s;k,\ell}^{\left(j\right)},Q_{s^{\prime};k^{\prime},\ell^{\prime}}^{\left(j^{\prime}\right)}\right\}^{\left(1\right)} = \delta_{j1}\delta_{j^{\prime}1}\mathcal{O}\left(g_{\text{YM}}\right) + \left(1-\delta_{j1}\delta_{j^{\prime}1}\right)\mathbf{1}_{c}^{-1}\times \\
	&\left[\left(k+\tfrac{j}{2}-1\right)\left(s^{\prime}+j^{\prime}-1\right)-\left(k^{\prime}+\tfrac{j^{\prime}}{2}-1\right)\left(s+j-1\right)\right]Q^{1\left(j+j^{\prime}-2\right)}_{s+s^{\prime}-1;k+k^{\prime}-1,\ell+\ell^{\prime}} \,,
\ea\ee
with
\be\ba
	\left\{Q_{s;k,\ell}^{\left(1\right)a},Q_{s^{\prime};k^{\prime},\ell^{\prime}}^{\left(1\right)b}\right\}^{\left(1\right)} = 2g_{\text{YM}}f_{cd}^{\,\,\,\,\,a}\delta^{bd}Q^{1\left(1\right)c}_{s+s^{\prime};k+k^{\prime}-\frac{1}{2},\ell+\ell^{\prime}-\frac{1}{2}} \,.
\ea\ee
In summary,
\be\ba
	\left\{Q_{s;k,\ell}^{\left(1\right)a},Q_{s^{\prime};k^{\prime},\ell^{\prime}}^{\left(1\right)b}\right\}^{\left(1\right)} &= 2g_{\text{YM}}f_{cd}^{\,\,\,\,\,a}\delta^{bd}Q^{1\left(1\right)c}_{s+s^{\prime};k+k^{\prime}-\frac{1}{2},\ell+\ell^{\prime}-\frac{1}{2}} \,, \\
	\left\{Q_{s;k,\ell}^{\left(2\right)},Q_{s^{\prime};k^{\prime},\ell^{\prime}}^{\left(1\right)a}\right\}^{\left(1\right)} &= \left[ks^{\prime}-\left(k^{\prime}-\tfrac{1}{2}\right)\left(s+1\right)\right]Q^{1\left(1\right)a}_{s+s^{\prime}-1;k+k^{\prime}-1,\ell+\ell^{\prime}} \,, \\
	\left\{Q_{s;k,\ell}^{\left(2\right)},Q_{s^{\prime};k^{\prime},\ell^{\prime}}^{\left(2\right)}\right\}^{\left(1\right)} &= \left[k\left(s^{\prime}+1\right)-k^{\prime}\left(s+1\right)\right]Q^{1\left(2\right)}_{s+s^{\prime}-1;k+k^{\prime}-1,\ell+\ell^{\prime}} \,.
\ea\ee
These can be brought in the $sw_{1+\infty}$ algebra as originally written in~\cite{Strominger:2021mtt} according to the matching
\be
	Q_{s;k}^{\left(1\right)a}\left(\bar{z}\right) = -2i\,s^{\frac{s+2}{2},a}_{\frac{s+1}{2}-k}\left(\bar{z}\right) \,,\quad Q_{s;k}^{\left(2\right)} = -2w^{\frac{s+3}{2}}_{\frac{s+1}{2}-k}\left(\bar{z}\right) \,,
\ee
where $Q_{s;k}^{\left(j\right)}\left(\bar{z}\right)$ refers to the partial modes
\be
	Q_{s;k}^{\left(j\right)}\left(\bar{z}\right) \defeq Q_{s}^{\left(j\right)}\left(Z_{s;k}^{\left(j\right)}\left(z,\bar{z}\right)\right) \,,\quad Z_{s;k}^{\left(j\right)}\left(z,\bar{z}\right) = z^{\frac{j}{2}+s-k}f\left(\bar{z}\right)
\ee
for arbitrary anti-holomorphic functions $f\left(\bar{z}\right)$. Then, one recovers
\be\ba\label{eq:sw}
	&\left\{s^{p,a}_{m},s^{q,b}_{n}\right\} = ig_{\text{YM}}f_{cd}^{\,\,\,\,\,a}\delta^{bd} s^{p+q-1,c}_{m+n} \,, \\
	&\left\{w^{p}_{m},s^{q,a}_{n}\right\} = \left(m\left(q-1\right)-n\left(p-1\right)\right)s^{p+q-2,a}_{m+n} \,, \\
	&\left\{w^{p}_{m},w^{q}_{n}\right\} = \left(m\left(q-1\right)-n\left(p-1\right)\right)w^{p+q-2}_{m+n} \,.
\ea\ee

\section{Summary and outlook}
\label{ref:summary}
In this paper, we have shown how the $sw_{1+\infty}$ algebra \eqref{eq:sw}, which organizes the current algebra soft sector of celestial CFT \cite{Guevara:2021abz,Strominger:2021lvk}, emerges from the structure of subleading equations of motion in Einstein--Yang-Mills theory. The underlying analysis, which was first reported in \cite{Freidel:2021ytz,Freidel:2023gue,Geiller:2024bgf}, is based on the identification of a truncation of the phase space which allows to write recursion relations for an infinite tower of higher spin weight charges. 
In this work, we have established the closure of renormalized higher spin charge algebra at linear order. A natural next step is to investigate the extent to which this construction can be carried out beyond the linear order, as has been explored in the context of Yang-Mills theory \cite{Freidel:2023gue} and gravity \cite{Cresto:2024fhd,Cresto:2024mne}.

One of the striking lessons one can learn from this work is that it would have seemed a priori impossible to predict the emergence of the $sw_{1+\infty}$ algebra solely from an asymptotic spacetime analysis without the insights from the celestial current algebra (or twistor theory). This is one of many examples highlighting the value of bridging different approaches to flat space holography: On one hand, the lowest-spin generators of this algebra (supertranslations and superrotations), identified long ago as asymptotic symmetries of flat spacetimes, were at the foundation of celestial holography. In return, insights from celestial OPEs subsequently revealed the presence of an infinite extension of these symmetries. This is of course far from the full story. Below, we present a non-exhaustive list of topics to which the existence of $w_{1+\infty}$ symmetries can be directly connected and explore potential future directions and open questions.\\

\noindent \textbf{Twistors}\quad
As already pointed out, the celestial $w_{1+\infty}$ algebra have a very natural geometric realization in twistor space, where they act as Poisson diffeomorphisms \cite{Penrose:1976js,Penrose:1976jq}. The connection between the twistor space realization of these symmetries and celestial OPE of conformally soft gravitons was explained in \cite{Adamo:2021lrv}. From a Carrollian perspective, the representation of  $w_{1+\infty}$ symmetries on gravitational data at $\mathscr I$ was derived explicitly from twistor space through the Penrose transform in \cite{Donnay:2024qwq}. 
In particular, this construction provides a concrete way of seeing how the correspondence between twistor space and $\mathscr I$ gives rise to a non-local spacetime representation: for a Carrollian zero-rest-mass fields $\phi_j$ of any spin $j \in \mathbb Z$,  Eq. (3.8) of \cite{Donnay:2024qwq} gives\footnote{Comparing with the notations of \cite{Donnay:2024qwq}, we have $s=n-1$ and the symmetry generators are $\tau_{n-1} = g_{\alpha(n)}\bar{\lambda}^{\dot{\alpha}(n)}$.}
\begin{equation}
    \begin{split}\nonumber
        \delta_{\tau_s}\phi_j
        &=  \sum_{\ell=0}^s \frac{(\ell+1)}{(s-\ell)!}( \eth_z^{s-\ell}\tau_s)\partial_u^{1+j}(u^{s-\ell}\partial_u^{-\ell-j}\eth_{z}^\ell \phi_j(u,z))\,.
    \end{split}
\end{equation}
One can check that the above expression exactly matches with the canonical bracket for the smeared charges we obtained here, i.e.  $\delta_{\tau_s}\phi_j=\{Q_s^{2,gr}(\tau_s), \phi_j(u,z)\}$, the latter being easily derived from the spin $j$ expressions given in  \eqref{eq:compact1}, \eqref{eq:compact2}. In \cite{Kmec:2024nmu}, the gravitational canonical celestial charges of \cite{Freidel:2021ytz,Geiller:2024bgf} were related to twistor space by means of a BF twistor action for self-dual gravity. It would be interesting to derive the full Einstein--Yang-Mills set of charges and canonical brackets we obtained in \eqref{eq:qs2bra} from a twistor construction as well.\\

\noindent \textbf{Scattering amplitudes}\quad
From an $S$-matrix perspective, the symmetries we studied in this work are related to the existence of an infinite set of soft photon and soft graviton theorems at tree level. The latter arise as a consequence of on-shell gauge invariance on certain projected pieces of the scattering amplitude \cite{Hamada:2018vrw,Li:2018gnc}. However, these `infinite-order soft theorems' are expected to be modified once accounting for loop corrections \cite{Bern:2014oka,Laddha:2018myi,Sahoo:2018lxl,Krishna:2023fxg,Ciafaloni:2018uwe,DiVecchia:2023frv,Alessio:2024onn,Alessio:2024wmz,Campiglia:2019wxe,Agrawal:2023zea,Choi:2024ygx} as well as when including higher-dimensional operators \cite{Bianchi:2014gla,Elvang:2016qvq,Laddha:2017vfh} which take into account the details of the interaction. In the context of celestial holography, studies of loop corrections and deformations of celestial soft algebras include Refs. \cite{Ball:2021tmb,Monteiro:2022xwq,Bhardwaj:2022anh,Krishna:2023ukw,Bhardwaj:2024wld,Bissi:2024brf,Melton:2022fsf,Ren:2022sws,Costello:2022upu,Costello:2022wso,Bittleston:2022jeq}, which provide important lessons for the fate of $w_{1+\infty}$ symmetry algebras beyond the context studied in this work. \\

\noindent \textbf{Multipoles and memories}\quad
An interesting relationship between $w_{1+\infty}$ celestial charges and multipole expansions of the gravitational field close to null and spatial infinity was pointed out in \cite{Compere:2022zdz}. It was shown that, in the linear theory, there is a correspondence between the tower of celestial charges and canonical multipole moments; see also~\cite{Blanchet:2020ngx,Blanchet:2023pce,Flanagan:2018yzh,Grant:2021hga,Siddhant:2024nft} for related studies. These works suggest that the infinite tower of celestial charges could be related to novel, subleading in magnitude, type of gravitational wave memory effects. Further exploration in this direction holds promising prospects, as potential links between higher memories, $w_{1+\infty}$ symmetries, and scattering amplitudes might unveil a new infrared web of relationships for gravity and gauge theories in asymptotically flat spacetimes. 

\begin{acknowledgments}
S.A., P.C. and L.D. are supported by the European Research Council (ERC) Project 101076737 -- CeleBH. Views and opinions expressed are however those of the authors only and do not necessarily reflect those of the European Union or the European Research Council. Neither the European Union nor the granting authority can be held responsible for them.
The authors also thank the INFN Iniziativa Specifica ST\&FI and the Programme ``Carrollian Physics and Holography'' at the Erwin-Schrödinger International Institute for Mathematics and Physics for their support. L.D.'s research was also supported in part by the Simons Foundation through the Simons Foundation Emmy Noether Fellows Program at Perimeter Institute. Research at Perimeter Institute is supported in part by the Government of Canada through the Department of Innovation, Science and Economic Development and by the Province of Ontario through the Ministry of Colleges and Universities. 
\end{acknowledgments}

\appendix

\section{Elements of the Newman-Penrose formalism}
\label{app:NPFormalism}

The Newman-Penrose (NP) formalism~\cite{Newman:1961qr,Geroch:1973am} is based on the vierbein formulation of gravitational dynamics, in the special case where the tetrad vectors are null. In particular, the NP formalism introduces two real null vectors, $\ell^{\mu}$ and $n^{\mu}$, and a doublet of complex null vectors, $m^{\mu}$ and $\bar{m}^{\mu}$, normalized such that
\be
	\ell_{\mu}n^{\mu} = -1 \,,\quad m_{\mu}\bar{m}^{\mu} = +1 \,,
\ee
with all other inner products evaluating to zero. The metric tensor is then reconstructed as follows
\be
	g^{\mu\nu} = 2\left(-\ell^{(\mu}n^{\nu)} + m^{(\mu}\bar{m}^{\nu)}\right) \,.
\ee

After the introduction of a set tetrad vectors, one may reformulate a theory built up from spacetime tensorial fields by projecting them onto the various local frame directions and working directly with the resulting spacetime scalar fields. For instance, the $6$ independent components of the field strength tensor $F_{\mu\nu}$ of a vector gauge field are traded for the $3$ complex Maxwell-NP scalars
\be
	\Phi_0 \defeq F_{\ell m} \,,\quad \Phi_1 \defeq \frac{1}{2}\left(F_{\ell n} - F_{m\bar{m}}\right) \,,\quad \Phi_2 \defeq F_{\bar{m}n} \,,
\ee
where we are using the notation of labeling the frame indices with the symbol of the corresponding tetrad vector, e.g. $F_{\ell m} \defeq \ell^{\mu}m^{\nu}F_{\mu\nu}$.

Similarly, the $10$ independent components of the Weyl tensor $C_{\rho\sigma\mu\nu}$ are rearranged into the $5$ complex Weyl-NP scalars,
\be
	\Psi_0 \defeq -C_{\ell m \ell m} \,,\quad \Psi_1 \defeq -C_{\ell m \ell n} \,,\quad \Psi_2 \defeq -C_{\ell m \bar{m}n} \,,\quad \Psi_3 \defeq -C_{\ell n\bar{m} n} \,,\quad \Psi_4 \defeq -C_{\bar{m}n\bar{m}n} \,,
\ee
while the $10$ independent components of the Ricci tensor $R_{\mu\nu}$ are rearranged into $4$ real and $3$ complex Ricci-NP scalars,
\be
	\begin{gathered}
		\Phi_{00} \defeq \frac{1}{2}R_{\ell\ell} \,,\quad \Phi_{11} \defeq \frac{1}{4}\left(R_{\ell n} + R_{m\bar{m}}\right) \,,\quad \Phi_{22} \defeq \frac{1}{2}R_{nn} \,,\quad \Lambda_{\text{R}} \defeq \frac{R}{24} \,, \\
		\Phi_{01} \defeq \frac{1}{2}R_{\ell m} = \bar{\Phi}_{10} \,,\quad \Phi_{12} \defeq \frac{1}{2}R_{nm} = \bar{\Phi}_{21} \,,\quad \Phi_{02} \defeq \frac{1}{2}R_{mm} = \bar{\Phi}_{20} \,.
	\end{gathered}
\ee

Finally, the spacetime covariant derivatives $\nabla_{\mu}$ are traded for the directional derivatives
\be
	\begin{pmatrix}
		D \\
		\triangle \\
		\delta \\
		\bar{\delta}
	\end{pmatrix} \defeq
	\begin{pmatrix}
		\ell^{\mu} \\
		n^{\mu} \\
		m^{\mu} \\
		\bar{m}^{\mu}
	\end{pmatrix}\nabla_{\mu} \quad\Leftrightarrow\quad \nabla_{\mu} = -\ell_{\mu}\triangle - n_{\mu}D +m_{\mu}\bar{\delta} + \bar{m}_{\mu}\delta \,,
\ee
while the Christoffel symbols are rearranged into $12$ complex spin coefficients
\be
	\begin{gathered}
		\begin{pmatrix}
			\kappa \\ \tau \\ \sigma \\ \rho
		\end{pmatrix} \defeq -m^{\mu}
		\begin{pmatrix}
			D \\ \triangle \\ \delta \\ \bar{\delta}
		\end{pmatrix}\ell_{\mu} \,,\quad
		\begin{pmatrix}
			\pi \\ \nu \\ \mu \\ \lambda
		\end{pmatrix} \defeq +\bar{m}^{\mu}
		\begin{pmatrix}
			D \\ \triangle \\ \delta \\ \bar{\delta}
		\end{pmatrix}n_{\mu} \,, \\
		\begin{pmatrix}
			\epsilon \\ \gamma \\ \beta \\ \alpha
		\end{pmatrix} \defeq +\frac{1}{2}\bigg(\bar{m}^{\mu}
		\begin{pmatrix}
			D \\ \triangle \\ \delta \\ \bar{\delta}
		\end{pmatrix}m_{\mu} - n^{\mu}
		\begin{pmatrix}
			D \\ \triangle \\ \delta \\ \bar{\delta}
		\end{pmatrix}\ell_{\mu} \bigg) \,.
	\end{gathered}
\ee

\subsection{Type-III transformations and the GHP derivatives}
The local Lorentz transformations are parameterized by $6$ local real parameters. These can be classified into $3$ types of transformations: the type-I and type-II transformations, each characterized by $1$ complex parameter, and the type-III transformations characterized by $2$ real parameters. The type-I and type-II transformations refer to local Lorentz transformations that preserve the null vector $\ell^{\mu}$ or the null vector $n^{\mu}$ respectively, but act non-homogeneously on the other tetrad vectors. On the other hand, type-III transformations correspond to homogeneous transformations of the tetrad vectors and are controlled by $2$ real parameters, $\upsilon$ and $\chi$. The former are reciprocal rescalings of the vectors $\ell^{\mu}$ and $n^{\mu}$, while leaving $m^{\mu}$ and $\bar{m}^{\mu}$ intact, while the latter are angles that capture the $SO\left(2\right)$ rotations in the $m-\bar{m}$ plane, while leaving the vectors $\ell^{\mu}$ and $n^{\mu}$ unchanged,
\be
	\begin{pmatrix}
		\ell^{\mu} \\
		n^{\mu} \\
		m^{\mu} \\
		\bar{m}^{\mu}
	\end{pmatrix} \xrightarrow[\upsilon,\chi]{\text{III}}
	\begin{pmatrix}
		\tilde{\ell}^{\mu} = e^{+\upsilon}\ell^{\mu} \\
		\tilde{n}^{\mu} = e^{-\upsilon}n^{\mu} \\
		\tilde{m}^{\mu} = e^{+i\chi}m^{\mu} \\
		\bar{\tilde{m}}^{\mu} = e^{-i\chi}\bar{m}^{\mu}
	\end{pmatrix} \,.
\ee
The homogeneity of these transformations allows to characterize the NP scalars one constructs by projecting spacetime tensors onto null directions. More specifically, an NP scalar $\psi_{b,s}$ is said to have boost-weight $b$ and spin weight $s$ if under type-III local Lorentz transformations it transforms as
\be
	\psi_{b,s} \xrightarrow[\upsilon,\chi]{\text{III}} \tilde{\psi}_{b,s} = e^{w\upsilon}e^{is\chi}\psi_{b,s} \,.
\ee
In Table~\ref{tbl:sbPsis} below, we summarize the boost-weights and spin weights of the various fundamental NP scalars that are obtained by projecting curvature tensors of various spacetime spins onto null directions. For antisymmetric spacetime tensors, such as the $2$-form field strength of a $1$-form gauge field, the resulting NP scalars are, in fact, uniquely characterized by their boost-weights and spin weights.

\begin{table}[H]
	\centering
	\begin{tabular}{|c|c|c|}
		\hline
		Fundamental NP scalar & $b$ & $s$ \\
		\hline\hline
		$\Phi$ & $0$ & $0$ \\
		\hline
		$\Phi_0$ & $+1$ & $+1$ \\
		$\Phi_1$ & $0$ & $0$ \\
		$\Phi_2$ & $-1$ & $-1$ \\
		\hline
		$\Psi_0$ & $+2$ & $+2$ \\
		$\Psi_1$ & $+1$ & $+1$ \\
		$\Psi_2$ & $0$ & $0$ \\
		$\Psi_3$ & $-1$ & $-1$ \\
		$\Psi_4$ & $-2$ & $-2$ \\
		\hline
	\end{tabular}
	\caption{Boost-weights $b$ and spin weights $s$ of the fundamental NP scalars for the curvature tensors associated with scalar, electromagnetic and gravitational fields.}
	\label{tbl:sbPsis}
\end{table}

A useful observation that will allow us to adopt a compact notation is that all of these fundamental NP scalars have equal boost-weights and spin weights and, more importantly, we remark that the spin weight of the Maxwell-NP and Weyl-NP scalars can be read directly from the $\Phi_{1-s}$ and $\Psi_{2-s}$ respectively.

Last, we remark that some of the spin coefficients also transform homogeneously under type-III transformations. These are the so-called ``good'' spin coefficients and their boost-weights and spin weights are given in Table~\ref{tbl:sbGammas} below.
\begin{table}[H]
	\centering
	\begin{tabular}{|c|c|c|}
		\hline
		Spin coefficient & $b$ & $s$ \\
		\hline\hline
		$\kappa$ & $+2$ & $+1$ \\
		$\tau$ & $0$ & $+1$ \\
		$\sigma$ & $+1$ & $+2$ \\
		$\rho$ & $+1$ & $0$ \\
		\hline
		$\pi$ & $0$ & $-1$ \\
		$\nu$ & $-2$ & $-1$ \\
		$\mu$ & $-1$ & $0$ \\
		$\lambda$ & $-1$ & $-2$ \\
		\hline
	\end{tabular}
	\caption{Boost-weights $b$ and spin weights $s$ of the ``good'' spin coefficients.}
	\label{tbl:sbGammas}
\end{table}
On the other hand, the ``bad'' spin coefficients $\epsilon$, $\gamma$, $\beta$ and $\alpha$ do not have definite boost-weights and spin weights, namely,
\be
	\begin{pmatrix}
		\epsilon \\
		\gamma \\
		\beta \\
		\alpha
	\end{pmatrix} \xrightarrow[\upsilon,\chi]{\text{III}}
	\begin{pmatrix}
		\tilde{\epsilon} = e^{+\upsilon}\left(\epsilon + \frac{1}{2}D\mathscr{A}\right) \\
		\tilde{\gamma} = e^{-\upsilon}\left(\gamma + \frac{1}{2}\triangle\mathscr{A}\right) \\
		\tilde{\beta} = e^{+i\chi}\left(\beta + \frac{1}{2}\delta\mathscr{A}\right) \\
		\tilde{\alpha} = e^{-i\chi}\left(\alpha + \frac{1}{2}\bar{\delta}\mathscr{A}\right)
	\end{pmatrix} \,,\quad \mathscr{A} \defeq \upsilon + i\chi \,.
\ee
This is to be contrasted to the inhomogeneous transformations of the directional derivatives when acting on a boost-weight-$b$ and spin weight-$s$ NP scalar,
\be
	\begin{pmatrix}
		D \\
		\triangle \\
		\delta \\
		\bar{\delta}
	\end{pmatrix}\psi_{b,s} \xrightarrow[\upsilon,\chi]{\text{III}} e^{b\upsilon}e^{is\chi}
	\begin{pmatrix}
		e^{+\upsilon}\left[D + \left(D\mathscr{B}_{b,s}\right)\right] \\
		e^{-\upsilon}\left[\triangle + \left(\triangle\mathscr{B}_{b,s}\right)\right] \\
		e^{+i\chi}\left[\delta + \left(\delta\mathscr{B}_{b,s}\right)\right] \\
		e^{-i\chi}\left[\bar{\delta} + \left(\bar{\delta}\mathscr{B}_{b,s}\right)\right]
	\end{pmatrix}\psi_{b,s} \,,\quad \mathscr{B}_{b,s} \defeq b\upsilon + is\chi \,.
\ee
From these, one can construct the GHP ``eth'' operators as
\be\label{eq:GHPethOps}
	\eth = \delta - \left(b-s\right)\bar{\alpha} - \left(b+s\right)\beta \,,\quad \eth^{\prime} = \bar{\delta} - \left(b+s\right)\alpha - \left(b-s\right)\bar{\beta} \,,
\ee
such that they act homogeneously on NP scalars of definite boost-weights and spin weights. In particular, $\eth$ raises the spin weight by one unit, while $\eth^{\prime}$ lowers the spin weight by one unit,
\be\ba
	\eth\psi_{b,s} &\xrightarrow[\upsilon,\chi]{\text{III}} e^{b\upsilon}e^{i\left(s+1\right)\chi}\eth\psi_{b,s} \,, \\
	\eth^{\prime}\psi_{b,s} &\xrightarrow[\upsilon,\chi]{\text{III}} e^{b\upsilon}e^{i\left(s-1\right)\chi}\eth^{\prime}\psi_{b,s} \,.
\ea\ee
Similarly, one can construct the GHP ``thorn'' operators as
\be\label{eq:GHPthronOps}
	\thorn = D - \left(b+s\right)\epsilon - \left(b-s\right)\bar{\epsilon} \,,\quad \thorn^{\prime} = \triangle - \left(b+s\right)\gamma - \left(b-s\right)\bar{\gamma} \,,
\ee
such that $\thorn$ raises the boost-weight by one unit, while $\thorn^{\prime}$ lowers the boost-weight by one unit,
\be\ba
	\thorn\psi_{b,s} &\xrightarrow[\upsilon,\chi]{\text{III}} e^{\left(b+1\right)\upsilon}e^{is\chi}\thorn\psi_{b,s} \,, \\
	\thorn^{\prime}\psi_{b,s} &\xrightarrow[\upsilon,\chi]{\text{III}} e^{\left(b-1\right)\upsilon}e^{is\chi}\thorn^{\prime}\psi_{b,s} \,.
\ea\ee

\subsection{Equations of motion in NP formalism}

\subsubsection{Maxwell fields}
\label{sec:NPeomEM}
The equations of motion for electromagnetic fields, with action\footnote{In this appendix, we are using the canonical field variables, as opposed to the main text of this work where the encountered equations are simplified by the performance of a field redefinition in the form of global rescalings, e.g. $A_{\mu}^{\text{main text}}=\frac{1}{\sqrt{8\pi G}}A_{\mu}^{\text{appendix}}$.}
\be
	S_{\text{Maxwell}} = \int d^{4}x\sqrt{-g}\,\left[-\frac{1}{4}F_{\mu\nu}F^{\mu\nu} - J^{\mu}A_{\mu}\right] \,,
\ee
are the Maxwell field equations $\nabla^{\nu}F_{\nu\mu} = J_{\mu}$ and the Maxwell Bianchi identities $\nabla_{[\rho}F_{\mu\nu]} = 0$. In the NP formalism, these are rearranged into the following $4$ complex equations of motion
\begin{subequations}\label{eq:MaxwelFE1}
	\begin{gather}
		\left(\triangle+1\mu-2\gamma\right)\Phi_0 - \left(\delta-2\tau+0\beta\right)\Phi_1 - \sigma\Phi_2 = - \frac{1}{2}J_{m} \,, \\
		\left(\triangle+2\mu+0\gamma\right)\Phi_1 - \left(\delta-1\tau+2\beta\right)\Phi_2 - \nu\Phi_0 = - \frac{1}{2}J_{n} \,,
	\end{gather}
\end{subequations}
\begin{subequations}\label{eq:MaxwelFE2}
	\begin{gather}
		\left(\bar{\delta}+1\pi-2\alpha\right)\Phi_0 - \left(D-2\rho+0\epsilon\right)\Phi_1  - \kappa\Phi_2 = - \frac{1}{2}J_{\ell} \,, \\
        \left(\bar{\delta}+2\pi+0\alpha\right)\Phi_1 - \left(D-1\rho+2\epsilon\right)\Phi_2 - \sigma\Phi_0 = - \frac{1}{2}J_{\bar{m}} \,.
	\end{gather}
\end{subequations}

These can be written more compactly as
\begin{subequations}
	\begin{equation}
		\begin{split}
			{}&\left[\triangle+\left(2-s\right)\mu-2s\gamma\right]\Phi_{1-s} - \left[\delta-\left(1+s\right)\tau+2\left(1-s\right)\beta\right]\Phi_{2-s} \\
			&\quad- s\sigma\Phi_{3-s} - \left(1-s\right)\nu\Phi_{-s} = J^{\left(1\right)}_{s-1,s} \,,
		\end{split}
	\end{equation}
	\begin{equation}
		\begin{split}
			{}&\left[\bar{\delta}+\left(2-s\right)\pi-2s\alpha\right]\Phi_{1-s} - \left[D-\left(1+s\right)\rho+2\left(1-s\right)\epsilon\right]\Phi_{2-s} \\
			&\quad- s\kappa\Phi_{3-s} - \left(1-s\right)\lambda\Phi_{-s} = J^{\left(1\right)}_{s,s-1} \,,
		\end{split}
	\end{equation}
\end{subequations}
where it is understood that only the range $0 \le s \le +1$ gives non-trivial equations. The terms ``$J^{\left(1\right)}_{b,s}$'' refer to terms of boost-weight $b$ and spin weight $s$ that are switched on in the presence of sources, and can be read directly from~\eqref{eq:MaxwelFE1}-\eqref{eq:MaxwelFE2}, e.g. $J_{0,1}^{\left(1\right)} = -\frac{1}{2}J_{m}$.

Furthermore, the stress-energy momentum tensor can be decomposed in NP scalars in a fashion similar to the Ricci tensor. In particular,
\be
\begin{gathered}
	\text{T}_{00} \defeq \frac{1}{2}T_{\ell\ell} \,,\quad \text{T}_{11} \defeq \frac{1}{4}\left(T_{\ell n} + T_{m\bar{m}}\right) \,,\quad \text{T}_{22} \defeq \frac{1}{2}T_{nn} \,,\quad \Lambda_{\text{T}} \defeq \frac{T}{24} \,, \\
	\text{T}_{01} \defeq \frac{1}{2}T_{\ell m} = \bar{\text{T}}_{10} \,,\quad \text{T}_{12} \defeq \frac{1}{2}T_{nm} = \bar{\text{T}}_{21} \,,\quad \text{T}_{02} \defeq \frac{1}{2}T_{mm} = \bar{\text{T}}_{20} \,.
\end{gathered}
\ee
For electromagnetism, in the absence of sources,
\be
	T_{\mu\nu} = F_{\mu}^{\,\,\,\,\rho}F_{\nu\rho} - \frac{1}{4}g_{\mu\nu}F_{\rho\sigma}F^{\rho\sigma}
\ee
and the stress-energy-momentum-NP scalars acquire the remarkably simple expression in terms of the Maxwell-NP scalars
\be
	\Lambda_{\text{T}} = 0 \,,\quad \text{T}_{\text{a}\text{b}} = \Phi_{\text{a}}\bar{\Phi}_{\text{b}} \,,\quad \text{a},\text{b}\in\left\{0,1,2\right\} \,.
\ee
Then, for general-relativistic gravity with a cosmological constant $\Lambda$, the Einstein-Hilbert equations of motion, $R_{\mu\nu} - \frac{1}{2}g_{\mu\nu}R + \Lambda g_{\mu\nu} = 8\pi GT_{\mu\nu}$, for a purely electromagnetic stress-energy momentum tensor in the absence of electromagnetic sources reduce to the following equations in the NP formalism
\be\label{eq:EFE_Max}
	\Lambda_{\text{R}} = \frac{\Lambda}{6} \,,\quad \Phi_{\text{a}\text{b}} = 8\pi G\text{T}_{\text{a}\text{b}} = 8\pi G\Phi_{\text{a}}\bar{\Phi}_{\text{b}} \,.
\ee
We also remark that if one redefines the Maxwell fields according to $A_{\mu} \rightarrow \frac{1}{\sqrt{8\pi G}}A_{\mu}$, such that the Einstein-Hilbert-Maxwell action in the absence of electromagnetic sources reads $S = \frac{1}{16\pi G}\int d^4x\sqrt{-g}\left[R - 2\Lambda - \frac{1}{2}F_{\mu\nu}F^{\mu\nu}\right]$, then the Einstein-Hilbert equations of motion in the NP formalism reduce to $\Phi_{\text{a}\text{b}} = \Phi_{\text{a}}\bar{\Phi}_{\text{b}}$.

\subsubsection{Yang-Mills fields}
\label{sec:NPeomYM}
For the case of a non-abelian $1$-form gauge fields $A = A_{\mu}^{a}dx^{\mu}\otimes T_{a}$, where $T_{a}\in\mathfrak{g}$ generate the Lie algebra $\mathfrak{g}$ that corresponds to the gauge symmetry Lie group $G$,
\be
	\left[T_{a},T_{b}\right] = if_{ab}^{\,\,\,\,\,c}T_{c} \,,\quad a,b,c \in \left\{1,2,\dots,\dim G\right\} \,,
\ee
the field strength tensor $F = \frac{1}{2}F_{\mu\nu}^{a}dx^{\mu}\wedge dx^{\nu}\otimes T_{a}$ is related to the gauge $1$-form field according to $F = \text{D}\wedge A = \text{d}A - ig_{\text{YM}}\left[A,A\right]$, where $\text{D} = \text{d} - ig_{\text{YM}}\rho\left(A\right)$ is the gauge covariant derivative, with $g_{\text{YM}}$ the gauge coupling constant and $\rho\left(A\right) = A_{\mu}^{a}dx^{\mu}\otimes\rho\left(T_{a}\right)$, $\rho\left(T_{a}\right)$ being the representation of the algebra generators under which the field on which the covariant derivative acts transforms. In components,
\be
	F_{\mu\nu}^{a} = 2\partial_{[\mu}A_{\nu]}^{a} + g_{\text{YM}}f_{bc}^{\,\,\,\,\,a}A_{\mu}^{b}A_{\nu}^{c} \,.
\ee
In the following we will also be working with spacetime tensors that are Lie-algebra-valued, for instance,
\be
	A_{\mu} \defeq A_{\mu}^{a}T_{a} \,,\quad F_{\mu\nu} \defeq F_{\mu\nu}^{a}T_{a} = 2\partial_{[\mu}A_{\nu]} - ig_{\text{YM}}\left[A_{\mu},A_{\nu}\right] \,.
\ee
The action describing the dynamics of the non-abelian gauge field is the Yang-Mills (YM) action\footnote{As with the abelian case, in this appendix we are using the canonical field variables, while in the main text we have simplified the form of equations by rescaling the gauge fields, as well as the gauge coupling constant, according to $A_{\mu}^{\text{main text},a}=\frac{1}{\sqrt{8\pi G}}A_{\mu}^{\text{appendix},a}$ and $g_{\text{YM}}^{\text{main text}} = \sqrt{8\pi G}g_{\text{YM}}^{\text{appendix}}$.},
\be
	S_{\text{YM}} = \int d^{4}x\sqrt{-g}\,\tr\left(-\frac{1}{4}F_{\mu\nu}F^{\mu\nu} - J^{\mu}A_{\mu}\right) \,.
\ee
The trace entering the YM action refers to the trace in the Lie algebra vector space, which can be used to define a notion of ``gauge metric'' $c_{ab}$ according to
\be
	c_{ab} \defeq \tr\left(T_{a}T_{b}\right)  \,,
\ee
such that
\be
	S_{\text{YM}} = \int d^{4}x\sqrt{-g}\,c_{ab}\left[-\frac{1}{4}F_{\mu\nu}^{a}F^{b,\mu\nu} - J^{a,\mu}A_{\mu}^{b}\right] \,.
\ee
One can always choose the gauge metric $c_{ab}$ to be diagonal while, if the gauge group is a compact and simple Lie group, one can always normalize the generators $T_{a}$ in the fundamental representation such that $c_{ab} = \delta_{ab}$, and such that the structure constants $f_{abc} \defeq f_{ab}^{\,\,\,\,\,d}c_{cd}$ are totally antisymmetric in their gauge indices. Here, we will try to be as generic as possible, the only assumption being that there is a Lie algebra $\mathfrak{g}$ associated with the gauge symmetry group $G$.

The equations of motion for the YM field are the following
\be
	D^{\nu}F_{\nu\mu} = J_{\mu} \,,\quad D_{[\rho}F_{\mu\nu]} = 0 \,,
\ee
where $D_{\mu} = \nabla_{\mu} - ig_{\text{YM}}A_{\mu}^{a}\rho\left(T_{a}\right)$ are the spacetime components of the covariant derivative we introduced before. In components form, these read
\be
	\nabla^{\nu}F_{\nu\mu}^{a} + g_{\text{YM}}f_{bc}^{\,\,\,\,\,a}g^{\nu\rho}A_{\rho}^{b}F_{\nu\mu}^{c} = J_{\mu}^{a} \,,\quad \nabla_{[\rho}F_{\mu\nu]}^{a} + g_{\text{YM}}f_{bc}^{\,\,\,\,\,a}A_{[\rho}^{b}F_{\mu\nu]}^{c} = 0 \,.
\ee

To translate these in the NP formalism, the first step is to introduce the $3\times\dim G$ complex YM-NP scalars $\Phi_{1-s}^{a}$, with $s=-1,0,+1$, according to
\be
	\Phi_0^{a} \defeq F_{\ell m}^{a} \,,\quad \Phi_1^{a} \defeq \frac{1}{2}\left(F_{\ell n}^{a}-F_{m\bar{m}}^{a}\right) \,,\quad \Phi_2^{a} \defeq F_{\bar{m}n}^{a} \,.
\ee
It will also be necessary to deal with projections of the gauge field itself, so we also define
\be
	\begin{pmatrix} A_0^{a} \\ A_1^{a} \\ A_2^{a} \\ A_3^{a} \end{pmatrix} \defeq \begin{pmatrix} \ell^{\mu} \\ n^{\mu} \\ m^{\mu} \\ \bar{m}^{\mu} \end{pmatrix}A_{\mu}^{a} \,.
\ee
A technical difference compared to the abelian case is that the gauge field $A_{\mu}$ is in general complex-valued and, hence, $A_3^{a} \ne \left(A_2^{a}\right)^{\ast}$, where ``$\phantom{}^{\ast}$'' refers to complex conjugation of the entire quantity. Nevertheless, there is a sense of complex conjugacy relation in that $A_2^{a}$ and $A_3^{a}$ are obtained from $A_{\mu}^{a}$ by projecting onto $m^{\mu}$ or its complex conjugate, $\bar{m}^{\mu}$, respectively. We will then say that $\bar{A}_2^{a} \defeq A_3^{a}$, but always be mindful that the ``$\bar{\phantom{a}}$'' operation refers to complex conjugacy acting on the tetrad fields, not the gauge fields spacetime vectors. In the following we will often work with the Lie-algebra-valued YM-NP scalars,
\be
	\Phi_{1-s} \defeq \Phi_{1-s}^{a}T_{a} \,,
\ee
and similarly for the gauge field NP scalars. Then, for instance, from the definition of the field strength tensor in terms of the gauge $1$-form field,
\be\ba
	\Phi_0 &= \left[D-\bar{\rho}-\epsilon+\bar{\epsilon}\right]A_2 - \left[\delta+\bar{\pi}-\bar{\alpha}-\beta\right]A_0 + \kappa A_1 - \sigma\bar{A}_2 - ig_{\text{YM}}\left[A_0,A_2\right] \,, \\
	2\Phi_1 &= \left[D+\rho-\bar{\rho}+\epsilon+\bar{\epsilon}\right]A_1 - \left[\triangle-\mu+\bar{\mu}-\gamma-\bar{\gamma}\right]A_0 - ig_{\text{YM}}\left[A_0,A_1\right]\\
	&\quad+ \left[\bar{\delta}-\bar{\tau}-\pi-\alpha+\bar{\beta}\right]A_2 - \left[\delta+\tau+\bar{\pi}-\bar{\alpha}+\beta\right]\bar{A}_2 + ig_{\text{YM}}\left[A_2,\bar{A}_2\right] \,, \\
	\Phi_2 &= \left[\bar{\delta}-\bar{\tau}+\alpha+\bar{\beta}\right]A_1 - \left[\triangle+\bar{\mu}+\gamma-\bar{\gamma}\right]\bar{A}_2 + \nu A_0 - \lambda A_2 - ig_{\text{YM}}\left[A_1,\bar{A}_2\right] \,.
\ea\ee

The YM equations of motion in the NP formalism can be constructed in the same way as for the Maxwell field and can be collectively written as
\begin{subequations}
	\begin{align}
		\begin{split}
			{}&\left[\triangle+\left(2-s\right)\mu-2s\gamma\right]\Phi_{1-s} - \left[\delta-\left(1+s\right)\tau+2\left(1-s\right)\beta\right]\Phi_{2-s} \\
			&-s\sigma\Phi_{3-s} - \left(1-s\right)\nu\Phi_{-s} - ig_{\text{YM}}\left(\left[A_1,\Phi_{1-s}\right]-\left[A_2,\Phi_{2-s}\right]\right) = J^{\left(1\right)}_{s-1,s} \,, \\\\
		\end{split} \label{eq:NPEvolYM}\\
		\begin{split}
			{}&\left[\bar{\delta}+\left(2-s\right)\pi-2s\alpha\right]\Phi_{1-s} - \left[D-\left(1+s\right)\rho+2\left(1-s\right)\epsilon\right]\Phi_{2-s} \\
			&-s\kappa\Phi_{3-s} - \left(1-s\right)\lambda\Phi_{-s} - ig_{\text{YM}}\left(\left[\bar{A}_2,\Phi_{1-s}\right]-\left[A_0,\Phi_{2-s}\right]\right) = J^{\left(1\right)}_{s,s-1} \,,
		\end{split} \label{eq:NPHypersYM}
	\end{align}
\end{subequations}
where $J^{\left(1\right)}_{b,s}$ is (minus one-half of) the boost-weight $b$ and spin weight-$s$ projection of the source $J_{\mu} = J_{\mu}^{a}T_{a}$, and it is understood that only the range $0\le s\le +1$ gives non-trivial equations.

As for the stress-energy-momentum tensor that enters the Einstein field equations, this is functionally the same as for the Maxwell fields, modulo an overall Lie algebra vector space trace. The Einstein field equations are then simply
\be
	\Lambda_{\text{R}} = \Lambda \,,\quad \Phi_{\text{a}\text{b}} = 8\pi G\,\tr\left(\Phi_{\text{a}}\bar{\Phi}_{\text{b}}\right) = 8\pi G\,c_{ab}\Phi_{\text{a}}^{a}\bar{\Phi}_{\text{b}}^{b} \,,
\ee
where it is understood that $\text{a},\text{b}\in\left\{0,1,2\right\}$. As with the non-abelian case, one can always rescale the gauge fields and the gauge coupling constant such that the Einstein field equations simplify to $\Phi_{\text{a}\text{b}} = \tr\left(\Phi_{\text{a}}\bar{\Phi}_{\text{b}}\right) = c_{ab}\Phi_{\text{a}}^{a}\bar{\Phi}_{\text{b}}^{b}$; this is the convention we are using in the main text.

\subsubsection{Gravity}
For the gravitational case, the equations of motion of interest are purely mathematical identities that do not rely on some particular theory gravity, as long as gravity is geometric. These are the differential Bianchi identities $\nabla_{[\lambda}R_{\rho\sigma]\mu\nu} = 0$. They are repackaged into the following $8$ independent Binachi identities within the NP formalism\footnote{Some of these expressions can also be re-expressed by using the twice-contracted Bianchi identities, $\nabla^{\nu}R_{\mu\nu} -\frac{1}{2}\nabla_{\mu}R = 0$ which in the NP formalism read:
	\begin{equation*}
		\begin{split}
			{}&\left(\delta + \bar{\pi} - 2\tau - 2\bar{\alpha}\right)\Phi_{10} - \left(D-2\rho-2\bar{\rho}\right)\Phi_{11} + \left(\bar{\delta}+\pi-2\bar{\tau}-2\alpha\right)\Phi_{01} - \left(\triangle+\mu+\bar{\mu}-2\gamma-2\bar{\gamma}\right)\Phi_{00} \\
			&\quad\quad\quad-3D\Lambda_{\text{R}} -\bar{\kappa}\Phi_{12} - \kappa\Phi_{21} + \bar{\sigma}\Phi_{02} + \sigma\Phi_{20} = 0 \,, \\
			&\left(\delta+2\bar{\pi}-\tau+2\beta\right)\Phi_{21} - \left(D-\rho-\bar{\rho}+2\epsilon+2\bar{\epsilon}\right)\Phi_{22} + \left(\bar{\delta}+2\pi-\bar{\tau}+2\bar{\beta}\right)\Phi_{12} - \left(\triangle+2\mu+2\bar{\mu}\right)\Phi_{11} \\
			&\quad\quad\quad-3\triangle\Lambda_{\text{R}} + \nu\Phi_{01} + \bar{\nu}\Phi_{10} - \lambda\Phi_{02} - \bar{\lambda}\Phi_{20} = 0 \,, \\
			&\left(\delta+2\bar{\pi}-2\tau\right)\Phi_{11} - \left(D-2\rho-\bar{\rho}+2\bar{\epsilon}\right)\Phi_{12} + \left(\bar{\delta}+\pi-\bar{\tau}-2\alpha+2\bar{\beta}\right)\Phi_{02} - \left(\triangle+\mu+2\bar{\mu}-2\gamma\right)\Phi_{01} \\
			&\quad\quad\quad-3\delta\Lambda_{\text{R}} - \kappa\Phi_{22} + \bar{\nu}\Phi_{00} - \bar{\lambda}\Phi_{10} + \sigma\Phi_{21} = 0 \,.
		\end{split}
\end{equation*}}:
\begin{subequations}\label{eq:BianchiIds1}
	\begin{align}
		\begin{split}
			{}&\left(\triangle+1\mu-4\gamma\right)\Psi_0 - \left(\delta-4\tau-2\beta\right)\Psi_1 - 3\sigma\Psi_2 \phantom{+0\Psi_0\quad}  = \\
			&\,\,+\left(D-\bar{\rho}-2\epsilon+2\bar{\epsilon}\right)\Phi_{02} - \left(\delta+2\bar{\pi}-2\beta\right)\Phi_{01} + \bar{\lambda}\Phi_{00} - 2\sigma\Phi_{11} + 2\kappa\Phi_{12} \,,
		\end{split} \\
		\begin{split}
			{}&\left(\triangle+2\mu-2\gamma\right)\Psi_1 - \left(\delta-3\tau+0\beta\right)\Psi_2 - 2\sigma\Psi_3 + 1\nu\Psi_0 = -2\delta\Lambda_{\text{R}} \\
			&\,\,+\left(\bar{\delta}-\bar{\tau}-2\alpha+2\bar{\beta}\right)\Phi_{02} - \left(\triangle+2\bar{\mu}-2\gamma\right)\Phi_{01} -2\tau\Phi_{11}+\bar{\nu}\Phi_{00} + 2\rho\Phi_{12} \,,
		\end{split} \\
		\begin{split}
			{}&\left(\triangle+3\mu+0\gamma\right)\Psi_2 - \left(\delta-2\tau+2\beta\right)\Psi_3 - 1\sigma\Psi_4 - 2\nu\Psi_1 = + 2\triangle\Lambda_{\text{R}} \\
			&\,\,+\left(D-\bar{\rho}+2\epsilon+2\bar{\epsilon}\right)\Phi_{22}-\left(\delta+2\bar{\pi}+2\beta\right)\Phi_{21} - 2\pi\Phi_{12} + 2\mu\Phi_{11} + \bar{\lambda}\Phi_{20} \,,
		\end{split} \\
		\begin{split}
			{}&\left(\triangle+4\mu+2\gamma\right)\Psi_3 - \left(\delta-1\tau+4\beta\right)\Psi_4 \phantom{- 1\sigma\Psi_4\,\,\,} - 3\nu\Psi_2 = \\
			&\,\,+\left(\bar{\delta}-\bar{\tau}+2\alpha+2\bar{\beta}\right)\Phi_{22} - \left(\triangle+2\bar{\mu}+2\gamma\right)\Phi_{21} + 2\nu\Phi_{11} - 2\lambda\Phi_{12} + \bar{\nu}\Phi_{20} \,,
		\end{split}
	\end{align}
\end{subequations}
\begin{subequations}\label{eq:BianchiIds2}
	\begin{align}
		\begin{split}
			{}&\left(\bar{\delta}+1\pi-4\alpha\right)\Psi_0 - \left(D-4\rho-2\epsilon\right)\Psi_1 - 3\kappa\Psi_2 \phantom{+0\Psi_4\quad} = \\
			&\,\,+\left(D-2\bar{\rho}-2\epsilon\right)\Phi_{01} - \left(\delta+\bar{\pi}-2\bar{\alpha}-2\beta\right)\Phi_{00} +2\kappa\Phi_{11} - 2\sigma\Phi_{10} + \bar{\kappa}\Phi_{02} \,,
		\end{split} \\
		\begin{split}
			{}&\left(\bar{\delta}+2\pi-2\alpha\right)\Psi_1 - \left(D-3\rho+0\epsilon\right)\Psi_2 - 2\kappa\Psi_3 - 1\lambda\Psi_0 = - 2D\Lambda_{\text{R}} \\
			&\,\,+\left(\bar{\delta}-2\bar{\tau}-2\alpha\right)\Phi_{01} -\left(\triangle+\bar{\mu}-2\gamma-2\bar{\gamma}\right)\Phi_{00} + 2\rho\Phi_{11} - 2\tau\Phi_{10} + \bar{\sigma}\Phi_{02} \,,
		\end{split} \\
		\begin{split}
			{}&\left(\bar{\delta}+3\pi-0\alpha\right)\Psi_2 - \left(D-2\rho+2\epsilon\right)\Psi_3 - 1\kappa\Psi_4 - 2\lambda\Psi_1 = + 2\bar{\delta}\Lambda_{\text{R}} \\
			&\,\,+\left(D-2\bar{\rho}+2\epsilon\right)\Phi_{21} - \left(\delta+\bar{\pi}-2\bar{\alpha}+2\beta\right)\Phi_{20} + \bar{\kappa}\Phi_{22} - 2\pi\Phi_{11} + 2\mu\Phi_{10} \,,
		\end{split} \\
		\begin{split}
			{}&\left(\bar{\delta}+4\pi+2\alpha\right)\Psi_3 - \left(D-1\rho+4\epsilon\right)\Psi_4 \phantom{- 2\kappa\Psi_3\,\,\,} - 3\lambda\Psi_2 = \\
			&\,\,+\left(\bar{\delta}-2\bar{\tau}+2\alpha\right)\Phi_{21} -\left(\triangle+\bar{\mu}+2\gamma-2\bar{\gamma}\right)\Phi_{20} + 2\nu\Phi_{10} + \bar{\sigma}\Phi_{22} - 2\lambda\Phi_{11} \,,
		\end{split}
	\end{align}
\end{subequations}

The above equations can also be written more compactly as
\begin{subequations}
	\begin{equation}
		\begin{split}
			{}&\left[\triangle+\left(3-s\right)\mu-2s\gamma\right]\Psi_{2-s} - \left[\delta-\left(2+s\right)\tau+2\left(1-s\right)\beta\right]\Psi_{3-s} \\
			&\quad- \left(1+s\right)\sigma\Psi_{4-s} - \left(2-s\right)\nu\Psi_{1-s} = J^{\left(2\right)}_{s-1,s} \,,
		\end{split}
	\end{equation} \\
	\begin{equation}
		\begin{split}
			{}&\left[\bar{\delta}+\left(3-s\right)\pi-2s\alpha\right]\Psi_{2-s} - \left[D-\left(2+s\right)\rho+2\left(1-s\right)\epsilon\right]\Psi_{3-s} \\
			&\quad- \left(1+s\right)\kappa\psi_{4-s} - \left(2-s\right)\lambda\Psi_{1-s} = J^{\left(2\right)}_{s,s-1} \,,
		\end{split}
	\end{equation}
\end{subequations}
where it is understood that only the range $-1 \le s \le +2$ gives non-trivial equations and ``$J^{\left(2\right)}_{b,s}$'' refers to terms of boost-weight $b$ and spin weight $s$ that are switched on in the presence of gravitational sources, i.e. for non-Ricci-flat geometries. These sources can also be written in a collective form, after splitting the above equations into two overlapping branches, one for $0 \le s \le +2$ and one for $-1 \le s \le +1$ (with overlap for $0 \le s \le +1$). For $0 \le s \le +2$,
\begin{subequations}
	\begin{align}
		\begin{split}
			{}&J_{s-1,s}^{\left(2\right)} = s\left[\kappa\Phi_{3-s,2}-\sigma\Phi_{3-s,1}\right] - \left(s-2\right)\left[\nabla_{s-1s,s}\Lambda_{\text{R}} - \pi\Phi_{1-s,2} + \mu\Phi_{1-s,1}\right] \\
			&\quad\quad+\left[D-\bar{\rho}+2\left(1-s\right)\epsilon+2\bar{\epsilon}\right]\Phi_{2-s,2} - \left[\delta+2\bar{\pi}+2\left(1-s\right)\beta\right]\Phi_{2-s,1} + \bar{\lambda}\Phi_{2-s,0} \,,
		\end{split} \\
		\begin{split}
			{}&J_{s,s-1}^{\left(2\right)} = s\left[\kappa\Phi_{3-s,1}-\sigma\Phi_{3-s,0}\right] - \left(s-2\right)\left[\nabla_{s,s-1}\Lambda_{\text{R}} - \pi\Phi_{1-s,1} + \mu\Phi_{1-s,0}\right] \\
			&\quad\quad+\bar{\kappa}\Phi_{2-s,2} + \left[D-2\bar{\rho}+2\left(1-s\right)\epsilon\right]\Phi_{2-s,1} - \left[\delta+\bar{\pi}-2\bar{\alpha}+2\left(1-s\right)\beta\right]\Phi_{2-s,0} \,,
		\end{split}
	\end{align}
\end{subequations}
while, for $-1 \le s \le +1$,
\begin{subequations}
	\begin{align}
		\begin{split}
			{}&J_{s-1,s}^{\left(2\right)} = \left(s-1\right)\left[\lambda\Phi_{-s,2}-\nu\Phi_{-s,1}\right] - \left(s+1\right)\left[\nabla_{s-1,s}\Lambda_{\text{R}} - \rho\Phi_{2-s,2} + \tau\Phi_{2-s,1}\right] \\
			&\quad\quad+\left[\bar{\delta}-\bar{\tau}-2s\alpha+2\bar{\beta}\right]\Phi_{1-s,2} - \left[\triangle+2\bar{\mu}-2s\gamma\right]\Phi_{1-s,1} + \bar{\nu}\Phi_{1-s,0} \,,
		\end{split} \\
		\begin{split}
			{}&J_{s,s-1}^{\left(2\right)} = \left(s-1\right)\left[\lambda\Phi_{-s,1}-\nu\Phi_{-s,0}\right] - \left(s+1\right)\left[\nabla_{s,s-1}\Lambda_{\text{R}}-\rho\Phi_{2-s,1}+\tau\Phi_{2-s,0}\right] \\
			&\quad\quad+\bar{\sigma}\Phi_{1-s,2} + \left[\bar{\delta}-2\bar{\tau}-2s\alpha\right]\Phi_{1-s,1} - \left[\triangle+\bar{\mu}-2s\gamma-2\bar{\gamma}\right]\Phi_{1-s,0} \,.
		\end{split}
	\end{align}
\end{subequations}
In the above equations, $\nabla_{b,s}$, with $-1 \le s,b \le +1$ and $\left|s\right|\ne\left|b\right|$, is the directional derivative of boost-weight $b$ and spin weight $s$, namely, $\nabla_{+1,0} = D$, $\nabla_{-1,0} = \triangle$, $\nabla_{0,+1} = \delta$ and $\nabla_{0,-1} = \bar{\delta}$.

\section{Various identities in pseudo-differential calculus}
\label{app:MathIds}

In this appendix, we collect a number of mathematical identities that have proved to be useful in deriving the $w_{1+\infty}$ charge algebra. First of all, we are using the iterated anti-derivative operator $\partial_{u}^{-n}$, $n\ge0$, defined as the $n$'th repeated integral with base point $+\infty$~\cite{Freidel:2021dfs,Geiller:2024bgf}
\be
	\left(\partial_{u}^{-n}\mathcal{F}\right)\left(u\right) \defeq \int_{+\infty}^{u}du_1\int_{+\infty}^{u_{1}}du_{2}\dots \int_{+\infty}^{u_{n-1}}du_{n}\,\mathcal{F}\left(u_{n}\right)
\ee
for any scalar function $\mathcal{F}\left(u\right)$. This is a well-defined operation as long as
\be
	\mathcal{F}\left(u\right) = o\left(u^{-n}\right) \quad\text{as $u\rightarrow+\infty$} \,.
\ee
The action of the iterated anti-derivative operator can be equivalently written as the single integral
\be
	\left(\partial_{u}^{-n}\mathcal{F}\right)\left(u\right) = \int_{+\infty}^{u}du^{\prime}\frac{\left(u-u^{\prime}\right)^{n-1}}{\left(n-1\right)!}\mathcal{F}\left(u^{\prime}\right)
\ee
by virtue of the Cauchy formula for repeated integration. In the following, and also in the main text, we loosen our notation and write $\partial_{u}^{-n}\mathcal{F}\left(u\right)$ in place of $\left(\partial_{u}^{-n}\mathcal{F}\right)\left(u\right)$. The anti-derivative operator obeys the generalized integral Leibniz rule,
\be\label{eq:IntLR1}
	\partial_{u}^{-1}\left(f\left(u\right)g\left(u\right)\right) = \sum_{n=0}^{\infty}(-1)^{n}\left(\partial_{u}^{n}f\left(u\right)\right)\left(\partial_{u}^{-(n+1)}g\left(u\right)\right) \,,
\ee
which, in particular, implies that
\be\label{eq:IntLR2}
	\partial_{u}^{-1}\left(\frac{(-u)^{s}}{s!}f\left(u\right)\right) = \sum_{n=0}^{s}\frac{(-u)^{n}}{n!}\partial_{u}^{-\left(s-n+1\right)}f\left(u\right) \,.
\ee
This can be generalized to
\be\label{eq:AntiDervProdL}
    \partial_{u}^{-\ell}\left(\frac{(-u)^{s}}{s!}f\left(u\right)\right) = \sum_{n=0}^{s}\binom{s-n+\ell-1}{\ell-1}\frac{(-u)^{n}}{n!}\partial_{u}^{-\left(s-n+\ell\right)}f\left(u\right) \,,
\ee
for all $\ell\in\mathbb{N}$.

We will also make frequent use of the following distributional identities involving the $\delta$-functional~\cite{Freidel:2021ytz},
\begin{subequations}
	\begin{align}
		\partial_{u}^{n}\delta(u-u^{\prime}) &= (-1)^{n}\partial_{u^{\prime}}^{n}\delta(u-u^{\prime}) \,, \label{eq:DeltaId1} \\
		f\left(u\right)\partial_{u}^{n}\delta(u-u^{\prime}) &= (-1)^{n}\partial_{u^{\prime}}^{n}f\left(u^{\prime}\right)\delta(u-u^{\prime}) \,, \label{eq:DeltaId2} \\
        \begin{split}
		  \partial_{u}^{-n}\delta(u-u^{\prime}) &= \frac{\left(u-u^{\prime}\right)^{n-1}}{\left(n-1\right)!}\partial_{u}^{-1}\delta(u-u^{\prime}) \\
		  &= -\frac{\left(u-u^{\prime}\right)^{n-1}}{\left(n-1\right)!}\,\theta\left(u^{\prime}-u\right) \,,
        \end{split} \label{eq:DeltaId3}
	\end{align}
\end{subequations}
with $n\in\mathbb{N}$ and where $\theta\left(x\right)$ is the Heaviside step function. Using~\eqref{eq:DeltaId1}-\eqref{eq:DeltaId2} and the distributional property of the derivative operator, one can then show that
\be\ba
	f\left(u\right)\partial_{u}^{n}\delta(u-u^{\prime}) &= (-1)^{n}f\left(u\right)\partial_{u^{\prime}}^{n}\delta(u-u^{\prime}) \\
	&= (-1)^{n}\partial_{u^{\prime}}^{n}\left(f\left(u\right)\delta(u-u^{\prime})\right) \\
	&= (-1)^{n}\partial_{u^{\prime}}^{n}\left(f\left(u^{\prime}\right)\delta(u-u^{\prime})\right) \\
	&= (-1)^{n}\sum_{m=0}^{n}\binom{n}{m}\partial_{u^{\prime}}^{m}f\left(u^{\prime}\right)\partial_{u^{\prime}}^{n-m}\delta(u-u^{\prime}) \\
	&= \sum_{m=0}^{n}(-1)^{m}\binom{n}{m}\partial_{u^{\prime}}^{m}f\left(u^{\prime}\right)\partial_{u}^{n-m}\delta(u-u^{\prime}) \,.
\ea\ee
This will be applied in the main text in the form
\be\label{eq:DiracId4}
	\eth_{z}^{s-\ell}\left(f(z)\eth_{z}^{\ell-m}\delta\left(z-z^{\prime}\right)\right) = \sum_{n=m}^{\ell}(-1)^{n-m}\binom{\ell-m}{n-m}\eth_{z^{\prime}}^{n-m}f(z^{\prime})\eth_{z}^{s-n}\delta\left(z-z^{\prime}\right)
\ee
to find the action of the celestial charges on the gravitational shear and the gluon field.

Last, the following manipulations have proved to be particularly useful~\cite{Freidel:2021ytz}
\begin{subequations}\label{eq:manip}
	\begin{align}
		u^{n}\partial_{u}^{n} &= (\Delta_{u}-1)_{n} \,, \\
		\partial_{u}^{n}u^{n} &= (\Delta_{u}+n-1)_{n} \,, \\
		u^{-n}\partial_{u}^{-n} &= (\Delta_{u}+n-1)_{n}^{-1} \,, \\
		\partial_{u}^{k}(\Delta_{u}+\alpha)_{n} &= (\Delta_{u}+\alpha+k)_{n}\partial_{u}^{k} \,, \\
		u^{k}(\Delta_{u}+\alpha)_{n}^{\pm1} &= (\Delta_{u}+\alpha-k)_{n}^{\pm1}u^{k} \,,
	\end{align}
\end{subequations}
with $n\in\mathbb{N}$, $k\in\mathbb{Z}$ and $\alpha\in\mathbb{R}$, and where $(x)_{n} \defeq x\left(x-1\right)\dots\left(x-n+1\right)$, $(x)_0=1$, is the falling factorial, while $\Delta_{u} \defeq u\partial_{u} + 1$. A nice property of the operator $\Delta_{u}$ is that it, as well as any analytic function of it, integrate to zero, when integrated over the entire real line,
\be
	\int_{-\infty}^{+\infty}du \, \Delta_{u}g\left(u\right) = 0 \,,
\ee
assuming the function $g\left(u\right)$ falls off sufficiently fast at large $u$, namely, $g\left(u\right)=o\left(u^{-1}\right)$ above. For the cases encountered in this work, namely, for the fall-conditions of the radiative fields $\bar{N}$ and $\bar{F}$, this is indeed the case for all the integrals involved in deriving the action of the celestial charges on the sub$^{s}$-leading soft operators, and allows to simply set $\Delta_{u} = 0$ under such expressions,
\be
	\int_{-\infty}^{+\infty}du\,f\left(\Delta_{u}\right)g\left(u\right) = f\left(0\right)\int_{-\infty}^{+\infty}du\,g\left(u\right) \,.
\ee

\section{Proof of quasi-conservation equation}
\label{app:tqsCons}
In this appendix we derive the evolution equation~\eqref{eq:EvolutionstqsEYM} for $\tilde{q}_{s}^{\left(j\right)}$, where we are adopting the $j$-index notation introduced in Section~\ref{sec:w1pInftyEYM} to treat all types of fields at the same time.

To do this, let us first see how the following renormalized charges
\be
	\hat{q}_{s}^{\left(j\right)} \defeq \sum_{n=0}^{s}\alpha_{n}\eth^{n}\mathcal{Q}_{s-n}^{\left(j\right)}
\ee
evolve, where $\alpha_{n}(u) \defeq \frac{(-u)^{n}}{n!}$. These are the charges that were first introduced in~\cite{Freidel:2021ytz} and they are linearly conserved in the absence of radiation. Indeed, using the evolution equations~\eqref{eq:EvolutionQsEYM}, we have
\be\ba
	\partial_{u}\hat{q}_{s}^{\left(j\right)} &= -\sum_{n=0}^{s-1}\alpha_{n}\eth^{n+1}\mathcal{Q}_{s-n-1}^{\left(j\right)} + \sum_{n=0}^{s}\alpha_{n}\eth^{n+1}\mathcal{Q}_{s-n-1}^{\left(j\right)} \\
	&\quad+ \sum_{n=0}^{s}\alpha_{n}\eth^{n}\left(- ig_{\text{YM}}\left[A,\mathcal{Q}_{s-n-1}^{\left(j\right)}\right] + \left(j+s-n-1\right)\sum_{i=0}^{j-1}\sigma^{\left(2-i\right)}\bullet\mathcal{Q}_{s-n-2+i}^{\left(j-i\right)}\right) \\
	&= \alpha_{s}\eth^{s}\partial_{u}\mathcal{Q}_0^{\left(j\right)} -ig_{\text{YM}}\sum_{\ell=1}^{s}\alpha_{s-\ell}\eth^{s-\ell}\left[A,\mathcal{Q}_{\ell-1}^{\left(j\right)}\right] \\
	&\quad+\sum_{\ell=1}^{s}\left(\ell+j-1\right)\alpha_{s-\ell}\eth^{s-\ell}\sum_{i=0}^{j-1}\sigma^{\left(2-i\right)}\bullet\mathcal{Q}_{\ell-2+i}^{\left(j-i\right)} \,,
\ea\ee
where, in the first equality, $\partial_{u}$ has acted on $\alpha_{n}$ in the first term (where we have changed the summation index from $n$ to $n+1$) and on $\mathcal{Q}_{s-n}^{\left(j\right)}$ in the remaining terms (where we have used the evolution equations~\eqref{eq:EvolutionQsEYM} and isolated the linear piece in the first line, with the remaining quadratic terms displayed in the second line). In the second equality, we have isolated the $n=s$ term from the first equality (which we restored to its initial expression as $\partial_{u}\mathcal{Q}_0^{\left(j\right)}$ for space economy) to take advantage of the cancellation occurring among the $\eth^{n+1}\mathcal{Q}_{s-n-1}^{\left(j\right)}$ terms for $0\le n\le s-1$, while we have also changed the summation index from $n$ to $\ell = s-n$ in the remaining terms that are quadratic in the fields. Since $\partial_{u}\mathcal{Q}_0^{\left(j\right)} \overset{\text{non-rad}}{=} 0$, we see then that $\hat{q}_{s}^{\left(j\right)}$ is linearly quasi-conserved, i.e. it is linearly conserved in the absence of radiation. Next, we consider the following quadratically corrected renormalized charges
\be\ba
	{}&\tilde{q}_{s}^{\left(j\right)} = \hat{q}_{s}^{\left(j\right)} -ig_{\text{YM}}\sum_{\ell=1}^{s}\eth^{s-\ell}\sum_{n=0}^{\ell-1}\left[\left(-\partial_{u}\right)^{-(n+1)}\left(\alpha_{s-\ell}A\right),\eth^{n}\mathcal{Q}_{\ell-n-1}^{\left(j\right)}\right] \\
	&\quad+ \sum_{i=0}^{j-1}\sum_{\ell=2-i}^{s}\left(\ell+j-1\right)\eth^{s-\ell}\sum_{n=0}^{\ell-2+i}\left(-\partial_{u}\right)^{-(n+1)}\left(\alpha_{s-\ell}\sigma^{\left(2-i\right)}\right)\bullet\eth^{n}\mathcal{Q}_{\ell-n-2+i}^{\left(j-i\right)} \,.
\ea\ee
These evolve according to
\be\ba
	{}&\partial_{u}\tilde{q}_{s}^{\left(j\right)} = \partial_{u}\hat{q}_{s}^{\left(j\right)} +ig_{\text{YM}}\sum_{\ell=1}^{s}\eth^{s-\ell}\sum_{n=-1}^{\ell-2}\left[\left(-\partial_{u}\right)^{-(n+1)}\left(\alpha_{s-\ell}A\right),\eth^{n+1}\mathcal{Q}_{\ell-n-2}^{\left(j\right)}\right] \\
	&\quad- \sum_{i=0}^{j-1}\sum_{\ell=2-i}^{s}\left(\ell+j-1\right)\eth^{s-\ell}\sum_{n=-1}^{\ell-3+i}\left(-\partial_{u}\right)^{-(n+1)}\left(\alpha_{s-\ell}\sigma^{\left(2-i\right)}\right)\bullet\eth^{n+1}\mathcal{Q}_{\ell-n-3+i}^{\left(j-i\right)} \\
	&\quad-ig_{\text{YM}}\sum_{\ell=1}^{s}\eth^{s-\ell}\sum_{n=0}^{\ell-1}\left[\left(-\partial_{u}\right)^{-(n+1)}\left(\alpha_{s-\ell}A\right),\eth^{n+1}\mathcal{Q}_{\ell-n-2}^{\left(j\right)}\right] \\
	&\quad+\sum_{i=0}^{j-1}\sum_{\ell=2-i}^{s}\left(\ell+j-1\right)\eth^{s-\ell}\sum_{n=0}^{\ell-2+i}\left(-\partial_{u}\right)^{-(n+1)}\left(\alpha_{s-\ell}\sigma^{\left(2-i\right)}\right)\bullet\eth^{n+1}\mathcal{Q}_{\ell-n-3+i}^{\left(j-i\right)} + \mathcal{O}\left(\mathbb{F}^3\right) \,,
\ea\ee
where, in the first two lines, $\partial_{u}$ has acted on $\left(-\partial_{u}\right)^{-(n+1)}$ and we have shifted the summation index from $n$ to $n+1$, while, in the third and fourth lines, it has acted on the $\mathcal{Q}$'s and we have used \eqref{eq:EvolutionQsEYM} to write $\partial_{u}\mathcal{Q}_{s}^{\left(j\right)} = \eth\mathcal{Q}_{s-1}^{\left(j\right)} + \mathcal{O}\left(\mathbb{F}^2\right)$. In the above, we are suppressing terms that are of cubic order in the fields. The $n=-1$ terms of the sums are precisely the ones that cancel with the non-radiative quadratic terms in $\partial_{u}\hat{q}_{s}^{\left(j\right)}$, while all of the remaining terms in the $n$-sums cancel with each other, except for $n=\ell-1$ and $n=\ell-2+i$ in the third and fourth lines respectively. In the end, we are left with
\be\ba
	{}&\partial_{u}\tilde{q}_{s}^{\left(j\right)} = \alpha_{s}\eth^{s}\partial_{u}\mathcal{Q}_0^{\left(j\right)} - ig_{\text{YM}}\sum_{\ell=1}^{s}\eth^{s-\ell}\left[\left(-\partial_{u}\right)^{-\ell}\left(\alpha_{s-\ell}A\right),\eth^{\ell}\mathcal{Q}_{-1}^{\left(j\right)}\right] \\
	&\quad+\sum_{\ell=1}^{s}\left(\ell+j-1\right)\eth^{s-\ell}\sum_{i=0}^{j-1}\left(-\partial_{u}\right)^{1-\ell-i}\left(\alpha_{s-\ell}\sigma^{\left(2-i\right)}\right)\bullet\eth^{\ell-1+i}\mathcal{Q}_{-1}^{\left(j-i\right)} + \mathcal{O}\left(\mathbb{F}^3\right) \,,
\ea\ee
which is the quasi-conservation equation we wanted to derive. In non-radiative configurations, all the terms that contribute above vanish identically, up to cubic corrections which are out of the scope of this work.

\section{Brackets between anti-derivatives of fundamental fields}
\label{app:CanBraAntiDer}

In this appendix, we collect a list of the Poisson brackets between the fundamental fields and their anti-derivatives that are needed for finding the action of the celestial charges onto the photon/gluon field and the gravitational shear.

From the Einstein-YM action functional~\eqref{eq:ActionEYM}, the symplectic structure on $\scri^{+}$ can be worked out to be
\be\ba
	\Omega_{\scri^{+}} &= \frac{1}{\kappa^2}\oint\int_{-\infty}^{+\infty}du\left[\frac{1}{4}\delta C_{AB}\wedge\delta N^{AB} + \tr\left(\delta A_{A}^{\left(0\right)}\wedge\delta\partial_{u}A^{\left(0\right)A}\right)\right] \\
	&= \frac{1}{\kappa^2}\oint\int_{-\infty}^{+\infty}du\left[\delta C\wedge\delta\bar{N} + \delta_{ab}\delta A^{a}\wedge\delta\bar{F}^{b} + \left(\text{c.c.}\right) \right] \\
	&=  \frac{1}{\kappa^2}\oint\int_{-\infty}^{+\infty}du\sum_{j=1}^{2}\left(\delta C^{\left(j\right)}\wedge\delta\bar{N}^{\left(j\right)} + \left(\text{c.c.}\right)\right) \,,
\ea\ee
from which we extract the following fundamental Poisson brackets
\be\ba
	&\left\{C(u,z),\bar{N}(u^{\prime},z^{\prime})\right\} = \kappa^2\,\delta(u-u^{\prime})\delta(z,z^{\prime}) \,, \\
	&\left\{A^{a}(u,z),\bar{F}^{b}(u^{\prime},z^{\prime})\right\} = \kappa^2\,\delta(u-u^{\prime})\delta(z,z^{\prime}) \delta^{ab}\,, 
\ea\ee
or, more compactly, using the spacetime-spin-$j$ notation developed in Section~\ref{sec:w1pInftyEYM},
\be
	\left\{C^{\left(j\right)}(u,z),\bar{N}^{\left(j^{\prime}\right)}(u^{\prime},z^{\prime})\right\} = \kappa^2\,\delta(u-u^{\prime})\delta(z,z^{\prime})\delta^{j,j^{\prime}}\mathbf{1}_{\text{c}}^{-1} \,,
\ee
where $\mathbf{1}_{\text{c}}^{-1}$ refers to the color space inverse metric structure as dictated by the color of the objects involved in the bracket. The above brackets also come with their complex conjugate pair for the opposite helicity quantities,
\be
	\left\{\bar C^{\left(j\right)}(u,z),N^{\left(j^{\prime}\right)}(u^{\prime},z^{\prime})\right\} = \kappa^2\,\delta(u-u^{\prime})\delta(z,z^{\prime})\delta^{j,j^{\prime}}\mathbf{1}_{\text{c}}^{-1}  \,.
\ee
It will also be useful to have at hand the Poisson bracket among $C^{\left(j\right)}$'s. Integrating the above, we get
\be
	\left\{C^{\left(j\right)}(u,z),\bar{C}^{\left(j^{\prime}\right)}(u^{\prime},z^{\prime})\right\} = -\frac{\kappa^2}{2}\,\Theta(u-u^{\prime})\delta(z,z^{\prime})\delta^{j,j^{\prime}}\mathbf{1}_{\text{c}}^{-1} \,,
\ee
where $\Theta\left(t\right) = \frac{1}{\pi i}\int_{-\infty}^{+\infty}\frac{d\omega}{\omega}e^{i\omega t}$, with $\Theta\left(t\ne0\right)=\text{sign}\left\{t\right\}$\footnote{Notice it satisfies $\frac{d}{dt}\Theta\left(t\right) = 2\delta\left(t\right)$.}. More generally,
\be\ba
    \left\{C^{\left(j\right)}(u,z),\partial_{u^{\prime}}^{-n}\bar{N}^{\left(j^{\prime}\right)}(u^{\prime},z^{\prime})\right\} &= \frac{\kappa^2}{2}\left(\partial_{u^{\prime}}^{-n}+\left(-1\right)^{n}\partial_{u}^{-n}\right)\delta(u-u^{\prime})\delta(z,z^{\prime})\delta^{j,j^{\prime}}\mathbf{1}_{\text{c}}^{-1} \\
    &= -\frac{\kappa^2}{2}\frac{\left(u^{\prime}-u\right)^{n-1}}{\left(n-1\right)!}\Theta(u-u^{\prime})\delta(z,z^{\prime})\delta^{j,j^{\prime}}\mathbf{1}_{\text{c}}^{-1} \,,
\ea\ee
where in the last line we have used \eqref{eq:DeltaId3}.

To derive this bracket, let us focus without loss of generality to the $j=j^{\prime}=2$ case. Our first step is then to notice that, for generic $n\ge0$,
\be\ba
    \left\{C(u,z),\bar{N}(u^{\prime},z^{\prime})\right\} &= \left\{C(u,z),\partial_{u^{\prime}}^{n}\partial_{u^{\prime}}^{-n}\bar{N}(u^{\prime},z^{\prime})\right\} \\
    &= \partial_{u^{\prime}}^{n}\left\{C(u,z),\partial_{u^{\prime}}^{-n}\bar{N}(u^{\prime},z^{\prime})\right\} \,,
\ea\ee
which can be integrated to get
\be\ba
    \left\{C(u,z),\partial_{u^{\prime}}^{-n}\bar{N}(u^{\prime},z^{\prime})\right\} &= \partial_{u^{\prime}}^{-n}\left\{C(u,z),\bar{N}(u^{\prime},z^{\prime})\right\} + \kappa^2c_{n}\left(u-u^{\prime}\right)\delta(z,z^{\prime}) \\
    &= \kappa^2\delta(z,z^{\prime})\left(\partial_{u^{\prime}}^{-n}\delta\left(u-u^{\prime}\right) + c_{n}\left(u-u^{\prime}\right)\right) \\
    &= -\kappa^2\delta(z,z^{\prime})\left(\frac{\left(u^{\prime}-u\right)^{n-1}}{\left(n-1\right)!}\theta\left(u-u^{\prime}\right) - c_{n}\left(u-u^{\prime}\right)\right) \,,
\ea\ee
for some integration functions $c_{n}\left(u-u^{\prime}\right)$ such that $\partial_{u^{\prime}}^{n}c_{n}\left(u-u^{\prime}\right) = 0$. Acting with $\partial_{u^{\prime}}$ on the above equation, one then sees that the integration functions satisfy the recursion relation
\be
    \partial_{u^{\prime}}c_{n}\left(u-u^{\prime}\right) = c_{n-1}\left(u-u^{\prime}\right) \,,
\ee
with $c_0\left(u-u^{\prime}\right) = 0$ set by the already known canonical bracket. Furthermore, we know from the antisymmetry of the $\left\{C(u,z),\bar{C}(u^{\prime},z^{\prime})\right\}$ bracket that
\be
    c_1\left(u-u^{\prime}\right) = \frac{1}{2} = \frac{1}{2}\left(\theta\left(u-u^{\prime}\right)+\theta\left(u^{\prime}-u\right)\right) \,.
\ee
The general solution to the above recursion relation is then
\be
    c_{n}\left(u-u^{\prime}\right) = \frac{1}{2}\frac{\left(u^{\prime}-u\right)^{n-1}}{\left(n-1\right)!} + \sum_{m=0}^{n-2}a_{m}\frac{\left(u^{\prime}-u\right)^{m}}{m!} \,,
\ee
where $a_{m}$ is the integration constant that enters when integrating the recursion relation for $n=m$. We choose to set all of these integration constants to be zero, e.g. by imposing the parity condition $c_{n}\left(u^{\prime}-u\right) = \left(-1\right)^{n-1}c_{n}\left(u-u^{\prime}\right)$,
\be
    c_{n}\left(u-u^{\prime}\right) = \frac{1}{2}\frac{\left(u^{\prime}-u\right)^{n-1}}{\left(n-1\right)!} = \frac{1}{2}\left(\partial_{u^{\prime}}^{-n}-\left(-1\right)^{n}\partial_{u}^{-n}\right)\delta\left(u-u^{\prime}\right) \,,
\ee
which results into
\be\ba
    \left\{C(u,z),\partial_{u^{\prime}}^{-n}\bar{N}(u^{\prime},z^{\prime})\right\} &= \frac{\kappa^2}{2}\left(\partial_{u^{\prime}}^{-n}+\left(-1\right)^{n}\partial_{u}^{-n}\right)\delta(u-u^{\prime})\delta(z,z^{\prime}) \\
    &= -\frac{\kappa^2}{2}\frac{\left(u^{\prime}-u\right)^{n-1}}{\left(n-1\right)!}\Theta(u-u^{\prime})\delta(z,z^{\prime})
\ea\ee
as promised. Fixing these integration constants as above is consistent with the resulting action of the celestial charges on the negative helicity gravitational shear that were reported in~\cite{Freidel:2021ytz,Geiller:2024bgf}.

\addcontentsline{toc}{section}{References}
\bibliographystyle{JHEP}
\bibliography{references}

\end{document}